\documentclass[12pt]{article} 
\usepackage{graphicx,subfigure,latexsym}
\usepackage{amsmath,amssymb,amsfonts}
\usepackage{bm}

\newcommand{\be}{\begin{equation}}
\newcommand{\ee}{\end{equation}}
\newcommand{\bear}{\begin{eqnarray}}
\newcommand{\eear}{\end{eqnarray}}
\newcommand{\ba}{\begin{array}}
\newcommand{\ea}{\end{array}}

\DeclareMathOperator{\sech}{sech}

\def \be {\begin{equation}}
\def \ee {\end{equation}}

\def \bes {\begin{subequations}}
\def \ees {\end{subequations}}

\def \<{\langle}
\def \>{\rangle}
\def \+{\dagger}
\def \({\left(}
\def \){\right)}
\def \[{\left[}
\def \]{\right]}

\def \vp {\bm{p}}
\def \vq {\bm{q}}

\def \vk {\bm{k}}

\def \vsigma{\bm{\sigma}}

\def \Im {\text{Im}}

\begin{document}

\begin{titlepage}
\vfill
\begin{flushright}
{\normalsize RBRC-1129\\\normalsize IFT-UAM/CSIC-15-035}\\
\end{flushright}

\vfill
\begin{center}
{\Large\bf  Second Order Transport Coefficient \\ from Chiral Anomaly at Weak Coupling: Diagrammatic Resummation }
\vskip 0.3in

\vskip 0.3in
Amadeo Jimenez-Alba$^{1}$\footnote{e-mail: {\tt amadeo.j@gmail.com}} and 
Ho-Ung Yee$^{2,3}$\footnote{e-mail:
{\tt hyee@uic.edu}}
\vskip 0.15in
{\it $^{1}$ Instituto de Fisica Teorica IFT-UAM/CSIC, Universidad Autonoma de Madrid,}\\{\it  28049
Cantoblanco, Spain}\\
[0.15in]{\it $^{2}$ Department of Physics, University of Illinois, Chicago, Illinois 60607}\\[0.15in]
{\it $^{3}$ RIKEN-BNL Research Center, Brookhaven National Laboratory,}\\
{\it Upton, New York
11973-5000}
\\[0.15in]
{\normalsize  2015}

\end{center}

\vfill

\begin{abstract}

We compute one of the second order transport coefficients arising from the chiral anomaly in a high temperature weakly coupled regime of quark-gluon plasma. This transport coefficient is responsible for the CP-odd current that is proportional to the time derivative of the magnetic field, and can be considered as a first correction to the chiral magnetic conductivity at finite, small frequency. We observe that this transport coefficient has a non-analytic dependence on the coupling as $\sim 1/(g^4\log(1/g))$ at weak coupling regime, which necessitates a re-summation of infinite ladder diagrams with leading pinch singularities to get a correct leading log result: a feature quite similar to that one finds in the computation of electric conductivity. We formulate and solve the relevant CP-odd Schwinger-Dyson equation in real-time perturbation theory that reduces to a coupled set of second order differential equations at leading log order. Our result for this second order transport coefficient
indicates that chiral magnetic current has some resistance to the time change of magnetic field, which may be called ``chiral induction effect''.
We also discuss the case of color current induced by color magnetic field.

\end{abstract}

\vfill

\end{titlepage}
\setcounter{footnote}{0}

\baselineskip 18pt \pagebreak
\renewcommand{\thepage}{\arabic{page}}
\pagebreak

\section{Introduction}

The chiral anomaly is an intriguing quantum mechanical phenomenon arising from an interplay between charge and chirality of massless particles such as chiral fermions.
It has recently been appreciated that the chiral anomaly may induce interesting parity-odd transport phenomena in the plasmas of such particles \cite{Kharzeev:2007tn,Kharzeev:2007jp,Fukushima:2008xe,Son:2004tq,Metlitski:2005pr}: at the lowest order in derivative expansion of hydrodynamics (that is, at the first order) it has been shown that the second law of thermodynamics dictates the existence of such phenomena \cite{Son:2009tf}:
the chiral magnetic effect \cite{Fukushima:2008xe} and the chiral vortical effect \cite{Erdmenger:2008rm,Banerjee:2008th}. Moreover, the magnitudes of these transport phenomena in the static, homogeneous limit are fixed by underlying anomaly coefficients and are not renormalized by interactions.
This universality has been confirmed explicitly in both weak \cite{Kharzeev:2009pj,Landsteiner:2011cp,Golkar:2012kb,Satow:2014lva} and strong coupling \cite{Yee:2009vw,Rebhan:2009vc,Gynther:2010ed,Hoyos:2011us,Amado:2011zx} computations, and there are also evidences in favor of them in lattice simulations \cite{Buividovich:2009wi,Abramczyk:2009gb,Yamamoto:2011gk,Buividovich:2013hza,Bali:2014vja}. Recent results from heavy-ion experiments at RHIC \cite{Abelev:2009ac,Wang:2012qs,Ke:2012qb,Shou:2014cua,Adamczyk:2015kwa} and LHC \cite{Selyuzhenkov:2011xq} seem consistent with the predictions from chiral magnetic and vortical effects \cite{Kharzeev:2010gr,Jiang:2015cva} (as well as chiral magnetic wave \cite{Kharzeev:2010gd,Newman:2005hd,Burnier:2011bf,Gorbar:2011ya,Yee:2013cya}), and quite interestingly, there is a successful experimental test of chiral magnetic effect in Dirac/Weyl semimetals which feature chiral fermionic excitations \cite{Li:2014bha} (the spin degree of freedom in this case arises from an internal degeneracy, in other words, it is a pseudo-spin).
Therefore, the existence and the magnitudes of these transport phenomena at lowest order (i.e. first order) in derivative
expansion seem by now quite robust. One can also generalize them to all even higher space-time dimensions than four \cite{Kharzeev:2011ds,Loganayagam:2011mu,Loganayagam:2012pz,Yee:2014dxa}.

As we go beyond the lowest order in derivatives, the possible anomaly induced transport phenomena become numerous:
in four dimensions there are thirteen possible second order anomalous transport coefficients in the current and energy-momentum tensor in a conformal plasma \cite{Kharzeev:2011ds} (and more in non-conformal plasma \cite{Bhattacharyya:2013ida}), while the second law of thermodynamics seems to constrain only eight combinations of them \cite{Kharzeev:2011ds}. Some of these constraints have been confirmed in a strong coupling computation \cite{Megias:2013joa}.
The interesting fact is that the values of these anomalous second order transport coefficients, although they are proportional to anomaly coefficients, do depend on the dynamics of the microscopic theory up to the mentioned constraints, so computing them in weak and strong coupling regimes is a non-trivial, but worthwhile task in any theoretical model.   

The purpose of this work is to take a small step in computing these second order anomalous transport coefficients in weakly coupled gauge theories, having in mind QCD and electro-weak theory.
Our current study will be based on diagrammatic techniques, and we hope to address a similar computation in chiral kinetic theory framework \cite{Son:2012wh,Stephanov:2012ki,Gao:2012ix,Chen:2014cla,Chen:2015gta} in a separate work.
We will show that one particular second order anomalous transport coefficient in the charge current has a non-analytic dependence on the coupling constant, $\sim 1/g^4 \log(1/g)$, which is similar to that one finds in the shear viscosity and electric conductivity \cite{Baym:1990uj,Arnold:2000dr} (and also in the chiral electric separation conductivity \cite{Huang:2013iia})\footnote{Our definition of transport coefficients does not include a trivial $e^2$ factor from the definition of electromagnetic current which is $e$ times of the fermion number current. Therefore, all quantities in our work are defined with the fermion number current. For example, the electric conductivity will be $\sim 1/(e^4\log(1/e))$ and the chiral magnetic conductivity at zero frequency for a single right-handed Weyl fermion is $\sigma_\chi={\mu\over 4\pi^2}$.}.
This transport coefficient appears in the second derivative correction to the current constitutive relation as, 
\be
\nu^\mu_{(2)}\sim \xi_5 \,\,\epsilon^{\mu\nu\alpha\beta} u_\nu {\cal D}_\alpha E_\beta\,,
\ee
where $E_\mu=F_{\mu\nu}u^\nu$ is the electric field strength in a local fluid rest frame defined by $u^\mu$, and we followed the notation introduced in Ref.\cite{Kharzeev:2011ds} to denote the transport coefficient $\xi_5$. Using the Bianchi identity, one can replace $\epsilon^{\mu\nu\alpha\beta} u_\nu {\cal D}_\alpha E_\beta$ with $u^\nu{\cal D}_\nu B^\mu$ with $B^\mu={1\over 2}\epsilon^{\mu\nu\alpha\beta} u_\nu F_{\alpha\beta}$ being the magnetic field strength in the local rest frame, which means that $\xi_5$ can be viewed as a first correction to the static chiral magnetic effect at finite frequency. More explicitly, it appears in the anomalous part of the current density as
\be
\vec J=\cdots+\sigma_\chi\vec B+\xi_5 {d\vec B\over dt}+\cdots\,,
\ee
where $\sigma_\chi$ is the topologically protected value of chiral magnetic conductivity at zero frequency.
$\xi_5$ is parity (P) and charge conjugation-parity (CP) odd, so must arise from chiral anomaly. 

As was observed first by Jeon \cite{Jeon:1994if}, in diagrammatic language, the non-analytic behavior in the coupling dependence is signaled by the presence of pinch singularities in multi-loop ladder diagrams of two point correlation functions,
which necessitates a resummation of all ladder graphs by solving a Schwinger-Dyson type equation to get a leading log result. Previously, it was observed in Ref.\cite{Satow:2014lva} that the zero frequency-momentum limit of the P-odd part of the 1-loop correlation function does not have pinch singularity, reproducing the correct static value of chiral magnetic conductivity. We first motivate our study by observing an appearance of pinch singularity in the P-odd part of 1-loop diagram at first order in frequency, which enters in a Kubo formula for $\xi_5$. Following intuitions from the computation of electric conductivity \cite{ValleBasagoiti:2002ir,Aarts:2002tn,Gagnon:2006hi}, we then identify multi-loop ladder graphs whose P-odd parts contain a chain of pinch singularities that have to be re-summed to get a correct leading log result for $\xi_5$. The emerging Schwinger-Dyson equation is more difficult to solve than that in the electric conductivity, because we need to keep finite external momentum $\vec k$ (up to first order in $\vec k$) to extract P-odd part of the correlation function. In section \ref{sec4}, we prove an important fact that all $\vec k$ dependence in the denominators of pinching propagators do not contribute to the P-odd part of our interest up to first order in $\vec k$,
allowing us to neglect them in the denominators of pinching propagators. The necessary $\vec k$ dependence for P-odd correlation functions arises only from the spinor projection part of the fermion propagators. With this important simplification, we are able to
reduce the leading log part of the P-odd Schwinger-Dyson equation into a coupled set of second order differential equations, which can be solved numerically. Along the way, we develop and use the sum rule for the P-odd part of Hard Thermal Loop (HTL) photon spectral density, which is summarized in the Appendix 1.  

For most part of our presentation, we will consider a single species of Weyl fermion in quantum electrodynamics (QED) for simplicity, and a generalization to finite number of species of Weyl and Dirac fermions as well as to a non-abelian $SU(N_c)$ gauge theory is trivial at our leading log order. We will describe this generalization in our discussion section at the end.
Our results are summarized as follows: for QED with a single right-handed Weyl fermion, we have
\be
\xi_5={-3.006 \over e^4\log(1/e)}{\mu\over T}\,.
\ee
For 2-flavor massless QCD ($N_c=3$) with $Q_u=3/2$ and $Q_d=-1/3$, our result is
\be
\xi_5^{\rm QCD}=(Q_u^2+Q_d^2){-3.6\over g^4\log(1/g)}{\mu_A\over T}={-2.003\over g^4\log(1/g)}{\mu_A\over T}\,.
\ee
where $\mu_A$ is an axial chemical potential.
The sign of $\xi_5$ compared to the zero-frequency value $\sigma_\chi$($={\mu\over 4\pi^2}$ for QED) is a meaningful dynamical result. A relative negative sign between the two means that the chiral magnetic current has some resistance to the change of the magnetic field. We may call this ``chiral induction effect''.

\section{Pinch singularity in P-odd part at 1-loop}

In this section, let us motivate our work by observing an appearance of pinch singularity in the P-odd part of 1-loop diagram at first order in frequency $\omega$. It will also serve to fix our notations and conventions. For simplicity, we will consider the case of single Weyl fermion species of unit charge in QED plasma at finite equilibrium temperature $T$, as the generalization to multi flavors or non-abelian gauge groups is simple (which will be summarized towards the end of the paper). Throughout our analysis, we will use the real-time Schwinger-Keldysh formalism in ``ra''-basis to compute the retarded current-current correlation function that contains the chiral magnetic conductivity $\sigma_\chi(k)$ in its P-odd part,
\be
\langle J^i(k) J^j(-k)\rangle^{\rm P-odd}_R=i\sigma_\chi(k)\epsilon^{ijl}k^l\,,
\ee
where italic letters run over the three spatial dimensions, and $k=(k^0,\vk)\equiv(\omega,\vk)$ is an external four momentum. Note that in ra-basis, the retarded two-point function is equal to 
\be
\langle J^i(k) J^j(-k)\rangle_R=(-i) \langle J^i_r(k) J^j_a(-k)\rangle_{\rm SK}\equiv (-i)G^{ij}_{(ra)}(k)\,,
\ee
where the subscript SK in the second term emphasizes that it is computed in the Schwinger-Keldysh path integral with $J_r\equiv 1/2(J_1+J_2)$ and $J_a\equiv J_1-J_2$ (1 and 2 denote the two time contours in the Schwinger-Keldysh formalism). We follow the notations in Ref.\cite{Satow:2014lva} for consistency\footnote{In literature it is often chosen to denote the retarded function by $G_{(ra)}$, which we find confusing.}. Explicitly,
\be
J^\mu_r=\psi_r^\dagger\sigma^\mu\psi_r +{1\over 4}\psi^\dagger_a\sigma^\mu\psi_a\,,\quad J^\mu_a=\psi^\dagger_r \sigma^\mu\psi_a +\psi^\dagger_a \sigma^\mu\psi_r\,,
\ee 
with $\sigma^\mu=({\bf 1}_{2\times 2},\vec\sigma)$ in terms of two component Weyl spinor field $\psi$.
Therefore, the task is to compute the P-odd part (or anti-symmetric part in $i,j$ indices) of the (ra)-correlator $G_{(ra)}^{ij}(k)$ for small frequency-momentum.

The zero frequency-momentum limit of $\sigma_\chi(k)$ has been shown to be universally
\be
\lim_{\vk\to 0}\lim_{\omega\to0} \sigma_\chi(k)={\mu\over 4\pi^2}\,,
\ee
and in particular, there appears no pinch singularity in this limit as shown in Ref.\cite{Satow:2014lva}. Our $\xi_5$ appears in
first order expansion in $\omega=k^0$ (while still taking zero momentum limit $\vk\to 0$),
\be
\lim_{\vk\to 0}\sigma_\chi(k)={\mu\over 4\pi^2}-i \xi_5 \omega+{\cal O}(\omega^2)\,.
\ee
However, since the P-odd part of $G^{ij}_{(ra)}$ contains a linear term in $\vk$ in defining $\sigma_\chi(k)$,
\be
G^{ij,{\rm P-odd}}_{(ra)}(k)=-\sigma_{\chi}(k)\epsilon^{ijl} k^l\,,
\ee
we have to keep $\vk$ dependence in $G^{ij}_{(ra)}(k)$ up to first order in $\vk$, and then take $\vk\to 0$ limit
of the P-odd coefficient $\sigma_\chi(k)$. This essentially means that we need to keep finite $\vk$ in the Schwinger-Dyson equation for ladder resummation, which is in contrast to the case of electric conductivity where one can put $\vk=0$ from the very outset which greatly simplifies the analysis. Despite this difficulty, we will be able to solve 
the Schwinger-Dyson equation for the P-odd part of $G^{ij}_{(ra)}(k)$, and extract the coefficient $\xi_5$.
\begin{figure}[h]
\centering
\includegraphics[width=460pt]{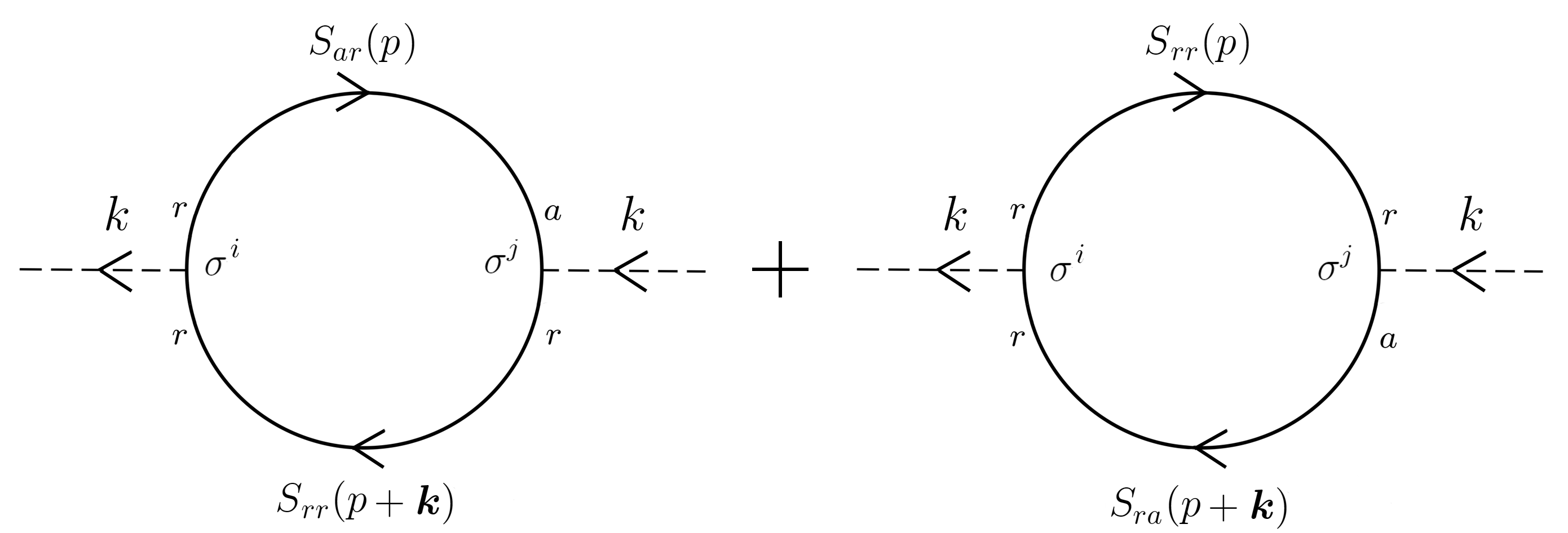}
\caption{Diagrams responsible for the retarded response at one loop in the ra-basis.\label{fig1}}
\end{figure}
At 1-loop there are two Feynman diagrams for $G_{(ra)}^{ij}(k)$ in real-time formalism as depicted in Fig.\ref{fig1}.
The fermion propagators are given by
\bear 
S_{ra}(p)&=&\sum_{s=\pm}{i\over p^0-s|\vp|+i\zeta/2}{\cal P}_s(\vp)
\,,\label{ra}\\
S_{ar}(p)&=&\sum_{s=\pm}{i\over p^0-s|\vp|-i\zeta/2}{\cal P}_s(\vp)
\,\label{ar},\\
S_{rr}(p)&=&\left({1\over 2}-n_+(p^0)\right)\rho(p)\,,\label{rr}
\eear
where $n_\pm(p^0)=1/(e^{\beta(p^0\mp\mu)}+1)$, and the spectral density $\rho$ is 
\be
\rho(p)=\sum_{t=\pm}{\zeta\over (p^0-t|\vp|)^2+(\zeta/2)^2}{\cal P}_t(\vp)
\,,
\ee
and we introduce the damping rate $\zeta\sim g^2\log(1/g)T$ in the propagators, which will be needed to regularize possible pinch singularities, that is essential to have a non-analytic dependence on the coupling constant.
It is important to observe a thermal relation 
\be
S_{rr}(p)=\left({1\over 2}-n_+(p^0)\right)\left(S_{ra}(p)-S_{ar}(p)\right)\,,\label{thermal}
\ee
which plays a central role in our analysis.
The projection operators ${\cal P}_\pm(\vp)$ are defined as,
\be
{\cal P}_\pm(\vp)\equiv{1\over 2}\left({\bf 1}\pm {\vsigma\cdot\vp\over |\vp|}\right)
= \mp\frac{\bar\sigma\cdot p_\pm}{2|\vp|}\,,\label{proj}
\ee
where $\bar\sigma^\mu=({\bf 1},-\vsigma)$ and $p_\pm^\mu\equiv (\pm|\vp|,\vp)$. Our metric convention is $(-,+,+,+)$.
The operators ${\cal P}_\pm(\vp)$ project onto particle and anti-particle states respectively with given momentum $\vp$,
and the ($s$,$t$) summation in the above physically represents distinctive contributions from particles and anti-particles.

The 1-loop expression for $G^{ij}_{(ra)}(k)$ from the two Feynman diagrams is
\be
(-1)\int {d^{4}p\over (2\pi)^{4}}
{\rm tr}\left[\sigma^i S_{ra}(p+k)\sigma^j S_{rr}(p)
+\sigma^i S_{rr}(p+k)\sigma^j S_{ar}(p)\right]\,.
\ee
where $(-1)$ in front comes from fermion statistics. In Ref.\cite{Satow:2014lva} it was shown that after extracting $\epsilon^{ijl}k^l$ for the P-odd part, the limit $\vk\to 0$ and $\omega\to 0$ commutes and produces the correct result $\mu/4\pi^2$, without featuring pinch singularity.
However, we will see that $\omega\to 0$ limit hides the pinch singularity appearing at first order in $\omega$ for $\xi_5$.
Using the thermal relation (\ref{thermal}) to replace $S_{rr}$ with $(S_{ra}-S_{ar})$, we have several combinations of $S_{ra}$ and $S_{ar}$. From the well-known fact (see Ref.\cite{Jeon:1994if}) that the pinch singularity appears only from the pair of $S_{ra}$ and $S_{ar}$ sharing a same momentum\footnote{This is because $S_{ra}$ ($S_{ar}$) has particle poles slightly below (above) the real axis by an amount $\pm i\zeta/2$, so that the residue of their product contains $1/\zeta$ factor which is the (regularized) pinch singularity. This also means that $s=\pm$ in (\ref{ra}) and (\ref{ar}) must be common in $S_{ra}$ and $S_{ar}$ pair causing a pinch singularity.}, let us select only terms that potentially contain pinch singularity, which results in
\be
G^{ij,{\rm Pinch}}_{(ra)}(k)=\int {d^{4}p\over (2\pi)^{4}}\,\,\left(n_+(p^0+\omega)-n_+(p^0)\right)
{\rm tr}\left[\sigma^i S_{ra}(p+k)\sigma^j S_{ar}(p)\right]\,.\label{1-loop2}
\ee
It is clear from this expression that $\omega\to 0$ limit does not produce a pinch singularity, because $\left(n_+(p^0+\omega)-n_+(p^0)\right)\approx (d n_+(p^0)/d p^0)\omega+{\cal O}(\omega^2)$ already gives a linear factor in $\omega$.
Moreover, one can put $\omega\to 0$ in the rest expression as we are only interested in the linear term in $\omega$ for $\xi_5$. In computing the above using (\ref{ra}) and (\ref{ar}) for $S_{ra}$ and $S_{ar}$, let us recall that the chosen $s=\pm$ representing particle or anti-particle from $S_{ra}$ must be the same $s$ chosen in $S_{ar}$ to have a pinch singularity in their product. Therefore, we have
\bear
G^{ij,{\rm Pinch}}_{(ra)}(k)&\approx&\omega\int {d^{4}p\over (2\pi)^{4}}\,\,\left(dn_+(p^0)\over dp^0\right)
{\rm tr}\left[\sigma^i S_{ra}(p+k)\sigma^j S_{ar}(p)\right]\nonumber\\
&\approx &\omega\int {d^{4}p\over (2\pi)^{4}}\,\,\left(dn_+(p^0)\over dp^0\right)\sum_{s=\pm}
{i^2 \,\,{\rm tr}\left[\sigma^i {\cal P}_s(\vp+\vk)\sigma^j{\cal P}_s(\vp)\right]\over (p^0-s|\vp+\vk|+i\zeta/2) ( p^0-s|\vp|-i\zeta/2)}\,.\nonumber\\\label{1-loop}
\eear

To identify the P-odd structure containing $\epsilon^{ijl}k^l$ in small $\vk\to 0$ limit, we first note that there are two possible sources of $\vk$-dependence: one is from the denominator and the other is from the projection operators in the numerator. Recall that we need only up to first order in $\vk$ since we take $\vk\to 0$ limit after extracting $\epsilon^{ijl}k^l$ piece. 
If one expands the denominator to linear order in $\vk$, we then need to put $\vk=0$ in the projection operators. The resulting trace using (\ref{proj}) gives 
\be
{\rm tr}\left[\sigma^i {\cal P}_s(\vp)\sigma^j{\cal P}_s(\vp)\right]={1\over 4|\vp|^2}{\rm tr}\left[\sigma^i (\bar\sigma\cdot p_s)
\sigma^j (\bar\sigma\cdot p_s)\right]={p^i p^j\over |\vp|^2}\,,
\ee
where we use
\be
{\rm tr}\left[\sigma^\mu\bar\sigma^\nu\sigma^\alpha\bar\sigma^\beta\right]=2(g^{\mu\nu}g^{\alpha\beta}-g^{\mu\alpha}g^{\nu\beta}+g^{\mu\beta}g^{\nu\alpha})+2i\,\epsilon^{\mu\nu\alpha\beta}\,,\label{sigtrace}
\ee
and $p_s^2=0$. It is clear that this contribution does not lead to a P-odd contribution which should be anti-symmetric in $i$ and $j$. Hence, we can ignore $\vk$ dependence in the denominator, which allows us to use the ordinary techniques dealing with pinch singularity in $\vk=0$ limit. In section \ref{sec4}, we will prove that this simplification generalizes to all order ladder diagrams, that is, the $\vk$-dependences in the denominators appearing in the ladder diagrams do not contribute to a P-odd part of the correlation function up to first order in $\vk$, and hence can be ignored. 

In computing P-odd $\vk$-dependence in ${\rm tr}\left[\sigma^i {\cal P}_s(\vp+\vk)\sigma^j{\cal P}_s(\vp)\right]$ using (\ref{proj}),
one can replace ${\cal P}_s(\vp+\vk)=-s \bar\sigma\cdot (p+k)_s/(2|\vp+\vk|)$ with $-s \bar\sigma\cdot (p+k)_s/(2|\vp|)$
by the same reason as above, and we have
\bear
{\rm tr}\left[\sigma^i {\cal P}_s(\vp+\vk)\sigma^j{\cal P}_s(\vp)\right]&\sim& {1\over 4|\vp|^2}{\rm tr}\left[\sigma^i(\bar\sigma\cdot (p+k)_s)\sigma^j (\bar\sigma\cdot p_s)\right]\nonumber\\
&\sim& {2i\over 4|\vp|^2}\epsilon^{i\mu j\nu}((p+k)_s)_\mu(p_s)_\nu
\eear
where in the second line, we use the fact that P-odd contribution can come only from the last P-odd term in the $\sigma$-matrix trace (\ref{sigtrace}). Our symbol $\sim$ cares only P-odd part linear in $\vk$. When $\mu=0$, $\nu=l$ we have
\be
s{i\over 2|\vp|^2}\epsilon^{ijl}|\vp+\vk| p^l\sim s{i\over 2|\vp|^3}\epsilon^{ijl}(\vp\cdot\vk)p^l\sim s{i\over 6|\vp|}\epsilon^{ijl}k^l\,,
\ee
where we use $|\vp+\vk|\approx |\vp|+(\vp\cdot\vk)/|\vp|+{\cal O}(\vk^2)$ and replace $p^mp^l\to (1/3)\delta^{ml}|\vp|^2$
since the angular $\vp$ integration in the final expression (\ref{1-loop}) is isotropic.
When $\mu=l$ and $\nu=0$ we have
\be
-s{i\over 2|\vp|^2}\epsilon^{ijl}(p^l+k^l)|\vp|\sim -s{i\over 2|\vp|}\epsilon^{ijl}k^l \,,
\ee
so, summing these two possibilities gives us the P-odd part of ${\rm tr}\left[\sigma^i {\cal P}_s(\vp+\vk)\sigma^j{\cal P}_s(\vp)\right]$ as
\be
{\rm tr}\left[\sigma^i {\cal P}_s(\vp+\vk)\sigma^j{\cal P}_s(\vp)\right] \sim -s{i\over 3|\vp|}\epsilon^{ijl}k^l \,,
\ee
and from (\ref{1-loop}) we have 
\be
G^{ij,{\rm Pinch}}_{(ra)}\sim i\omega\epsilon^{ijl}k^l \int {d^{4}p\over (2\pi)^{4}}\,\,\left(dn_+(p^0)\over dp^0\right){1\over 3|\vp|}\sum_{s=\pm}
{s  \over (p^0-s|\vp|+i\zeta/2) ( p^0-s|\vp|-i\zeta/2)}\,.\label{gij1}
\ee

The remaining computation is a standard procedure dealing with pinch singularity appearing in the denominators of (\ref{gij1}).
The $p^0$ integration can be done in the complex $p^0$ plane by closing the contour in either upper or lower half plane.
The leading singularity appears from the pole $p^0=s|\vp|\pm i{\zeta/2}$ where the residue contains a factor of $1/(\pm i\zeta)$: this gives a leading order contribution at weak coupling limit since $\zeta\sim g^2\log(1/g) T$. Once this $1/\zeta$ term is identified from the residue of the denominators in (\ref{gij1}), one can neglect $\zeta$ in the pole location $p^0\approx s|\vp|$ for all other terms as it engenders only higher order terms in $g$. This is because the $p$ integration has its dominant support in the region $|\vp|\sim T$ while $\zeta\sim g^2\log(1/g) T \ll |\vp|$. Therefore, one can effectively replace the two denominators in (\ref{gij1}) with
\be
{1\over (p^0-s|\vp|+i\zeta/2) ( p^0-s|\vp|-i\zeta/2)} \to {2\pi i\over i\zeta}\delta(p^0-s|\vp|)\,,
\ee
which will be used frequently in the following sections.
This gives us
\be
G^{ij,{\rm Pinch}}_{(ra)}\sim i\omega\epsilon^{ijl}k^l \,{1\over\zeta}\int {d^{3}\vp \over (2\pi)^{3}}\,\,{1\over 3|\vp|}\sum_{s=\pm} s \left(dn_+(p^0)\over dp^0\right)\bigg|_{p^0=s|\vp|}\,,
\ee
where we ignore the momentum dependence of $\zeta$ for now, which is not strictly valid (we will be more precise in our full ladder resummation in the next section).
From $dn_+(p^0)/  dp^0=-\beta n_+(p^0)(1-n_+(p^0))$ and $n_+(-|\vp|)=1-n_-(|\vp|)$, the integral becomes
\bear
&&\int {d^{3}\vp \over (2\pi)^{3}}\,\,{1\over 3|\vp|}\sum_{s=\pm}s\left(dn_+(p^0)\over dp^0\right)\bigg|_{p^0=s|\vp|}\\&=&-\beta\int {d^{3}\vp \over (2\pi)^{3}}\,\,{1\over 3|\vp|}\left(n_+(|\vp|)(1-n_+(|\vp|))-(n_-(|\vp|)(1-n_-(|\vp|))\right)\nonumber\\
&=&-{\beta\over 6\pi^2}\int^\infty_0 d|\vp| \,\,|\vp|\left(n_+(|\vp|)(1-n_+(|\vp|))-(n_-(|\vp|)(1-n_-(|\vp|))\right)=-{\mu\over 6\pi^2}\,,\nonumber
\eear
and we finally have a 1-loop expression
\be
G^{ij,{\rm Pinch}}_{(ra)}\sim -i\omega {\mu\over 6\pi^2 \zeta}\epsilon^{ijl}k^l \equiv i\omega\, \xi_5^{\rm 1-loop}\epsilon^{ijl} k^l\,,\qquad \xi_5^{\rm 1-loop}=-{\mu\over 6\pi^2\zeta}\,.
\ee
Although the overall sign of $\xi_5$ depends on the chirality, the relative negative sign compared to the static value of chiral magnetic conductivity $\sigma_\chi(0)={\mu/4\pi^2}$ doesn't depend on chirality and is a meaningful dynamical result. Holographic computations produce the same negative sign between $\sigma_\chi(0)$ and $\xi_5$.
\begin{figure}[h]
\centering
\includegraphics[width=250pt]{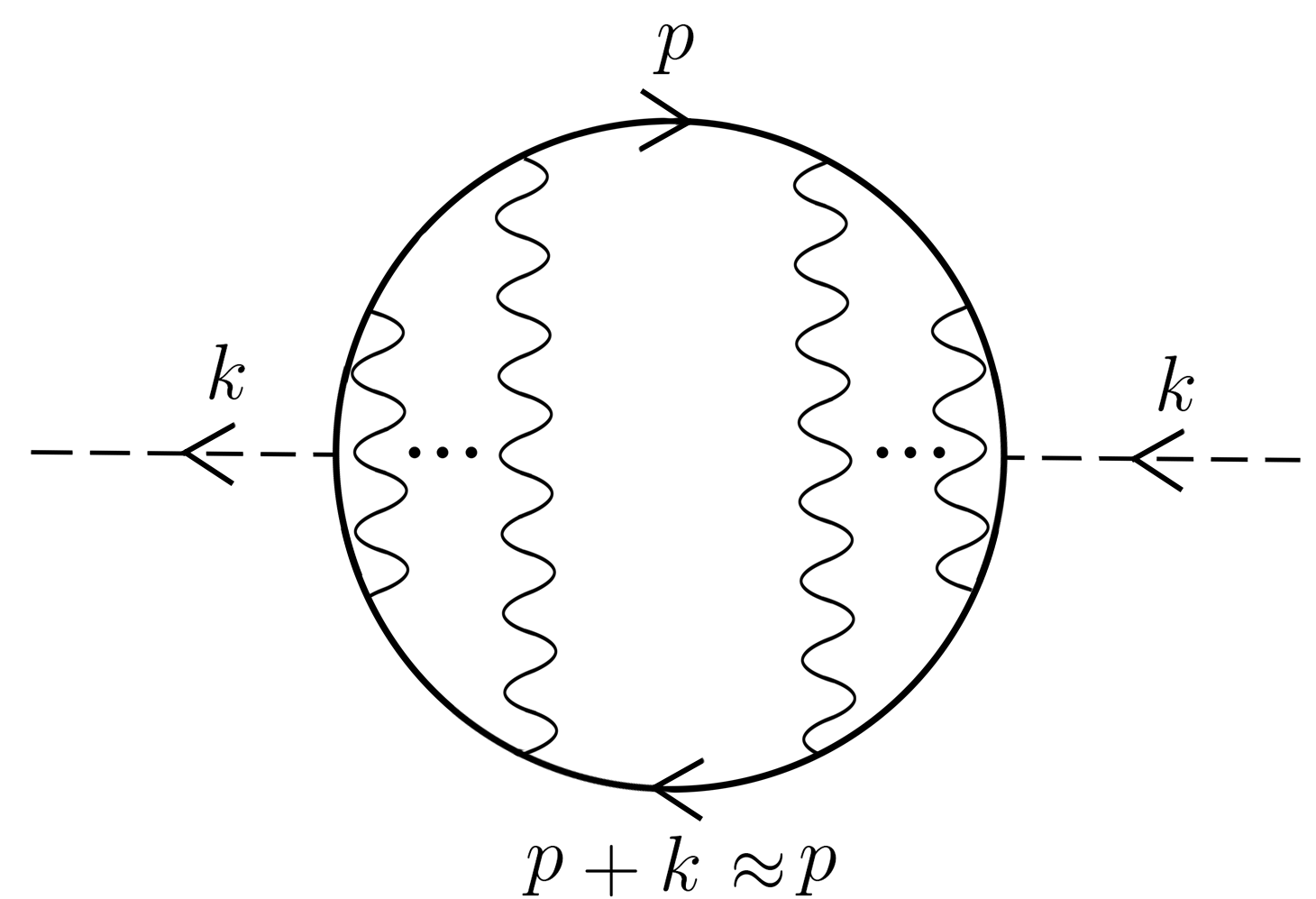}
\caption{A generic ladder diagram that contributes to the leading log result.\label{fig2}}
\end{figure}

The above exercise shows quite a similar feature to that one finds in the electric conductivity, and one can follow the lessons we have learned from the computation of electric conductivity. The $1/\zeta$ dependence from a pair of pinching propagators $S_{ra}(p)S_{ar}(p)$ signals a non-analytic dependence on the coupling constant. In a multi-loop ladder diagram shown in Fig.\ref{fig2} for example, each pair of pinching propagators sharing the same momentum produces a factor of $1/\zeta\sim 1/(g^2\log(1/g)T)$ which compensates a $g^2$ from an extra gauge boson exchange, making the diagram of the same order as the 1-loop diagram in the power counting of coupling constant. 
Hence, one needs to sum up all multi-loop ladder diagrams to get a correct leading order result for $\xi_5$, which can be achieved by solving a Schwinger-Dyson type integral equation that we will describe in the next section.
More elaborate power counting \cite{ValleBasagoiti:2002ir,Gagnon:2006hi} shows that the leading contribution comes from the soft region of gauge boson momentum $Q\sim gT$, so one needs a Hard Thermal Loop re-summed gauge boson propagator \cite{Pisarski:1988vd,Braaten:1989kk} for the internal gauge boson exchange lines. The fermion momentum stays hard $\sim T$, so fermion lines and all vertices are bare ones.

From the 1-loop result of $\xi_5^{\rm 1-loop}$ with $1/\zeta\sim 1/(g^2\log(1/g)T)$ dependence, the leading log result for $\xi_5$ from solving the Schwinger-Dyson equation might be expected to be $\sim 1/(g^2\log(1/g)T)$.
However, the correct dependence turns out to be $\sim 1/(g^4\log(1/g)T)$: this is also the same as in the electric conductivity. In both cases, a physics reason behind this is that small angle scatterings ($\theta\ll g$) by transverse space-like thermal gauge boson excitations (whose non-zero thermal spectral density is due to Landau damping physics) cannot affect the charge transport phenomena much, since they deflect charged fermion trajectories responsible for charge transports only slightly by small angles.
On the other hand, these small angle scatterings by ultra-soft ($p\sim g^2T$) transverse gauge bosons is the dominant source for the total decay rate $\zeta\sim g^2\log(1/g)T$,
where the log comes from $\log(m_D/\Lambda_{IR})\sim \log(1/g)$ with $\Lambda_{IR}\sim g^2 T$ being the non-perturbative IR cutoff for the transverse magnetic sector, and $m_D\sim gT$ is the characteristic soft scale.
This means that the effective IR regulator for the pinch singularities that is meaningful for the final conductivities is not given by the total damping rate $\zeta$, but is provided by larger angle scatterings ($\theta\gg g$) and fermion-conversion to gauge bosons, which are governed by $g^4\log(1/g)T$ rate. In the latter, the origin of the log is completely different: it is from $\log(T/m_D)\sim \log(1/g)$. 
In our diagrammatic approach of the Schwinger-Dyson equation, this physics manifests itself in a nice cancellation of leading log part of $\zeta$ in the equation that we will see in the following sections, and what remains is indeed something of $g^4 \log(1/g)T$ coming from the rate of fermion-conversion to gauge boson. 

We end this section by recalling that the situation is quite different for color conductivity where even small angle scatterings by thermal transverse gluons can change the color charge of charge carriers (either fermion or gauge boson) due to non-Abelian nature of color charges \cite{Selikhov:1993ns}, so that the same rate responsible for the leading log damping rate also governs the color conductivity, leading to its $1/(g^2\log(1/g))$ behavior \cite{Bodeker:1998hm,Arnold:1998cy}. 

\section{Ladder resummation of P-odd pinch singularities in ra-basis} 

In this section, we set-up the Schwinger-Dyson equation that sums up all-loop ladder diagrams with leading order pinch
singularities. The idea is essentially similar to the one in the diagrammatic computations of shear viscosity or electric conductivity \cite{ValleBasagoiti:2002ir,Aarts:2002tn}, except that we have to keep a finite external momentum $\vk$ up to first order in $\vk$ to extract a P-odd part (but, we can still put $\omega=0$ from the outset since one factor of $\omega$ comes out from kinematics, see (\ref{1-loop2}) and (\ref{1-loop})). We choose to work in real-time Schwinger-Keldysh formalism in ra-basis for our convenience, rather than the Euclidean formalism with subsequent analytic continuation as used in some previous literature. For electrical conductivity, we check that they produce the same result as they should.

\begin{figure}
\centering
\includegraphics[width=480pt]{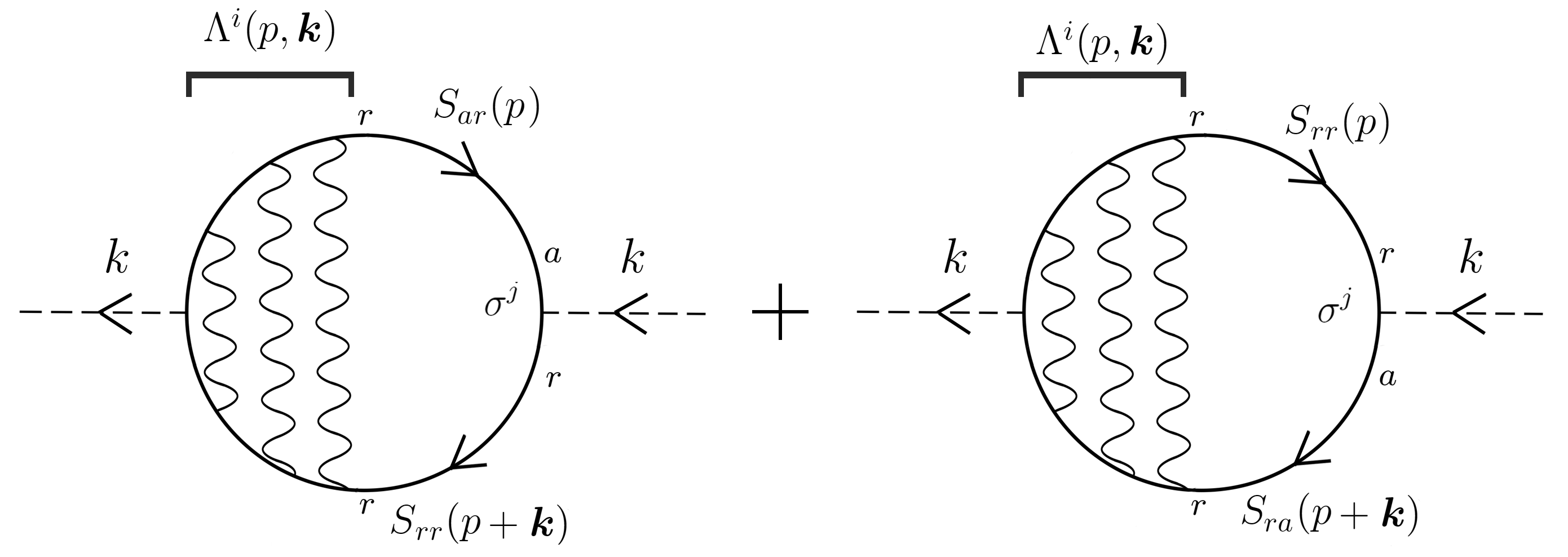}
\caption{Exemplar real time Feynman diagrams that can give leading pinch singularity.  \label{fig21}}
\end{figure}
Since we are computing $G_{(ra)}^{ij}(k)$, the vertex at the far right in any ladder diagram is an a-type one with one fermion leg r-type and the other a-type. Since there is no aa-propagator, the a-type leg should have r-type in the other end on its left. Because a pinch singularity can appear only from a pair of $S_{ra}$ and $S_{ar}$, the r-type leg from the vertex should have r-type on the other end on its left, since having a-type on the other end gives the same type of fermion propagator to the one from the former, and does not give a pinch singularity. See Fig.\ref{fig21} for an exemplar ladder diagram that can give a leading pinch singularity. In our convention, one reads ra-types of a fermion propagator along the reversed direction of its momentum arrow, which can be seen in Fig.\ref{fig21}. The reason why having a rr-type propagator in the diagram can give rise to a pinch singularity is the thermal relation (\ref{thermal}),
\be
S_{rr}(p)=\left({1\over 2}-n_+(p^0)\right)\left(S_{ra}(p)-S_{ar}(p)\right)\,,\label{thermal2}
\ee
so that one can pick either $S_{ra}$ or $S_{ar}$ piece from $S_{rr}$ to have a pair of $S_{ra}$ and $S_{ar}$ that gives
a pinch singularity. 
It is clear then that the rest of a ladder diagram on the left other than the far right vertex should have two final fermion legs of r-type on its right, in order to create a leading order pinch singularity: that is, it has to be an effective rr-type vertex. At 1-loop order, this was automatic since it is a bare $J_r^i$ vertex. What we have to do is to sum up all loop ladder diagrams for this effective rr-vertex that appears on the left side of the diagram.

\begin{figure}
\centering
\includegraphics[width=480pt]{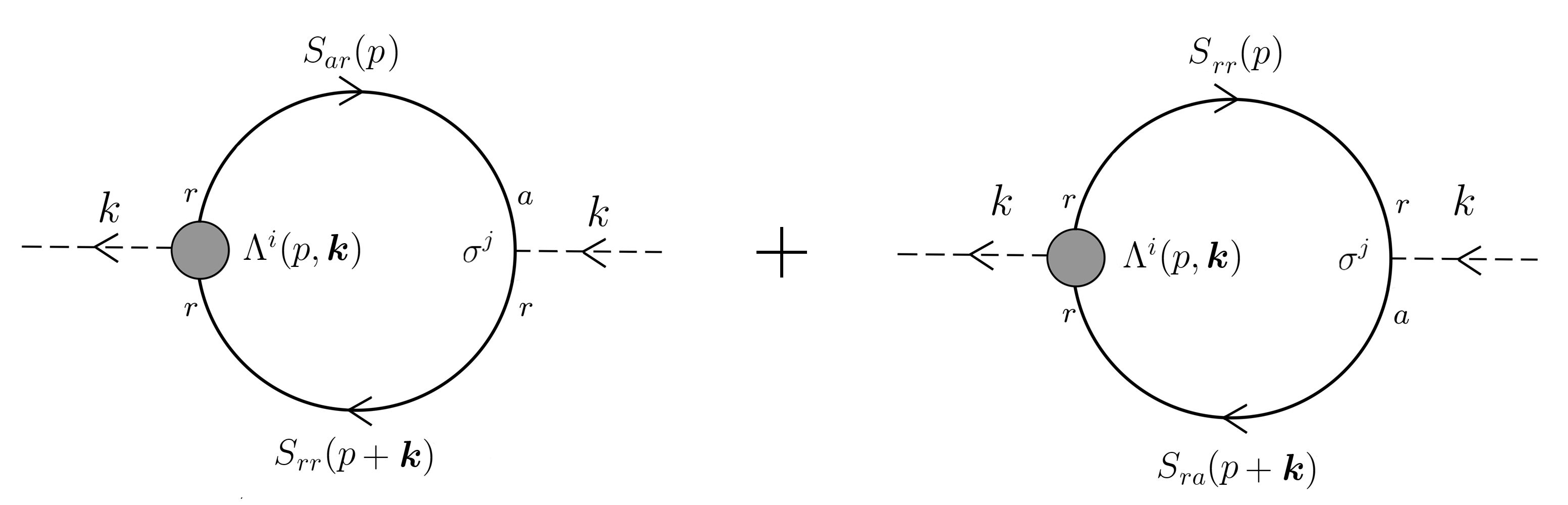}
\caption{The diagrams that need to be computed to obtain the retarded response function to leading log order. The effective vertex on the left includes infinite number of ladder diagrams.  \label{fig5}}
\end{figure}

Denoting the resulting summed vertex $\Lambda^i(p,k)$ that is a $2\times 2$ matrix acting on the spinor space, where $p$ is the loop momentum and $k$ is the (small) external momentum,
the final $G^{ij,{\rm Pinch}}_{(ra)}(k)$ is obtained from two possible Feynman diagrams in Fig.\ref{fig5} which look similar to those in Fig.\ref{fig1} except that the vertex on the left is now $\Lambda^i(p,k)$ instead of $\sigma^i$,
\be
G^{ij,{\rm Pinch}}_{(ra)}(k)=(-1)\int {d^{4}p\over (2\pi)^{4}}
{\rm tr}\left[\Lambda^i(p,k)\left( S_{ra}(p+k)\sigma^j S_{rr}(p)
+S_{rr}(p+k)\sigma^j S_{ar}(p)\right)\right]\,.
\ee
Using (\ref{thermal2}) picking up only pairs of $S_{ra}$ and $S_{ar}$ for a pinch singularity, and expanding it in $\omega$
with the same manipulation that led to (\ref{1-loop}) gives, up to first order in $\omega$,
\be
G^{ij,{\rm Pinch}}_{(ra)}(k)= -\omega\int {d^{4}p\over (2\pi)^{4}}\,\,\left(dn_+(p^0)\over dp^0\right)\sum_{s=\pm}
{ \,\,{\rm tr}\left[\Lambda^i(p,\vk) {\cal P}_s(\vp+\vk)\sigma^j{\cal P}_s(\vp)\right]\over (p^0-s|\vp+\vk|+i\zeta_{\vp+\vk,s}/2) ( p^0-s|\vp|-i\zeta_{\vp,s}/2)}\,,\label{finalgij}
\ee
where we put external frequency $\omega\equiv k^0=0$ in the effective vertex $\Lambda^i(p,\vk)$ and other places since we already have one $\omega$ factor in front. Note also that the damping rate $\zeta_{\vp,s}$ depends on the on-shell momentum as well as $s=\pm$ (that is, whether it is particle or antiparticle) as indicated in the expression\footnote{The dependence on $s$ comes via the combination $s\mu$ in the presence of chemical potential $\mu$ we are considering. See our Appendix 2 for a detailed discussion.}. We will be concerned with only this object $\Lambda^i(p,\vk)$ after putting $\omega=0$ in the following. 

\begin{figure}[h]
\centering
\includegraphics[width=440pt]{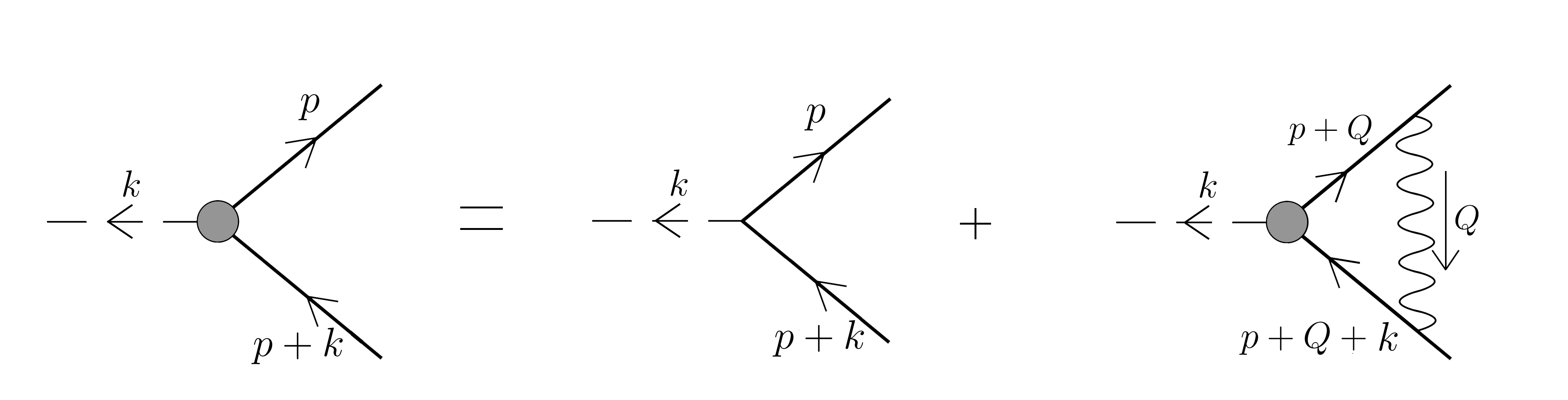}
\caption{The Schwinger-Dyson equation for the effective vertex $\Lambda^i(p,\vk)$ .\label{fig3}}
\end{figure}
\begin{figure}[t]
\centering
\includegraphics[width=500pt]{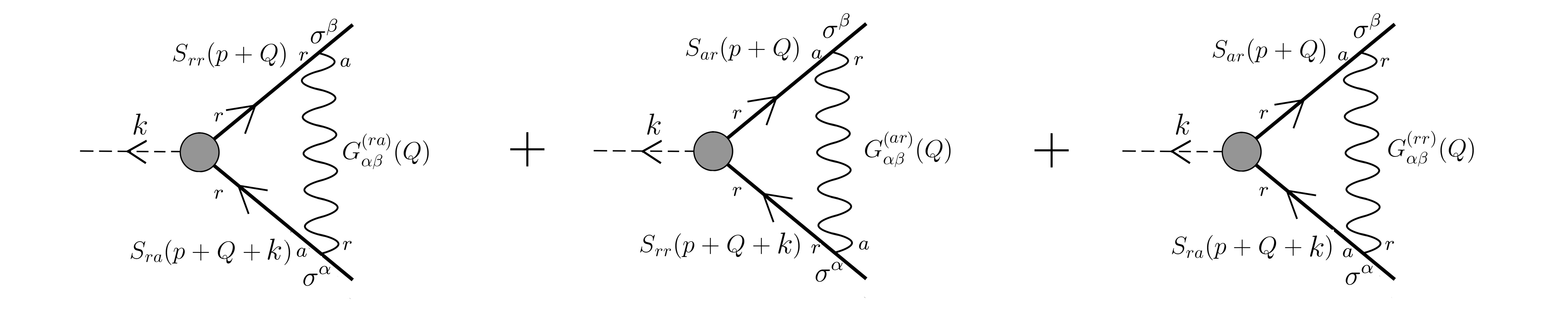}
\caption{The three real-time diagrams with leading pinch singularity for the kernel in the Schwinger-Dyson equation depicted in Fig. \ref{fig3}. The effective vertex connects only to r-type endings of fermion propagators. \label{fig9}}
\end{figure}

\setcounter{footnote}{0}
The summation of all multi-loop ladder diagrams for this effective rr-type vertex, starting from the bare one $J_r^i=\psi_r^\dagger \sigma^i\psi_r$ can be achieved by solving the associated Schwinger-Dyson type equation, which is depicted in Fig.\ref{fig3}. 
The ``kernel'' which is made of two internal fermion lines and one soft gauge boson (we call it photon) exchange can have three possible Feynman diagrams that can give a leading pinch singularity as shown in Fig.\ref{fig9}.  
The resulting Schwinger-Dyson equation \footnote{In \cite{Aarts:2002tn} it was shown that the Ward identity requires addition of an extra term in Fig.\ref{fig3} involving soft fermion lines. This diagram gives a sub-leading contribution to the electric conductivity and we expect the same for our $\xi_5$. We leave the explicit computation as a future work.} is written as (in the following we denote QED coupling constant by $e$ instead of $g$)

\bear
\Lambda^i(p,\vk)=\sigma^i&+&(ie)^2\int {d^4Q\over (2\pi)^4} \sigma^\beta S_{ar}(p+Q)\Lambda^i(p+Q,\vk) S_{ra}(p+Q+\vk)\sigma^\alpha G^{(rr)}_{\alpha\beta}(Q)\nonumber\\
&+&(ie)^2\int {d^4Q\over (2\pi)^4} \sigma^\beta S_{ar}(p+Q)\Lambda^i(p+Q,\vk) S_{rr}(p+Q+\vk)\sigma^\alpha G^{(ar)}_{\alpha\beta}(Q)\nonumber\\
&+&(ie)^2\int {d^4Q\over (2\pi)^4} \sigma^\beta S_{rr}(p+Q)\Lambda^i(p+Q,\vk) S_{ra}(p+Q+\vk)\sigma^\alpha G^{(ra)}_{\alpha\beta}(Q)\,,\nonumber\\
\eear
where $G_{\alpha\beta}^{(ab)}$ ($a,b=r$ or $a$) are the photon propagators in Schwinger-Keldysh contour
\be
G^{(ab)}_{\alpha\beta}\equiv \langle A^{(a)}_\alpha(Q) A^{(b)}_\beta(-Q)\rangle_{SK}=\int d^4 x\,\,e^{-iQx}\langle A^{(a)}_\alpha(x) A^{(b)}_\beta(0)\rangle_{SK}\,,
\ee
 including Hard Thermal Loop (HTL) photon self energy (that is $JJ$ correlation functions) since the photon momentum $Q$ is soft. We will work in the Coulomb gauge which separates longitudinal and transverse modes in a clear way. A summary of $G_{\alpha\beta}^{(ab)}$ in this gauge including the P-odd part coming from the P-odd part of HTL photon self-energy is given in the Appendix 1, where we also find
some useful sum rules for the P-odd part of their spectral density, which will be used importantly later. 
Using (\ref{thermal2}) and a similar thermal relation for photons (see the Appendix 1),
\be
G^{(rr)}_{\alpha\beta}(Q)=\left({1\over 2}+n_B(q^0)\right)\left(G^{(ra)}_{\alpha\beta}(Q)-G^{(ar)}_{\alpha\beta}(Q)\right)= \left({1\over 2}+n_B(q^0)\right) \rho_{\alpha\beta}^{\rm ph}(Q)\,,
\ee
where the photon spectral density is defined by \be\rho_{\alpha\beta}^{\rm ph}(Q)\equiv (G^{(ra)}_{\alpha\beta}(Q)-G^{(ar)}_{\alpha\beta}(Q))\,,\ee and $n_B(q^0)=1/(e^{\beta q^0}-1)$, the pinch singularity part of the integral equation becomes
\bear
\Lambda^i(p,\vk)=\sigma^i&+&(ie)^2\int {d^4 Q\over (2\pi)^4}[ \sigma^\beta S_{ar}(p+Q)\Lambda^i(p+Q,\vk)S_{ra}(p+Q+\vk)\sigma^\alpha \nonumber\\ &&\times \rho^{\rm ph}_{\alpha\beta}(Q)\,\,(n_+(p^0+q^0)+n_B(q^0))]\,.\label{mastereq}
\eear
Note that the photon spectral density $\rho^{\rm ph}_{\alpha\beta}(Q)$ is hermitian in $(\alpha,\beta)$ indices, but not necessarily real. In fact, the P-odd self-energy leads to a purely imaginary, anti-symmetric contribution to the spectral density. We refer the readers to the Appendix 1 for a detailed exposition.

From the pair $S_{ar}(p+Q)S_{ra}(p+Q+\vk)$ in (\ref{mastereq}) for small $\vk$ limit, one can extract the leading pinch singularity,
\bear
&&S_{ar}(p+Q)\Lambda^i(p+Q,\vk)S_{ra}(p+Q+k)\nonumber\\&&\to \sum_{t=\pm}{i^2 {\cal P}_t(\vp+\vq)\Lambda^i(p+Q,\vk){\cal P}_t(\vp+\vq+\vk)\over (p^0+q^0-t|\vp+\vq|-i\zeta_{\vp+\vq,t}/2)(p^0+q^0-t|\vp+\vq+\vk|+i\zeta_{\vp+\vq+\vk,t}/2)}\,.\label{pinchladder}
\eear
Since the photon momentum $Q$ is soft, $Q\ll |\vp|\sim T$, and the pinch singularity in the final equation for $G^{ij,{\rm Pinch}}_{(ra)}(k)$ in (\ref{finalgij}) necessitates the loop-momentum $p$ to be on-shell, $p^0=s|\vp|$, the only possible way to have a pinch singularity for soft $q^0$ integration in (\ref{mastereq}) is to pick only $t=s$ piece in the expression (\ref{pinchladder}): it means that in a ladder diagram the leading pinch singular contribution comes from a particle loop
or an anti-particle loop without any ``transition'' between particle and anti-particle throughout a diagram. Physically, it is obvious that a nearly on-shell particle (anti-particle) can not change to an anti-particle (particle) with soft photon scatterings.
Therefore, for a given choice of $s$ in (\ref{finalgij}), we keep only $t=s$ piece of (\ref{pinchladder}) in the integral equation (\ref{mastereq}), and the solution of the resulting integral equation we also label by $s$: $\Lambda_s(p,\vk)$.
The more correct expression for (\ref{finalgij}) is then
\be
G^{ij,{\rm Pinch}}_{(ra)}(k)= -\omega\int {d^{4}p\over (2\pi)^{4}}\,\,\left(dn_+(p^0)\over dp^0\right)\sum_{s=\pm}
{ \,\,{\rm tr}\left[\Lambda_s^i(p,\vk) {\cal P}_s(\vp+\vk)\sigma^j{\cal P}_s(\vp)\right]\over (p^0-s|\vp+\vk|+i\zeta_{\vp+\vk,s}/2) ( p^0-s|\vp|-i\zeta_{\vp,s}/2)}\,,\label{finalgij2}
\ee
where $\Lambda^i_s(p,\vk)$ satisfies the integral equation
\bear
\Lambda_s^i(p,\vk)=\sigma^i&+&e^2\int {d^4 Q\over (2\pi)^4}\bigg[ \sigma^\beta {\cal P}_s(\vp+\vq)\Lambda_s^i(p+Q,\vk){\cal P}_s(\vp+\vq+\vk)\sigma^\alpha \nonumber\\ &\times&{ \rho^{\rm ph}_{\alpha\beta}(Q)\,\,(n_+(p^0+q^0)+n_B(q^0))\over 
(p^0+q^0-s|\vp+\vq|-i\zeta_{\vp+\vq,s}/2)(p^0+q^0-s|\vp+\vq+\vk|+i\zeta_{\vp+\vq+\vk,s}/2)}\bigg]\,.\nonumber\\\label{mastereq2}
\eear
The rest of the paper is about solving the integral equation (\ref{mastereq2}) in leading logarithmic order in the coupling constant $e$. 

Since our transport coefficient $\xi_5$ is obtained from the P-odd part of $G^{ij,{\rm Pinch}}_{(ra)}(k)$ by
\be
G^{ij,{\rm Pinch, P-odd}}_{(ra)}(\omega,\vk)=i\omega\xi_5 \epsilon^{ijl} k^l+{\cal O}(\omega^2,\vk^2)\,,
\ee
we would like to expand (\ref{finalgij2}) in $\vk$ up to first order, focusing only on the P-odd $\epsilon^{ijl}k^l$ structure at the same time. Since $\xi_5$ is CP-odd which shares the same quantum number with the (axial) chemical potential $\mu$, it can only contain odd powers in $\mu$, as seen in the 1-loop computation in the preceding section. In our work, we will only compute $\xi_5$ up to linear order in $\mu$ in small $\mu$ limit, neglecting higher order terms of $\mu^3$ and beyond. Therefore, we will only be interested in a linear $\mu$ dependence of (\ref{finalgij2}) and (\ref{mastereq2}) in the following.

\section{An important simplification \label{sec4}}

In solving (\ref{mastereq2}) up to linear order in $\vk$, and using it to compute (\ref{finalgij2}), there are various sources of $\vk$ dependence appearing in the equations. The problematic source is the $\vk$ dependence in the denominators 
of the equations (\ref{finalgij2}) and (\ref{mastereq2}). For example in (\ref{finalgij2}), we have
\be
p^0-s|\vp+\vk|+i\zeta_{\vp+\vk,s}/2 \approx p^0-s|\vp|+i\zeta_{\vp,s}/2-s\hat{\vp}\cdot\vk+i(\partial \zeta_{\vp,s}/\partial \vp)\cdot\vk/2+\cdots\,\label{deno}
\ee
giving rise to up to linear in $\vk$,
\be
{1\over p^0-s|\vp+\vk|+i\zeta_{\vp+\vk,s}/2}\approx {1\over p^0-s|\vp|+i\zeta_{\vp,s}/2}+{s\hat{\vp}\cdot\vk-i(\partial \zeta_{\vp,s}/\partial \vp)\cdot\vk/2\over (p^0-s|\vp|+i\zeta_{\vp,s}/2)^2}+\cdots\,\label{deno1}
\ee
The second term is a double pole, and when used in (\ref{finalgij2}) it engenders a $\sim 1/\zeta^2$ dependence which is
larger than the usual $1/\zeta$ pinch singularity. The same is true for the $\vk$-dependence in the denominators of the integral equation (\ref{mastereq2}). For P-even part, this may be what one encounters when trying to include a finite $\vk$ in the current correlation functions, which seems to be related to the expected appearance of diffusion pole structure
\be
{\sigma_{el} \over \omega-iD k^2}\approx {\sigma_{el}\over\omega}+i{\sigma_{el} D }{k^2\over\omega^2}+\cdots\,,
\ee
since the Einstein relation gives $D=\sigma_{el}/\chi$ ($\chi$ is the charge susceptibility) and the $\vk$ dependence is quadratic in electric conductivity $\sigma_{el}\sim T^3/e^4\log(1/e)$. However, such a diffusion pole structure is not expected in the P-odd part of our interest \cite{Satow:2014lva}\footnote{There could arise a diffusion pole structure in the P-odd part if one considers coupling to energy-momentum sector of the theory leading to ``chiral magnetic energy flow''. However, when expanded in $\vk$ it would give a term of $\vk^3$ or higher \cite{Matsuo:2009xn,Sahoo:2009yq}. Also, the coupling to energy-momentum sector is of order $\mu^2$, and the resulting P-odd effect is of order $\mu^3$ \cite{Matsuo:2009xn,Sahoo:2009yq}. Therefore, we can ignore this possibility in our work.}, and it is natural to expect that these $\vk$-dependences from the denominators in (\ref{finalgij2}) and (\ref{mastereq2}) do not contribute to our P-odd structure $\epsilon^{ijl} k^l$.
Let us show this important simplification in the following. As a consequence, one can ignore all $\vk$'s in the denominators of the equations, and the only interesting $\vk$ dependence comes from the projection operators in the numerators. 
\begin{figure}[h]
\centering
\includegraphics[width=260pt]{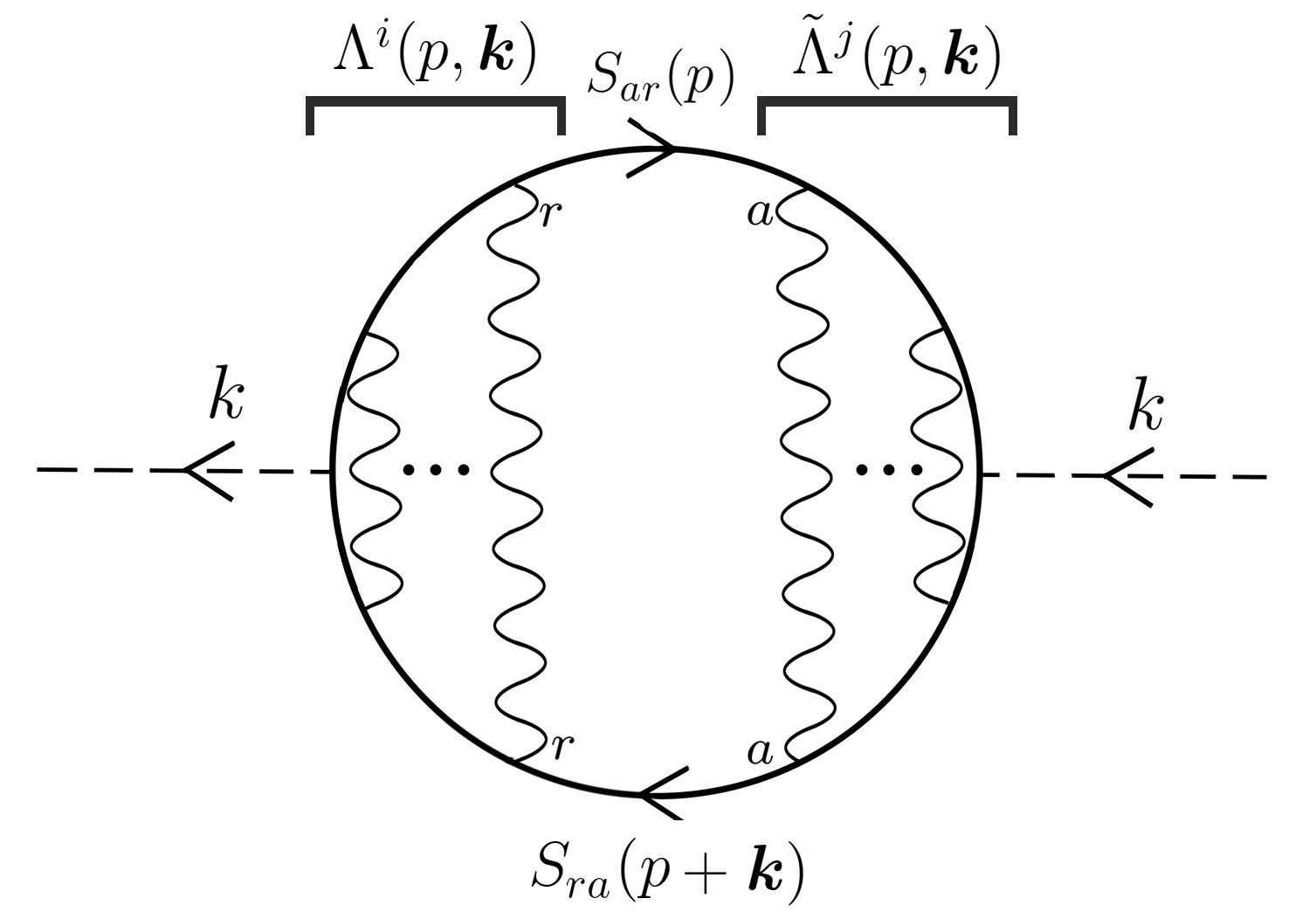}
\caption{Generic ladder diagram that gives rise to a leading log contribution. The ladders can be codified in effective vertices $\Lambda$,$\Lambda '$\label{fig6}}
\end{figure}

Let us consider a generic multi-loop ladder diagram depicted in Fig.\ref{fig6}, and let's choose an arbitrary internal $S_{ra}S_{ar}$ pair from the ``side rail'' that can give rise to a pinch singularity in small $\vk$ limit. By shifting loop momentum $p$, the momentum that flows in $S_{ra}$ can be put to $p+\vk$, then the momentum of $S_{ar}$ is $p$. 
The denominator of $S_{ra}(p+\vk)$ contains a $\vk$-linear piece as in (\ref{deno}). We would like to show that this $\vk$ dependence does not lead to any P-odd structure $\epsilon^{ijl} k^l$.

Once we get a term like (\ref{deno1}) that is linear in $\vk$ from expanding the denominator of $S_{ra}(p+\vk)$, we should put $\vk=0$ in all other parts of the diagram since it already saturates the linear $\vk$ dependence we are looking at.
These include projection operators in the numerators of $S_{ra}(p+\vk)$ and $S_{ar}(p)$: ${\cal P}_s(\vp+\vk)$ and ${\cal P}_s(\vp)$ (recall that we need to have a same $s$ throughout the diagram for the leading pinch singularity), as well as the remaining parts of the diagram other than the chosen $S_{ra}(p+\vk) S_{ar}(p)$ pair, which we call effective vertices: the part on the left let us call $\Lambda^i(p,\vk)$ and the part on the right $\tilde\Lambda^j(p,\vk)$. See Fig.\ref{fig6}.
The value of the diagram is then proportional to (the $p$ integral) of
\be
{\rm tr}\left[\Lambda^i(p){\cal P}_s(\vp)\tilde\Lambda^j(p){\cal P}_s(\vp)\right] \propto
{\rm tr}\left[\Lambda^i(p)(\bar\sigma\cdot p_s)\tilde\Lambda^j(p)(\bar\sigma\cdot p_s)\right] \,,\label{laddertrace}
\ee
where $\Lambda^i(p)\equiv \Lambda^i(p,\vk=0)$, $\tilde\Lambda^j(p)\equiv\tilde\Lambda^j(p,\vk=0)$, etc.
All $ij$ index structure comes from this spinor trace.

The effective vertex $\Lambda^i(p)$ is a $2\times 2$ matrix in the spinor space and since $\sigma^\mu=({\bf 1},\vec\sigma)$ forms a basis for any $2\times 2$ matrices, we write $\Lambda^i(p)=\sigma^\mu \lambda_\mu^i(p)$.
By invoking rotational invariance, we generally have
\be
\lambda_0^i(p)=f_0(p^0,|\vp|)\, \vp^i \,,\quad \lambda_l^i(p)=f_1(p^0,|\vp|)\,\delta_l^i+f_2(p^0,|\vp|)\, \vp_l\vp^i+f_3(p^0,|\vp|)\,\epsilon_l^{\,\,im}\vp_m\,,
\ee
where $f_i$ are functions only on $p^0$ and $|\vp|$. Similarly, we have for $\tilde\Lambda^j(p)=\sigma^\nu \tilde\lambda_\nu^j(p)$ with
\be
\tilde\lambda_0^j(p)=\tilde f_0(p^0,|\vp|)\, \vp^j \,,\quad \tilde\lambda_l^j(p)=\tilde f_1(p^0,|\vp|)\,\delta_l^j+\tilde f_2(p^0,|\vp|)\, \vp_l\vp^j+\tilde f_3(p^0,|\vp|)\,\epsilon_l^{\,\,jm}\vp_m\,.
\ee
Inserting these representations of $\Lambda^i(p)$ and $\tilde\Lambda^j(p)$ into the above (\ref{laddertrace}), and using the trace formula we repeat here
\be
{\rm tr}\left[\sigma^\mu\bar\sigma^\nu\sigma^\alpha\bar\sigma^\beta\right]=2(g^{\mu\nu}g^{\alpha\beta}-g^{\mu\alpha}g^{\nu\beta}+g^{\mu\beta}g^{\nu\alpha})+2i\,\epsilon^{\mu\nu\alpha\beta}\,,\label{sigtrace2}
\ee
we immediately see that the last $\epsilon$-tensor term in (\ref{sigtrace2}) does not contribute since $p_s$ appears twice in (\ref{laddertrace}), and we have
\be
{\rm tr}\left[\Lambda^i(p)(\bar\sigma\cdot p_s)\tilde\Lambda^j(p)(\bar\sigma\cdot p_s)\right]=4 (\lambda^i(p)\cdot p_s)(\tilde\lambda^j(p)\cdot p_s)\,,
\ee
where we use $p_s^2=0$. Then, we have
\be
\lambda^i(p)\cdot p_s=s|\vp|\lambda_0^i(p)+\vp^l\lambda^i_l(p)=(s|\vp|f_0(p^0,|\vp|)+f_1(p^0,|\vp|)+|\vp|^2 f_2(p^0,|\vp|) ) \,\vp^i\,,
\ee
which is proportional to $\vp^i$. Note that the piece involving $f_3$ drops out. The same conclusion is true, that is, $\tilde\lambda^j(p)\cdot p_s \sim \vp^j$, and hence the result for the trace in (\ref{laddertrace}) is proportional to $\vp^i\vp^j$.
Since it is symmetric with respect to $ij$, it is clear that the result can not contribute to the P-odd part of $\epsilon^{ijl}k^l$. In summary, we have shown that $\vk$ dependence from the denominator of any internal fermion line in leading pinch singularity limit does not contribute to the P-odd structure $\epsilon^{ijl}k^l$, and hence we can neglect all $\vk$'s appearing in the denominators, especially in our equations (\ref{finalgij2}) and (\ref{mastereq2}).

Once we remove all $\vk$'s from the denominators, we have
\be
G^{ij,{\rm Pinch}}_{(ra)}(k)= -\omega\int {d^{4}p\over (2\pi)^{4}}\,\,\left(dn_+(p^0)\over dp^0\right)\sum_{s=\pm}
{ \,\,{\rm tr}\left[\Lambda_s^i(p,\vk) {\cal P}_s(\vp+\vk)\sigma^j{\cal P}_s(\vp)\right]\over (p^0-s|\vp|+i\zeta_{\vp,s}/2) ( p^0-s|\vp|-i\zeta_{\vp,s}/2)}\,,\label{finalgij3}
\ee
and 
\bear
\Lambda_s^i(p,\vk)=\sigma^i&+&e^2\int {d^4 Q\over (2\pi)^4}\bigg[ \sigma^\beta {\cal P}_s(\vp+\vq)\Lambda_s^i(p+Q,\vk){\cal P}_s(\vp+\vq+\vk)\sigma^\alpha \nonumber\\ &\times&{ \rho^{\rm ph}_{\alpha\beta}(Q)\,\,(n_+(p^0+q^0)+n_B(q^0))\over 
(p^0+q^0-s|\vp+\vq|-i\zeta_{\vp+\vq,s}/2)(p^0+q^0-s|\vp+\vq|+i\zeta_{\vp+\vq,s}/2)}\bigg]\,.\nonumber\\\label{mastereq3}
\eear
The $p^0$ integration in (\ref{finalgij3}) can be computed in leading pinch singularity limit by replacing
\be
{1\over (p^0-s|\vp|+i\zeta_{\vp,s}/2) ( p^0-s|\vp|-i\zeta_{\vp,s}/2)} \to {2\pi\over \zeta_{\vp,s}}\delta(p^0-s|\vp|)\,,
\ee
which enforces the on-shell condition $p^0=s|\vp|$ on the $p$ appearing in the integral equation (\ref{mastereq3}).
We will assume this on-shell condition throughout our computation in the following sections.
Then, the integral equation becomes
\bear
\Lambda_s^i(\vp,\vk)=\sigma^i&+&e^2\int {d^4 Q\over (2\pi)^4}\bigg[ \sigma^\beta {\cal P}_s(\vp+\vq)\Lambda_s^i(\vp+\vq,\vk){\cal P}_s(\vp+\vq+\vk)\sigma^\alpha \nonumber\\ &\times&{ \rho^{\rm ph}_{\alpha\beta}(Q)\,\,(n_+(s|\vp|+q^0)+n_B(q^0))\over 
(q^0+s|\vp|-s|\vp+\vq|-i\zeta_{\vp+\vq,s}/2)(q^0+s|\vp|-s|\vp+\vq|+i\zeta_{\vp+\vq,s}/2)}\bigg]\,,\nonumber\\\label{mastereq3}
\eear
where $\Lambda_s^i(\vp,\vk)\equiv \Lambda_s^i(p,\vk)\big|_{p^0=s|\vp|}$. The reason why we can also replace
$\Lambda_s^i(p+Q,\vk)$ in the integral kernel with its on-shell value $\Lambda_s^i(\vp+\vq,\vk)$ is that the pinch singularity in the kernel of the integral equation 
\be
{1\over (p^0+q^0-s|\vp+\vq|-i\zeta_{\vp+\vq,s}/2)(p^0+q^0-s|\vp+\vq|+i\zeta_{\vp+\vq,s}/2)} \to {2\pi\over\zeta_{\vp+\vq,s}}\delta(p^0+q^0-s|\vp+\vq|)\,,\label{2ndpinch}
\ee
will impose the on-shell condition $p^0+q^0=s|\vp+\vq|$ as well. With this replacement of (\ref{2ndpinch}) in (\ref{mastereq3}), we finally have
\bear
\Lambda_s^i(\vp,\vk)=\sigma^i&+& e^2\int {d^4 Q\over (2\pi)^4}\bigg[ \sigma^\beta {\cal P}_s(\vp+\vq)\Lambda_s^i(\vp+\vq,\vk){\cal P}_s(\vp+\vq+\vk)\sigma^\alpha \nonumber\\ &\times& \rho^{\rm ph}_{\alpha\beta}(Q)\,\,(n_+(s|\vp|+q^0)+n_B(q^0))\,(2\pi)\delta(q^0+s|\vp|-s|\vp+\vq|)/\zeta_{\vp+\vq,s} \bigg]\,,\nonumber\\\label{mastereq4}
\eear
which is our starting point in solving the integral equation in leading logarithmic order in the next section.

\section{Leading-log computation}

The integral equation (\ref{mastereq4}) obtained in leading pinch singularity limit is a matrix equation, and it is desirable to transform it into a scalar equation. In fact, we don't need its full matrix structure: what we need at the end in (\ref{finalgij3}) is the trace 
\be
{\rm tr}\left[\Lambda_s^i(\vp,\vk) {\cal P}_s(\vp+\vk)\sigma^j{\cal P}_s(\vp)\right]={\rm tr}\left[{\cal P}_s(\vp)\Lambda_s^i(\vp,\vk) {\cal P}_s(\vp+\vk)\sigma^j\right]\,,
\ee
and it is obvious that we only need the component of $\Lambda_s^i(\vp,\vk)$ projected onto the eigenspace of the projection operator ${\cal P}_s(\vp)$ on the left and ${\cal P}_s(\vp+\vk)$ on the right, that is, \be{\cal P}_s(\vp)\Lambda_s^i(\vp,\vk) {\cal P}_s(\vp+\vk)\,.\ee
Since the spinor space is two dimensional, the above projected matrix is essentially a single number. This fact manifests itself by the following statement: for any $2\times 2$ matrix $A$, the projected matrix ${\cal P}_s(\vp)A{\cal P}_s(\vp+\vk)$ must be proportional to the matrix ${\cal P}_s(\vp) {\cal P}_s(\vp+\vk)$ composed only of the projection operators.
The proportionality constant, which contains the information on $A$, is easily found by comparing traces of the expected relation,
\bear
{\cal P}_s(\vp)A{\cal P}_s(\vp+\vk)&=&{{\rm tr}\left[{\cal P}_s(\vp)A{\cal P}_s(\vp+\vk)\right]\over {\rm tr}\left[{\cal P}_s(\vp){\cal P}_s(\vp+\vk)\right]} {\cal P}_s(\vp){\cal P}_s(\vp+\vk)\nonumber\\&=&{{\rm tr}\left[{\cal P}_s(\vp)A{\cal P}_s(\vp+\vk)\right]\over {1\over 2}(1+\widehat\vp\cdot\widehat{\vp+\vk})} {\cal P}_s(\vp){\cal P}_s(\vp+\vk)\,,
\eear
where $\widehat\vp\equiv {\vp}/|\vp|$, and ${\rm tr}\left[{\cal P}_s(\vp){\cal P}_s(\vp+\vk)\right]={1\over 2}(1+\widehat\vp\cdot\widehat{\vp+\vk})$. Using this, it is straightforward to convert our matrix integral equation into a scalar
equation, and for this purpose let us introduce three scalar functions ${\cal D}_s^i(\vp,\vk)$, $\Sigma_s^\mu(\vp,\vk)$, and $F_s(\vp;\vq,\vk)$ by
\bear
&&{\cal P}_s(\vp)\Lambda_s^i(\vp,\vk) {\cal P}_s(\vp+\vk)={\cal D}_s^i(\vp,\vk){\cal P}_s(\vp){\cal P}_s(\vp+\vk)\,,\nonumber\\
&& {\cal P}_s(\vp)\sigma^\mu {\cal P}_s(\vp+\vk)=\Sigma_s^\mu(\vp,\vk){\cal P}_s(\vp){\cal P}_s(\vp+\vk)\,,\nonumber\\
&& {\cal P}_s(\vp){\cal P}_s(\vp+\vq){\cal P}_s(\vp+\vq+\vk) {\cal P}_s(\vp+\vk)=F_s(\vp;\vq,\vk){\cal P}_s(\vp){\cal P}_s(\vp+\vk)\,.\label{functions}
\eear
The expressions for $\Sigma_s^\mu(\vp,\vk)$ and $F_s(\vp;\vq,\vk)$ can easily be found by computing the necessary traces involved, for example we have
\be
\Sigma_s^0(\vp,\vk)=1\,,\quad \Sigma_s^i(\vp,\vk)={s(\widehat{\vp}^i+\widehat{(\vp+\vk)}^i)-i\epsilon^{ijl}\widehat{\vp}_j\widehat{(\vp+\vk)}_l\over (1+\widehat{\vp}\cdot\widehat{\vp+\vk})}\,,\label{sigmaexp}
\ee
and the expression for $F_s(\vp;\vq,\vk)$ can be found in the Appendix 3. 
Using the fact that $\sigma^\mu$ is hermitian, we also have
\be
{\cal P}_s(\vp+\vk)\sigma^\mu {\cal P}_s(\vp)=(\Sigma_s^\mu(\vp,\vk))^*\,{\cal P}_s(\vp+\vk){\cal P}_s(\vp)\,.\label{fun2}
\ee
The scalar function ${\cal D}_s^i(\vp,\vk)$ is what we would like to find by solving the integral equation, and once it is found, the final expression for the correlation function $G^{ij,{\rm Pinch}}_{(ra)}(k)$ is given from (\ref{finalgij3}) by 
\bear
G^{ij,{\rm Pinch}}_{(ra)}(k)&=& -\omega\int {d^{4}p\over (2\pi)^{4}}\,\,\left(dn_+(p^0)\over dp^0\right)\sum_{s=\pm}
{ \,\,{\rm tr}\left[{\cal P}_s(\vp)\Lambda_s^i(\vp,\vk) {\cal P}_s(\vp+\vk)\sigma^j\right]\over (p^0-s|\vp|+i\zeta_{\vp,s}/2) ( p^0-s|\vp|-i\zeta_{\vp,s}/2)}\nonumber\\
&=& -\omega\int {d^{4}p\over (2\pi)^{4}}\,\,\left(dn_+(p^0)\over dp^0\right)\sum_{s=\pm}
{ \,\,{\cal D}^i_s(\vp,\vk){\rm tr}\left[ {\cal P}_s(\vp+\vk)\sigma^j{\cal P}_s(\vp)\right]\over (p^0-s|\vp|+i\zeta_{\vp,s}/2) ( p^0-s|\vp|-i\zeta_{\vp,s}/2)}\nonumber\\
&=& -\omega\int {d^{4}p\over (2\pi)^{4}}\,\,\left(dn_+(p^0)\over dp^0\right)\sum_{s=\pm}
{ \,\,{\cal D}^i_s(\vp,\vk) (\Sigma_s^j(\vp,\vk))^*\,{1\over 2}(1+\widehat{\vp}\cdot\widehat{\vp+\vk})\over (p^0-s|\vp|+i\zeta_{\vp,s}/2) ( p^0-s|\vp|-i\zeta_{\vp,s}/2)}\nonumber\\
&=&-{\omega}\int {d^{3}\vp\over (2\pi)^{3}}\,\,\sum_{s=\pm} \left(dn_+(p^0)\over dp^0\right)\bigg|_{p^0=s|\vp|}
{ \left(\Sigma_s^j(\vp,\vk)\right)^*}{{\cal D}^i_s(\vp,\vk)\over \zeta_{\vp,s}}\,,\label{finalgij4}
\eear
where we use (\ref{functions}), (\ref{fun2}) in the second and third lines, and replace ${1\over 2}(1+\widehat{\vp}\cdot\widehat{\vp+\vk})$ with unity in the last line, since $\widehat{\vp}\cdot\widehat{\vp+\vk}=1$ up to negligible ${\cal O}(\vk^2)$ corrections. As the last expression involves the combination ${\cal D}_s^i(\vp,\vk)/\zeta_{\vp,s}$, let us also define
\be
\phi^i_s(\vp,\vk)\equiv {{\cal D}_s^i(\vp,\vk)\over \zeta_{\vp,s}}\,.
\ee

Applying projection operators ${\cal P}_s(\vp)$ on the left and ${\cal P}_s(\vp+\vk)$ on the right to our integral equation (\ref{mastereq4}), and using (\ref{functions}), (\ref{fun2}) we finally get the following scalar integral equation to be solved for $\phi^i_s(\vp,\vk)$,
\bear
\zeta_{\vp,s}\, \phi_s^i(\vp,\vk)\,= \,\Sigma^i_s(\vp,\vk) &+& e^2\int{d^4Q\over (2\pi)^4} \,\,\Sigma^\beta_s(\vp,\vq)\phi^i_s(\vp+\vq,\vk) \left(\Sigma^\alpha_s(\vp+\vk,\vq)\right)^* F_s(\vp;\vq,\vk) \nonumber\\ &\times& \rho^{\rm ph}_{\alpha\beta}(Q)\,\,(n_+(s|\vp|+q^0)+n_B(q^0))\,(2\pi)\delta(q^0+s|\vp|-s|\vp+\vq|)\nonumber\\
\,=\,\Sigma^i_s(\vp,\vk) &+& e^2\int{d^4Q\over (2\pi)^4} \,\,{\cal K}_s(\vp,\vk;Q)\phi_s^i(\vp+\vq,\vk)(2\pi)\delta(q^0+s|\vp|-s|\vp+\vq|) \,
\,,\nonumber\\\label{scaleq}
\eear
with an integral kernel ${\cal K}_s(\vp,\vk;Q)$ defined as
\be
{\cal K}_s(\vp,\vk;Q)\equiv \Sigma^\beta_s(\vp,\vq) \left(\Sigma^\alpha_s(\vp+\vk,\vq)\right)^* \rho^{\rm ph}_{\alpha\beta}(Q) F_s(\vp;\vq,\vk)(n_+(s|\vp|+q^0)+n_B(q^0))\,.\label{kn}
\ee
Our task is to find a solution for $\phi_s^i(\vp,\vk)$
up to first order in $\vk$ which can give rise to the P-odd structure $\epsilon^{ijl}{\vk}^l$ in the expression (\ref{finalgij4}) for 
$G^{ij,{\rm Pinch}}_{(ra)}(k)$. We will be interested in only the leading logarithmic order in $e$.

The structure of $\phi^i_s(\vp,\vk)$ expanded up to first order in $\vk$ can be severely constrained by rotational invariance,
\be
\phi^i_s(\vp,\vk)=\chi^i_s(\vp)+f^{il}_s(\vp)\vk^l=\chi_s(|\vp|)\hat{\vp}^i+a_s(|\vp|)\vk^i+b_s(|\vp|)\hat{\vp}^i(\hat{\vp}\cdot\vk)+i{1\over|\vp|} c_s(|\vp|)\epsilon^{ilm}\hat{\vp}^l{\vk}^m\,,\label{expphi}
\ee
with four scalar functions $\chi_s, a_s, b_s, c_s$ which depend only on $|\vp|$. Using the expansion of $\left(\Sigma_s^j(\vp,\vk)\right)^*$ up to first order in $\vk$,
\be
\left(\Sigma_s^j(\vp,\vk)\right)^*=s\hat{\vp}^j+{s\over 2|\vp|}\left(\delta^{jl}-\hat{\vp}^j\hat{\vp}^l\right)\vk^l+{i\over 2|\vp|}\epsilon^{jlm}\hat{\vp}^l\vk^m\,,
\ee
the expression for $G^{ij,{\rm Pinch}}_{(ra)}(k)$ from (\ref{finalgij4}) can be computed to first order in $\vk$ to find the P-odd structure proportional to $\epsilon^{ijl}\vk^l$ as
\bear
G^{ij,{\rm Pinch}}_{(ra)}(k)&=&-{\omega}\int {d^{3}\vp\over (2\pi)^{3}}\,\,\sum_{s=\pm} \left(dn_+(p^0)\over dp^0\right)\bigg|_{p^0=s|\vp|}
{ \left(\Sigma_s^j(\vp,\vk)\right)^*}{\phi^i_s(\vp,\vk)}\nonumber\\&\sim&
-{i\over 3}\omega\int {d^{3}\vp\over (2\pi)^{3}}\,\,\sum_{s=\pm} \left(dn_+(p^0)\over dp^0\right)\bigg|_{p^0=s|\vp|}{1\over|\vp|}\left(s\, c_s(|\vp|)-{1\over 2}\chi_s(|\vp|)\right)\,\epsilon^{ijl}\vk^l\,,\nonumber\\\label{finalgijpodd}
\eear
where $\sim$ above only cares about P-odd terms. Note that the two functions $(a_s,b_s)$ do not contribute to our P-odd term, so we don't need to compute them. Therefore, we will focus on $\chi_s$ and $c_s$ only in the following.

\subsection{Computation of $\chi_s(|\vp|)$}

The function $\chi_s(|\vp|)$ is something that has already been known in previous computations of electric conductivity, although we will see that there is an important correction to it linear in $\mu$ that is relevant to our final value of $\xi_5$.
It satisfies the integral equation (\ref{scaleq}) after putting $\vk=0$
\be
\zeta_{\vp,s} \chi_s(|\vp|)\hat{\vp}^i=s\hat{\vp}^i+e^2\int {d^4 Q\over (2\pi)^4} {\cal K}_s(\vp,0;Q)\chi_s(|\vp+\vq|)\widehat{\vp+\vq}^i (2\pi)\delta(q^0+s|\vp|-s|\vp+\vq|)\,.\label{chieq}
\ee
Our treatment that follows for $\chi_s(|\vp|)$ is mostly the same to that one can find in Refs.\cite{ValleBasagoiti:2002ir,Aarts:2002tn}, and our computation for $\chi_s$ will confirm the previous results in literature. Let us however present some details along which we can introduce several key elements that will be needed in our next treatment for $c_s(|\vp|)$, which is new and more interesting to us.

First it is important to observe that the integral
\be
e^2\int {d^4 Q\over (2\pi)^4} {\cal K}_s(\vp,0;Q)(2\pi)\delta(q^0+s|\vp|-s|\vp+\vq|)\,,\label{sp}
\ee
that appears in the above equation is precisely equal to the contribution to the fermion damping rate from soft photon scatterings at full order, which contains the leading $\sim e^2\log(1/e)T$ part of the total damping rate. We reconfirm this fact explicitly in Appendix 2 including P-odd spectral density of soft photon propagator. Denoting this by $\zeta_{\vp,s}^{\rm sp}$ (following notations in Ref.\cite{Aarts:2002tn}), and writing the total damping rate as $\zeta_{\vp,s}=\zeta_{\vp,s}^{\rm sp}+\zeta_{\vp,s}^{\rm sf}$ where $\zeta_{\vp,s}^{\rm sf}$ is the other remaining contribution to the damping rate arising from soft fermion scatterings (or more precisely, hard fermions making conversion to hard photons and soft fermions) which is of order $e^4\log(1/e)T$, the integral equation (\ref{chieq}) takes a form
\bear
\zeta^{\rm sf}_{\vp,s}\chi_s(|\vp|)\hat{\vp}^i=s\hat{\vp}^i&+&e^2\int {d^4Q\over (2\pi)^4} {\cal K}_s(\vp,0;\vq)\left[\chi_s(|\vp+\vq|)\widehat{\vp+\vq}^i-\chi_s(|\vp|)\hat{\vp}^i\right]\nonumber\\
&\times&(2\pi)\delta(q^0+s|\vp|-s|\vp+\vq|)\,,\label{chieq2}
\eear
where we no longer have $\zeta_{\vp,s}^{\rm sp}\sim e^2\log(1/e)T$ explicitly in the equation, and what remains will be shown to be only of order $\sim e^4\log(1/e) T$. This cancellation of $e^2\log(1/e)T$ dependence due to the identity of (\ref{sp}) with $\zeta_{\vp,s}^{\rm sp}$ is the diagrammatic manifestation of the physics discussion at the end of section 2: the relevant part of damping rate that is responsible for fermionic charge transport phenomena is not the total damping rate governed by small-angle scatterings with ultra-soft transverse photons, but the part arising from fermion conversions to photons
with soft fermion scatterings, that is $\zeta_{\vp,s}^{\rm sf}$. In Appendix 2, we present a computation of $\zeta^{\rm sf}_{\vp,s}$ to leading log order (see Eq.(\ref{sfresult})), with a special care of its $s\mu$-dependence we will need later.

Introducing a variable $z=\cos\varphi$ where $\varphi$ is the angle between $\vp$ and $\vq$, one can show that \cite{Aarts:2002tn}
\be
\delta(q^0+s|\vp|-s|\vp+\vq|)=\left(|\vp|+s q^0\over |\vp||\vq|\right)\delta(z-z_0)\Theta\left(|\vq|^2-(q^0)^2\right)\,,
\ee where $\Theta(x)$ is the Heaviside theta function, and 
\be
z_0|\vq|=\hat{\vp}\cdot\vq=sq^0+{(q^0)^2-|\vq|^2\over 2|\vp|}\,,\label{pdotq}
\ee
and using this, one can transform the $Q$ integration into
\be
\int {d^4Q\over (2\pi)^4} (2\pi)\delta(q^0+s|\vp|-s|\vp+\vq|)=\int^\infty_0{d|\vq||\vq|\over (2\pi)}\int_{-|\vq|}^{|\vq|}{dq^0\over (2\pi)} \left(1+{s q^0\over |\vp|}\right)\bigg|_{\hat{\vp}\cdot\vq\to sq^0+{(q^0)^2-|\vq|^2\over 2|\vp|}}\,,\label{deltaid}
\ee
where one needs to replace any $\hat{\vp}\cdot\vq$ appearing in the integrand by $sq^0+((q^0)^2-|\vq|^2)/(2|\vp|)$.
As $Q=(q^0,\vq)\sim eT$ is soft and $\vp\sim T$ is hard for leading log contribution (which can be seen in retrospect), we expand the integrand in the integral equation (\ref{chieq2}) in powers of $(q^0,\vq)/|\vp|$.
For example, after some algebra we have up to ${\cal O}(Q^2)$,
\be
\chi_s(|\vp+\vq|)\widehat{\vp+\vq}^i-\chi_s(|\vp|)\hat{\vp}^i\approx \hat{\vp}^i\left({(q^0)^2-|\vq|^2\over 2|\vp|^2} \chi_s(|\vp|)
+sq^0 \chi_s'(|\vp|)+{1\over 2} (q^0)^2 \chi_s''(|\vp|)\right)\,,\label{secexp}
\ee
where we use the replacement $\hat{\vp}\cdot\vq\to sq^0+((q^0)^2-|\vq|^2)/(2|\vp|)$ in the middle of computation, and $\chi_s'(x)=d\chi_s(x)/dx$, etc.
Similarly, we need an expansion of ${\cal K}_s(\vp,0;\vq)$: with
\be
n_+(s|\vp|+q^0)+n_B(q^0)={1\over \beta q^0}+ s\left(n_s(|\vp|)-{1\over 2}\right)+{\cal O}(q^0)\,,
\ee
and 
\be
F_s(\vp;\vq,0)={1\over 2}\left(1+\hat{\vp}\cdot\widehat{\vp+\vq}\right)=1+{\cal O}({\vq}^2)\,,
\ee
what remains in ${\cal K}_s(\vp,0;\vq)$ is the polarization-contracted photon spectral density 
\be
\rho^{\rm ph}_{\alpha\beta}(Q) \left(\Sigma^\alpha_s(\vp,\vq)\right)^*\Sigma^\beta_s(\vp,\vq)\,,
\ee
where we need to expand the polarization part $\left(\Sigma^\alpha_s(\vp,\vq)\right)^*\Sigma^\beta_s(\vp,\vq)$ up to first order in $\vq$ for our leading log computation.
In Coulomb gauge, $\rho^{\rm ph}_{0i}(Q)=0$ ($i=1,2,3$) and $\rho^{\rm ph}_{00}(Q)\equiv \rho_L(Q)$ is the longitudinal part of spectral density. The transverse part is
\be
\rho^{\rm ph}_{ij}(Q)=\rho_T(Q)\left(\delta_{ij}-{\vq^i\vq^j\over |\vq|^2}\right)+i\rho_{\rm odd}(Q)\epsilon^{ijl}\vq_l\,,\label{spd}
\ee
where the second term is the P-odd contribution proportional to $\mu$ that arises from the P-odd part of current-current correlation function (or photon self-energy) in HTL limit, whose expression can be found in our Appendix 1. Note that it is purely imaginary, but anti-symmetric in $i,j$, so it is a hermitian matrix in $i,j$.
For our purpose, we would only need its sum rules derived in the Appendix 1, Eqns. (\ref{jnoddfirst}) and (\ref{jnoddsecond}),
\bear
&&\int_{-|\vq|}^{|\vq|} {dq^0\over (2\pi)} {1\over q^0}\rho_{\rm odd}(Q)=-{e^2\mu\over (2\pi)^2}{1\over |\vq|^4}+\cdots\,,\nonumber\\&&\int_{-|\vq|}^{|\vq|} {dq^0\over (2\pi)} {q^0}\rho_{\rm odd}(Q)={0\over|\vq|^2}+\cdots\label{odsum}
\eear
up to less singular terms in small $|\vq|\ll eT$ limit. All functions $(\rho_L,\rho_T,\rho_{\rm odd})$ are odd in $q^0\to -q^0$, so we need an extra odd power of $q^0$ in the final integrand to have a non-vanishing $q^0$ integral over $[-|\vq|,+|\vq|]$. The only $\mu$ dependence in the usual spectral densities $\rho_{L/T}(Q)$ is through the Debye mass, which is   $m_D^2=e^2 T^2/6+e^2\mu^2/(2\pi^2)$ for a single Weyl fermion. Since we are looking at only up to linear $\mu$ dependence, we can safely neglect $\mu^2$ corrections in $m_D^2$ and use the $\mu=0$ results for $\rho_{L/T}(Q)$.
After some algebra, we have up to ${\cal O}(\vq)$
\be
\rho^{\rm ph}_{\alpha\beta}(Q) \left(\Sigma^\alpha_s(\vp,\vq)\right)^*\Sigma^\beta_s(\vp,\vq)=\rho_L(Q)+\left(1-{(q^0)^2\over |\vq|^2}\right)\rho_T(Q)-{s\left((q^0)^2-|\vq|^2\right)\over |\vp|}\rho_{\rm odd}(Q)\,,
\ee
where the last contribution from the P-odd spectral density, although it is quadratic in $\vq$, is presented because its power counting is something new and different from those of $\rho_{L/T}(Q)$ as can be seen in (\ref{odsum}), and should be checked carefully. 

Let us first estimate this contribution from the P-odd spectral density in the integral equation (\ref{chieq2}).
Collecting everything presented above, the contribution from the P-odd spectral density to the integral in (\ref{chieq2}) becomes
\bear
&&-s{e^2\over |\vp|}\hat{\vp}^i\int_0^\infty{d|\vq||\vq|\over (2\pi)}\int_{-|\vq|}^{|\vq|}{dq^0\over (2\pi)}\left(1+{sq^0\over |\vp|}\right)\left({1\over \beta q^0}+s\left(n_s(|\vp|)-{1\over 2}\right)\right)\left((q^0)^2-|\vq|^2\right) \nonumber\\
&&\times \left({(q^0)^2-|\vq|^2\over 2|\vp|^2} \chi_s(|\vp|)
+sq^0 \chi_s'(|\vp|)+{1\over 2} (q^0)^2 \chi_s''(|\vp|)\right)\rho_{\rm odd}(Q)\,,
\eear
and using the sum rules derived in Appendix 1, (\ref{jnoddfirst}), (\ref{jnoddsecond}), it is easy to find that the result is at most of order $\sim e^4$ without any logarithmic enhancement. Note that the $|\vq|$ integration should have an UV cutoff $\sim T$ since we use HTL approximation for soft momentum $Q\ll T$.  In any case, these are of higher order than $e^4\log(1/e)$ of our interest, so can be neglected. Although we find $\rho_{\rm odd}(Q)$ does not affect the leading log equation for $\chi_s(|\vp|)$, we will find shortly that it does give an important contribution to the equation for $c_s(|\vp|)$ at leading log which is of our more interest.

The integral equation (\ref{chieq2}) then takes a form at leading log order as
\bear
\zeta_{\vp,s}^{\rm sf}\chi_s(|\vp|) =s&+&e^2\int_0^\infty{d|\vq||\vq|\over (2\pi)}\int_{-|\vq|}^{|\vq|}{dq^0\over (2\pi)}\left(1+{sq^0\over |\vp|}\right)\left({1\over \beta q^0}+s\left(n_s(|\vp|)-{1\over 2}\right)\right) \nonumber\\
&&\times \left(\rho_L(Q)+\left(1-{(q^0)^2\over |\vq|^2}\right)\rho_T(Q)\right)\nonumber\\
&&\times \left({(q^0)^2-|\vq|^2\over 2|\vp|^2} \chi_s(|\vp|)
+sq^0 \chi_s'(|\vp|)+{1\over 2} (q^0)^2 \chi_s''(|\vp|)\right)\,.\label{chieq3}
\eear
The remaining computation of various integrals of spectral densities is achieved at leading order using well-known sum rules of $\rho_{L/T}(Q)$ \cite{Blaizot:2001nr}. The leading log contribution will come from the region $m_D\ll |\vq| \lesssim T$, and following the notations in Ref.\cite{Aarts:2002tn} by defining 
\be
J^{L/T}_n\equiv \int_{-|\vq|}^{|\vq|}{dq^0\over (2\pi)} (q^0)^{2n-1} \rho_{L/T}(Q)\,,
\ee
we have for $|\vq|\gg m_D$ the sum rules\footnote{The sum rules for the case $|\vq|\ll m_D$ take different forms, and it can be checked that the ultra soft region $|\vq|\ll m_D$ does not give rise to logarithmic divergences.}
\bear
J^L_0\approx {m_D^2\over |\vq|^4}\,,\quad && J^T_0\approx {m_D^2\over 4 |\vq|^4}\left(\log {8 |\vq|^2\over m_D^2} -1\right)\,,\nonumber\\
J^L_1\approx {m_D^2\over 3|\vq|^2}\,,\quad && J^T_1\approx {m_D^2\over 4 |\vq|^2}\left(\log {8 |\vq|^2\over m_D^2} -3\right)\,,\nonumber\\
J^L_2\approx {m_D^2\over 5}\,,\quad && J^T_2\approx {m_D^2\over 4 }\left(\log {8 |\vq|^2\over m_D^2} -{11\over 3}\right)\,.\label{sr}
\eear
Using these in (\ref{chieq3}), one encounters a logarithmic divergence
\be
\int_{m_D}^ {T} {d|\vq|\over |\vq|}= \log(T/m_D)\sim \log(1/e)\,,
\ee
where the IR cutoff is $m_D$ since the sum rule expressions used are valid only for $|\vq|\gg m_D$ (see our footnote), and the UV cutoff is $T$
as we assume soft $Q\ll T$ throughout our treatment, and the modification for hard $Q$ will dampen away the UV divergences. Picking up the logarithmically enhanced terms in the integral, we finally get the differential equation
\be
\zeta_{\vp,s}^{\rm sf}\chi_s(|\vp|)=s-{e^2 m_D^2\log(1/e)\over 4\pi}\left({1\over \beta|\vp|^2}\chi_s(|\vp|)-\left({1\over \beta|\vp|}+n_s(|\vp|)-{1\over 2}\right)\chi_s'(|\vp|)-{1\over 2\beta}\chi_s''(|\vp|)\right)\,,\label{diffchi}
\ee
which 
agrees with the known result in literature in the case $\mu=0$. Note however the important $\mu$ dependence via $n_s(|\vp|)$ in the differential equation for $\chi_s(|\vp|)$, 
\be
n_s(|\vp|)={1\over e^{\beta(|\vp|-s\mu)}+1}\approx {1\over e^{\beta|\vp|}+1}+s\beta\mu {e^{\beta|\vp|}\over (e^{\beta|\vp|}+1)^2}+{\cal O}(\mu^2)\,,
\ee
as well as in $\zeta_{\vp,s}^{\rm sf}$ that we compute in Appendix 2 (see Eq.(\ref{sfresult})),
\be
\zeta_{\vp,s}^{\rm sf}={e^2 \over 4\pi}{m_f^2\log(1/e)\over |\vp|}\left(n_B(|\vp|)+n_{-s}(0)\right)\approx
{e^2 \over 4\pi}{m_f^2\log(1/e)\over |\vp|}\left(n_B(|\vp|)+{1\over 2}-s{\beta\mu\over 4}\right)+{\cal O}(\mu^2)   \,,
\ee
which give rise to a $s$-independent, linear $\mu$ part in $\chi_s(|\vp|)$ in addition to the usual $\mu$-independent part proportional to $s$. Here $m_f^2=(e^2/4)(T^2+\mu^2/\pi^2)$ is the asymptotic thermal mass of fermions.
The solution when expanded in $\mu$ then takes a form
\be
\chi_s(|\vp|)=s\chi_{(0)}(|\vp|)+\mu \chi_{(1)}(|\vp|)+{\cal O}(\mu^2)\,,\label{chimu}
\ee
and both $\chi_{(0)}$ and $\chi_{(1)}$ give separate contributions of the same leading order to the final 
expression for $\xi_5$ in (\ref{finalgijpodd}). In fact, this $\mu$ dependence via $n_s(|\vp|)$ and $\zeta_{\vp,s}^{\rm sf}$ in (\ref{diffchi}) (that is, the $\chi_{(1)}$ in the equation (\ref{chimu}))
also makes a contribution to the $\mu^2$ dependence of the electric conductivity in leading log order, which seems to have been missed in some previous literature.
Our analysis in the above (with full expressions for $m_D^2$ and $m_f^2$) contains all necessary elements that allow us to compute full $\mu^2$ correction to electric conductivity, and we present the correct computation of $\mu^2$ correction to the electric conductivity in Appendix 4.

\subsection{Computation of $c_s(|\vp|)$}

Let us next describe our analysis for $c_s(|\vp|)$, which appears as the P-odd component of the $\phi^i_s(\vp,\vk)\sim \phi^i_s(\vp,0)+i (c_s(|\vp|)/|\vp|)\epsilon^{ilm}\hat{\vp}^l\vk^m$ when expanded in linear $\vk$ (see (\ref{expphi})),
that satisfies our original integral equation (\ref{scaleq}) with a finite $\vk$.
Expanding the kernel function ${\cal K}_s(\vp,\vk;Q)$ defined by (\ref{kn}) up to linear in $\vk$,
\be
{\cal K}_s(\vp,\vk;Q)={\cal K}_s(\vp,0;Q)+{\cal K}_s^{(1)}(\vp,\vk;Q)+{\cal O}(\vk^2)\,,
\ee
where ${\cal K}_s(\vp,0;Q)$ is something we already use before (see Eq.(\ref{chieq}) and (\ref{sp})) to determine the zeroth order solution $\phi^i_s(\vp,0)=\chi_s(|\vp|)\hat{\vp}^i$, the part of integral equation in (\ref{scaleq}) that is linear in $\vk$
gives the integral equation for $c_s(|\vp|)$, which takes the form
\bear
&&\zeta_{\vp,s}{i c_s(|\vp|)\over |\vp|}\epsilon^{ilm}\hat{\vp}^l\vk^m\nonumber\\
&=&-{i\over 2|\vp|}\epsilon^{ilm}\hat{\vp}^l\vk^m + e^2 \int {d^4 Q\over (2\pi)^4} {\cal K}_s(\vp,0;Q){i c_s(|\vp+\vq|)\over |\vp+\vq|}\epsilon^{ilm}\widehat{\vp+\vq}^l\vk^m (2\pi)\delta(q^0+s|\vp|-s|\vp+\vq|)\nonumber\\
&+& e^2\int {d^4 Q\over (2\pi)^4} {\cal K}_s^{(1)}(\vp,\vk;Q) \chi_s(|\vp+\vq|)\widehat{\vp+\vq}^i (2\pi)\delta(q^0+s|\vp|-s|\vp+\vq|)\,,
\eear
where in the last term we understand that we extract only the P-odd term having the same structure of $\epsilon^{ilm}\hat{\vp}^l\vk^m$. The first term on the right arises from the fact that $\Sigma^i_s(\vp,\vk)$ given by (\ref{sigmaexp}) contains the P-odd term when expanded linear in $\vk$
\be
\Sigma^i_s(\vp,\vk) \sim -{i\over 2|\vp|}\epsilon^{ilm}\hat{\vp}^l\vk^m\,.
\ee
As before, it is important to use the fact that the integral
\be
 e^2 \int {d^4 Q\over (2\pi)^4} {\cal K}_s(\vp,0;Q) (2\pi)\delta(q^0+s|\vp|-s|\vp+\vq|)\,,
\ee 
which appears in the second term on the right side is precisely equal to the contribution to the fermion damping rate arising from soft photon scatterings, $\zeta_{\vp,s}^{\rm sp}$,
so that one can transform the above integral equation into the form
\bear
 &&\zeta^{\rm sf}_{\vp,s}{i c_s(|\vp|)\over |\vp|}\epsilon^{ilm}\hat{\vp}^l\vk^m=-{i\over 2|\vp|}\epsilon^{ilm}\hat{\vp}^l\vk^m\nonumber\\
&+& e^2 \int {d^4 Q\over (2\pi)^4} {\cal K}_s(\vp,0;Q)\left({i c_s(|\vp+\vq|)\over |\vp+\vq|}\epsilon^{ilm}\widehat{\vp+\vq}^l\vk^m-{i c_s(|\vp|)\over |\vp|}\epsilon^{ilm}\hat{\vp}^l\vk^m\right) \nonumber\\&\times&(2\pi)\delta(q^0+s|\vp|-s|\vp+\vq|)\nonumber\\
&+& e^2\int {d^4 Q\over (2\pi)^4} {\cal K}_s^{(1)}(\vp,\vk;Q) \chi_s(|\vp+\vq|)\widehat{\vp+\vq}^i (2\pi)\delta(q^0+s|\vp|-s|\vp+\vq|)\,,\label{cseq1}
\eear
where $\zeta_{\vp,s}^{\rm sf}$ that appears on the left is the damping rate contribution arising from soft fermion scatterings only which is of order $e^4\log(1/e)$ rather than $e^2\log(1/e)$.

The computation of the first integral on the right side of (\ref{cseq1}) at leading log order is almost identical to that of the previous integral in (\ref{chieq2}). Expanding up to quadratic in $Q$, with the
replacement $q^l\to (\hat{\vp}\cdot\vq)\hat{\vp}^l$ due to rotational invariance of the $\vq$ integral, we get after some algebra
\bear
&&{ c_s(|\vp+\vq|)\over |\vp+\vq|}\epsilon^{ilm}\widehat{\vp+\vq}^l\vk^m-{c_s(|\vp|)\over |\vp|}\epsilon^{ilm}\hat{\vp}^l\vk^m\nonumber\\
&=&\left({(q^0)^2-|\vq|^2\over 2|\vp|^2}\tilde c_s(|\vp|)+sq^0 \tilde c_s'(|\vp|)+{1\over 2}(q^0)^2\tilde c_s''(|\vp|)\right)\epsilon^{ilm}\hat{\vp}^l\vk^m+{\cal O}(Q^3)\,,
\eear
where $\tilde c_s(x)\equiv c_s(x)/x$.
Comparing this with the previous expansion (\ref{secexp}) for $\chi_s(|\vp|)$, we find the identical structure appearing, so that we can simply use the previous result of the integral in (\ref{chieq2}) (see Eq.(\ref{diffchi})) by replacing $\chi_s$ with $\tilde c_s(x)=c_s(x)/x$ to get 
\bear
&&e^2 \int {d^4 Q\over (2\pi)^4} {\cal K}_s(\vp,0;Q)\left({i c_s(|\vp+\vq|)\over |\vp+\vq|}\epsilon^{ilm}\widehat{\vp+\vq}^l\vk^m-{i c_s(|\vp|)\over |\vp|}\epsilon^{ilm}\hat{\vp}^l\vk^m\right) \nonumber\\&\times&(2\pi)\delta(q^0+s|\vp|-s|\vp+\vq|)\nonumber\\
&=&-i{e^2m_D^2\log(1/e)\over 4\pi}\left({1\over \beta|\vp|^2}\tilde c_s(|\vp|)-\left({1\over \beta|\vp|}+n_s(|\vp|)-{1\over 2}\right)\tilde c_s'(|\vp|)-{1\over 2\beta}\tilde c_s''(|\vp|)\right)\epsilon^{ilm}\hat{\vp}^l\vk^m\nonumber\\
&=&-i{e^2m_D^2\log(1/e)\over 4\pi|\vp|}\left({1\over \beta|\vp|^2}c_s(|\vp|)-\left(n_s(|\vp|)-{1\over 2}\right)\left(c_s'(|\vp|)-{c_s(|\vp|)\over |\vp|}\right)-{1\over 2\beta}c_s''(|\vp|)\right)\epsilon^{ilm}\hat{\vp}^l\vk^m\,,\nonumber\\\label{fstint}
\eear
up to leading log order. 

What is more complicated is the evaluation of the second integral in (\ref{cseq1}). 
Let us first look at the term $\chi_s(|\vp+\vq|)\widehat{\vp+\vq}^i$ in the integrand.
Defining $\tilde \chi_s(x)\equiv \chi_s(x)/x$ and using the fact that the $\delta$ function in the integrand imposes $|\vp+\vq|=|\vp|+sq^0$ we have
\be
 \chi_s(|\vp+\vq|)\widehat{\vp+\vq}^i=\tilde \chi_s(|\vp|+sq^0)(\vp^i+\vq^i)\,.
 \ee
Since what we need is the P-odd structure $\epsilon^{ilm}\hat{\vp}^l\vk^m$, it is clear that the first term proportional to $\vp^i$ can not possibly generate such structure, and therefore it is sufficient to consider only the second piece proportional to $\vq^i$,
\be
\chi_s(|\vp+\vq|)\widehat{\vp+\vq}^i\to \tilde \chi_s(|\vp|+sq^0)\vq^i\,,\label{ex3}
\ee
in the integral of (\ref{cseq1}). On the other hand,
 since ${\cal K}_s^{(1)}(\vp,\vk;Q)$ is a rotationally scalar function linear in $\vk$, rotational invariance dictates that it can only have three possible structures
\be
{\cal K}_s^{(1)}(\vp,\vk;Q)=\left(a_s^{(1)} \hat\vp^l + b_s^{(1)}\vq^l +c_s^{(1)}\epsilon^{lmn}\hat\vp^m\vq^n\right)\vk^l\,,\label{str2}
\ee
where $(a_s^{(1)},b_s^{(1)},c_s^{(1)})$ are some coefficient functions that depend only on $(p^0,|\vp|,q^0,|\vq|,\vp\cdot\vq)$.
Combining these two facts, and considering rotational invariance of $\vq$ integration, one can easily find that the only way to have the resulting P-odd structure $\epsilon^{ilm}\hat\vp^l \vk^m$ from the second integral in (\ref{cseq1}) is via the third term in (\ref{str2}), that is,
we only need to find the part of ${\cal K}_s^{(1)}(\vp,\vk;Q)$ that is proportional to $\epsilon^{lmn}\hat\vp^m\vq^n\vk^l=\epsilon^{lmn}\hat\vp^l\vq^m\vk^n$. This simplifies our computation by a great amount.

Since (\ref{ex3}) is already linear in $Q$, for a leading log contribution we only need to expand ${\cal K}_s^{(1)}(\vp,\vk;Q)$ up to linear in 
$Q$ which is already saturated by the wanted structure $\epsilon^{lmn}\hat\vp^l\vq^m\vk^n$. 
This in turn implies that one can neglect $sq^0$ correction in (\ref{ex3}) to have 
\be
\chi_s(|\vp+\vq|)\widehat{\vp+\vq}^i\to \tilde \chi_s(|\vp|)\vq^i={\chi_s(|\vp|)\over |\vp|}\vq^i\,,\label{ex4}
\ee
in the integral of (\ref{cseq1}).
 Given the expressions for $\Sigma_s^i(\vp,\vk)$ and $F_s(\vp;\vq,\vk)$ in (\ref{sigmaexp}) and Appendix 3, as well as the photon spectral density given in (\ref{spd}),
\be
\rho^{\rm ph}_{00}=\rho_L(Q)\,,\quad \rho^{\rm ph}_{ij}(Q)=\rho_T(Q)\left(\delta_{ij}-{\vq^i\vq^j\over |\vq|^2}\right)+i\rho_{\rm odd}(Q)\epsilon^{ijl}\vq_l\,,\label{spd2}
\ee
it is straightforward to find after some amount of algebra that
\bear
{\cal K}_s^{(1)}(\vp,\vk;Q)&=&\left(n_+(s|\vp|+q^0)+n_B(q^0)\right)\nonumber\\&\times&\left({is\over 2|\vp|^2}\left(\rho_L(Q)-\left(1+{(q^0)^2\over |\vq|^2}\right)\rho_T(Q)\right)+{i\over |\vp|}\rho_{\rm odd}(Q)\right) \epsilon^{lmn}\hat{\vp}^l\vq^m\vk^n\,,\nonumber\\
\eear
up to linear in $Q$, which will contribute to the leading log result of the integral in (\ref{cseq1}). Note that we have a non-negligible contribution from the P-odd part of the spectral density $\rho_{\rm odd}(Q)$: from its sum rule given in (\ref{odsum}) one can easily see that this term engenders a leading log contribution to the integral. When combining ${\cal K}_s^{(1)}(\vp,\vk;Q)$ with (\ref{ex4}) in the integral of (\ref{cseq1}), one has to replace $\vq^m\vq^i$ with 
\be
\vq^m\vq^i\to {1\over 2}\delta^{mi}|\vq_T|^2={1\over 2}\delta^{mi}\left(|\vq|^2-(\hat\vp\cdot\vq)^2\right)={1\over 2}\left(|\vq|^2-(q^0)^2\right)\,,
\ee
where $\vq_T$ is the perpendicular component of $\vq$ to $\vp$, and we use (\ref{pdotq}) in the last equality. This comes from the rotational invariance of $\vq$ integral around $\hat\vp$ axis. Collecting all these and following the same steps as in the leading log computation of $\chi_s(|\vp|)$ before, we finally have the second integral of (\ref{cseq1}) to be
given by at leading log order
\bear
&&e^2\int {d^4 Q\over (2\pi)^4} {\cal K}_s^{(1)}(\vp,\vk;Q) \chi_s(|\vp+\vq|)\widehat{\vp+\vq}^i (2\pi)\delta(q^0+s|\vp|-s|\vp+\vq|)\nonumber\\&=&-{e^2\over 2}{\chi_s(|\vp|)\over |\vp|}\int_0^\infty{d|\vq||\vq|\over (2\pi)}\int^{|\vq|}_{-|\vq|}{dq^0\over (2\pi)}\left(1+{sq^0\over |\vp|}\right)
\left({1\over \beta q^0}+s\left(n_s(|\vp|)-{1\over 2}\right)\right)\nonumber\\
&\times& \left({is\over 2|\vp|^2}\left(\rho_L(Q)-\left(1+{(q^0)^2\over |\vq|^2}\right)\rho_T(Q)\right)+{i\over |\vp|}\rho_{\rm odd}(Q)\right) \left(|\vq|^2-(q^0)^2\right) \epsilon^{ilm}\hat{\vp}^l\vk^m\nonumber\\
&=&-{e^2\over 2}{\chi_s(|\vp|)\over\beta |\vp|}\int^\infty_0{d|\vq||\vq|\over (2\pi)}\Bigg({is\over 2|\vp|^2}\left(|\vq|^2\left(J_0^L-J^T_0-{1\over|\vq|^2}J^T_1\right)-J^L_1+J_1^T+{1\over|\vq|^2}J_2^T\right)\nonumber\\&+&{i\over|\vp|}\left(-{e^2\mu\over (2\pi)^2|\vq|^2}\right)\Bigg)\epsilon^{ilm}\hat{\vp}^l\vk^m\nonumber\\
&=& i{e^2\over 4\pi}{\chi_s(|\vp|)\over \beta|\vp|^2}\log(1/e){e^2\mu\over 2\pi^2}\epsilon^{ilm}\hat{\vp}^l\vk^m\,,\label{sndint}
\eear
where we use the sum rules (\ref{sr}), (\ref{jnoddfirst}), (\ref{jnoddsecond}), and interestingly it turns out that the contributions from the P-even spectral densities $\rho_{L/T}$ cancel with each other exactly. We don't have a good understanding whether this has to be the case by some symmetry reason or it is just by accident. Therefore, the only contribution to the second integral in (\ref{cseq1}) at leading log order (that is, $e^4\log(1/e)$) comes from the P-odd part of the soft (HTL) photon spectral density $\rho_{\rm odd}(Q)$. Note that for this contribution, we have equal logarithmic contributions from both $e^2T\ll |\vq|\ll m_D$ and $T\gg|\vq|\gg m_D$ that add up together in the final result \footnote{Strictly speaking, QED does not possess ultra soft magnetic cutoff $\sim e^2T$. Since we have in mind the generalization to non-abelian QCD discussed in section 5, we simply assume this at this point.}.

From the integral equation (\ref{cseq1}) with (\ref{fstint}) and (\ref{sndint}), we finally obtain the sought-for second order differential equation for $c_s(|\vp|)$ as
\bear\label{eq:final}
&&\zeta_{\vp,s}^{\rm sf}c_s(|\vp|)=-{1\over 2}+{e^4\log(1/e) \mu\over (4\pi)2\pi^2}{\chi_s(|\vp|)\over \beta|\vp|}\nonumber\\&&
-{e^2m_D^2\log(1/e)\over 4\pi}\left({1\over \beta|\vp|^2}c_s(|\vp|)-\left(n_s(|\vp|)-{1\over 2}\right)\left(c_s'(|\vp|)-{c_s(|\vp|)\over |\vp|}\right)-{1\over 2\beta}c_s''(|\vp|)\right)\,,\nonumber\\
\eear
where the first line is an inhomogeneous source, especially the second term is in terms of $\chi_s(|\vp|)$ that should be obtained by solving the differential equation (\ref{diffchi}). We would need the expansion of $c_s(|\vp|)$ up to first order in chemical potential $\mu$,
\be
c_s(|\vp|)=c_{(0)}(|\vp|)+s\mu \,\,c_{(1)}(|\vp|)+{\cal O}(\mu^2)\,,
\ee
which can be found by solving the above differential equation order by order in $\mu$. 
We reemphasize that there are linear $\mu$ dependences coming from $\zeta^{\rm sf}_{\vp,s}$ and $n_s(|\vp|)$ in the above differential equation, which should not be missed to get a correct leading log answer.
After finding $\chi_s(|\vp|)$ and $c_s(|\vp|)$ from the above given differential equations, we compute our transport coefficient $\xi_5$ by (\ref{finalgijpodd}) with $G^{ij,{\rm Pinch, P-odd}}_{(ra)}(k)=i\omega\xi_5 \epsilon^{ijl}\vk^l$.

\subsection{Numerical evaluation}

As a first step to compute the explicit value of $\xi_5$, we solve numerically the equations for $(\chi_s(|\vp|),c_s(|\vp|))$ in (\ref{diffchi}) and (\ref{eq:final}). In order to do so we define  $\Psi_s(|\vp|/T)\equiv\alpha m_D^2 \log(1/e)  c_s(|\mathbf{p}|)/T$ and $\Phi_s(|\vp|/T)\equiv \alpha m_D^2 \log(1/e) \chi_s(|\mathbf{p}|)/T$ where $\alpha=e^2/4\pi$. Defining $y\equiv |\vp| /T$, the equation (\ref{eq:final}) can be rewritten as
\begin{align}\label{eq:finaleqparametricnice}
\Psi_s''(y) -\tanh(y/2) \Psi_s'(y) -  \frac{\Psi_s(y)}{y} \left( \frac{3}{2} \coth(y/2)-\tanh(y/2)  +\frac{2}{y}\right)\nonumber=\\1
-\frac{6 s\,\mu \,}{\pi^2T}\left(\frac{ s \Phi_s(y)}{y}+\frac{\pi^2 }{8}\frac{ \Psi_s(y)}{y}\right)+s\frac{\mu}{T}\frac{\sech^2(y/2)}{2}\left(\frac{ \Psi_s(y)}{y}- \Psi'_s(y)  \right)\,.
\end{align}
In addition we have the equation for the even vertex, $\chi_s(|\vp|)$ in (\ref{diffchi})
\begin{align}
\Phi_s''(y)+ \left( \frac{2}{y}-\tanh(y/2) \right) \Phi_s'(y)-\left(  \frac{2}{y^2} + \frac{3\coth(y/2)}{2y} \right)\Phi_s(y)=\nonumber\\-2s -s\frac{3\mu}{4T}\frac{\Phi_s(y)}{y}-s\frac{\mu }{2T }\sech^2(y/2) \Phi_s'(y)\,.
\end{align}
In the above, we expand $\mu$ dependence from $\zeta^{\rm sf}_{\vp,s}$ and $n_s(|\vp|)$ up to linear in $\mu$.
We then expand the solution to first order in $\mu$ as
\begin{align}
\Psi_s&= \Psi_A+ s \frac{\mu}{T} \Psi_B+\mathcal{O}(\mu^2)\,, \label{eq:decomposition}\\
\Phi_s&= s \Phi_A +\frac{\mu }{T}\Phi_B+\mathcal{O}(\mu^2)\,, \label{eq:decomposition2}
\end{align}
from which we have a coupled set of differential equations
\begin{align}\label{eq:finaleqparametricnice}
&\Psi_A''(y) -\tanh(y/2) \Psi_A'(y)  -  \frac{\Psi_A(y)}{y} \left( \frac{3}{2} \coth(y/2)-\tanh(y/2)  +\frac{2}{y}\right)=
1\,,\nonumber\\
&\Psi_B''(y) - \tanh(y/2) \Psi_B'(y)  -  \frac{\Psi_B(y)}{y} \left( \frac{3}{2} \coth(y/2)-\tanh(y/2)  +\frac{2}{y}\right)\nonumber=\\
&-\frac{6}{\pi^2}\frac{ \Phi_A(y)}{y}-\frac{3}{4}\frac{ \Psi_A(y)}{y}+\frac{\sech^2(y/2)}{2}\left(\frac{ \Psi_A(y)}{y}- \Psi'_A(y)  \right)\,,\nonumber\\
&\Phi_A''(y)+ \left( \frac{2}{y}-\tanh(y/2) \right) \Phi_A'(y)-\left(  \frac{2}{y^2} + \frac{3\coth(y/2)}{2y} \right)\Phi_A=-2 \,, \nonumber\\
&\Phi_B''(y)+ \left( \frac{2}{y}-\tanh(y/2) \right) \Phi_B'(y)-\left(  \frac{2}{y^2} + \frac{3\coth(y/2)}{2y} \right)\Phi_B(y)=\nonumber\\&-\frac{3}{4}\frac{\Phi_A(y)}{y}-\frac{1}{2}\sech^2(y/2) \Phi'_A(y)\,.
\end{align}

We solve the above equations by iterative method with vanishing boundary conditions at the IR ($y=0$) and the UV ($y=\infty$).
The last step is then to obtain the expression for the transport coefficient $\xi_5$ as an integral of the above quantities.
After computing the sum over $s=\pm$ and performing angular integrations in (\ref{finalgijpodd}), we obtain the result for the retarded propagator as an integral of $\Psi_{A/B}(y)$ and $\Phi_{A/B}(y)$,
\begin{align}
&G^{ij,{\rm Pinch}}_{(ra)}(k)=\frac{ 2i \omega  \mu  \epsilon^{ijl}\vk_l }{T\pi e^4 \log(1/e)}\times\nonumber\\
&\int_0^\infty y dy\Bigg(\left( \Psi_B(y) - \frac{ \Phi_B(y)   }{ 2}\right)\frac{1}{ \cosh^2(y/2)}
 +\left(\Psi_A(y) -  \frac{\Phi_A(y)}{2} \right) \frac{ \tanh(y/2)}{  \cosh^2(y/2)}
  \Bigg)\,.\nonumber\\\label{eq:finaln}
\end{align}
From the identification $G^{ij,{\rm Pinch, P-odd}}_{(ra)}(k)=i\omega\xi_5 \epsilon^{ijl}\vk^l$, we find
\begin{equation}
\xi_5= -\frac{3.006}{ e^4\log(1/e)}{\mu\over T}\,.
\end{equation}

\section{Discussion}

It is easy to generalize the above to the case of $N_F$ species of Dirac fermions in an $SU(N_c)$ gauge theory (but still the current and magnetic field are with respect to the global $U(1)$ flavor symmetry).
The chemical potential appearing in the integral equation is simply the axial chemical potential $\mu_A$. 
The Debye mass and asymptotic thermal mass are changed to
\be
m_D^2={g^2T^2\over 6}(2N_c+N_F)\,,\quad m_f^2={g^2 T^2\over 4}{N_c^2-1\over 2 N_c}\,.
\ee
The soft-fermion contribution to the hard photon damping rate is given by
\be
\zeta^{\rm sf}_{\vp,s}={g^2\over 4\pi}{N_c^2-1\over 2 N_c} {m_f^2\log(1/g)\over |\vp|}\left(n_B(|\vp|)+n_{-s}(0)\right)\,.
\ee
In the integral equations (\ref{diffchi}) and (\ref{eq:final}), the $e^2\log(1/e)$ in the kernel part should be replaced by 
\be g^2\log(1/g){N_c^2-1\over 2 N_c}\,.
\ee
In the second source term in the first line of (\ref{eq:final}) which is proportional to $e^4\log(1/e)$, this $e^4\log(1/e)$ should be replaced by
\be
g^4\log(1/g) N_F {N_c^2-1\over 2 N_c}\,.
\ee
This is because one factor of $e^2$ coming from the spectral density $\rho_{\rm odd}(Q)$ is replaced by $g^2 N_F$, while
the other $e^2$ coming from fermion-gluon couplings in the kernel is replaced by $g^2 (N_c^2-1)/(2N_c)$.
Finally, the expression for $G^{ij,{\rm Pinch}}_{(ra)}(k)$ from (\ref{finalgijpodd}), or equivalently our $\xi_5$ from $\chi_s(|\vp|)$ and $c_s(|\vp|)$ must be multiplied by the fermion degeneracy
\be
2N_c\times \sum_F Q_F^2\,,
\ee
where $Q_F$ are charges of $F$ flavors in units of $e$ (for (u,d)-quarks, it is $Q_u=2/3$ and $Q_d=-1/3$). 
For 2-flavor massless QCD ($N_c=3$) with $Q_u=2/3$ and $Q_d=-1/3$, the result for $\xi_5$ then becomes
\begin{equation}
\xi_5^{\rm QCD}=- \frac{2.003}{g^4\log(1/g)}{\mu\over T}\,.
\end{equation}
 
As mentioned at the end of section 2, the color conductivity $\sigma_c$ for non-Abelian
gauge theory that appears in the low energy effective theory at the scale $Q\lesssim g^2 T$,
\be
{\bm{J}}^a =\sigma_c {\bm{E}}^a + {\bm{\xi}}^a\,,
\ee
where $a$ denotes adjoint color charge and ${\bm{\xi}}^a$ is the thermal noise via fluctuation-dissipation relation to $\sigma_c$, is governed by scatterings with  ultra-soft transverse thermal gluons of momenta $Q\ll gT$ with the rate $\sim g^2\log(1/g)$, leading to $\sigma_c\sim g^2\cdot1/(g^2\log(1/g))\sim 1/\log(1/g)$.\footnote{Note that we now put an extra $g^2$ in the definition of color conductivities to follow the convention in literature.}  
 To get a correct leading log result for this, one also needs a similar diagrammatic resummation with essentially the same technique in our computation, except that charge carriers include gluons as well as quarks, and there is now no longer a precise cancellation of $g^2 \log(1/g)$ terms in the integral equation: this is due to the absence of U(1) Ward identity (replaced by non-Abelian version of Slavnov-Taylor identity) that ensures the cancellation of $g^2 \log(1/g)$ terms \cite{Aarts:2002tn}. The diagrammatic resummation (done in Ref.\cite{MartinezResco:2000pz}) for this is therefore somewhat simpler than that for
the electric conductivity, since one does not need to go to the next order of $g^4\log(1/g)$. In fact, one gets an algebraic equation to solve rather than a differential equation. The same resummation can also be achieved in the language of Bodecker's approach \cite{Bodeker:1998hm} as well as in kinetic theory \cite{Arnold:1998cy}. In the presence of axial charge $\mu_A$ breaking CP symmetry, Ref.\cite{Akamatsu:2014yza} recently obtained via Bodecker's approach a CP-odd contribution to the color current 
\be
{\bm{J}}^a =\sigma_c {\bm{E}}^a + {\bm{\xi}}^a+{\mu_A N_F g^2\over 4\pi^2}{\bm{B}}^a\,,
\ee
which is a colored analogue of chiral magnetic effect consistent with the $U(1)_A SU(3)_c^2$ triangle anomaly. Since triangle anomaly is topological, this contribution should be saturated at 1-loop diagrammatically without a need for resummation of ladder diagrams. From our computation in the text, the quantity that needs a ladder resummation and is sensitive to the same ultra soft scale dynamics of $g^2\log(1/g)$ rate that the color conductivity is also sensitive to, appears when one goes to the next order in derivative 
\be
{\bm{J}}^a =\sigma_c {\bm{E}}^a + {\bm{\xi}}^a+{\mu_A N_F g^2\over 4\pi^2}{\bm{B}}^a+\xi_5^c {d{\bm{B}}^a\over dt}\,.
\ee
It is clear that $\xi_5^c$, a colored analogue of our $\xi_5$, will be of order
\be
\xi_5^c\sim g^2\cdot 1/(g^2\log(1/g))(\mu_A/T)\sim (1/\log(1/g))(\mu_A/T) \,,
\ee 
 due to the absence of precise cancellation of $g^2\log(1/g)$ terms in the integral equations.
 The computation of $\xi_5^c$ at leading log order is doable, following the same steps we present in our work keeping only $g^2\log(1/g)$ terms in the integral equations (note that it receives contributions only from quarks, not from gluons). The leading order fermion damping rate from soft gluon scatterings is 
 \be
 \zeta_{\vp,s}^{\rm sp}={N_c^2-1\over 2 N_c} {g^2\log(1/g) T\over 2\pi}\,,
 \ee and the solution of the integral equations which become algebraic is 
 \be
 \chi_s(|\vp|)={s\over  \zeta_{\vp,s}^{\rm sp}+{1\over 2N_c}{g^2\log(1/g) T\over 2\pi}}= {4\pi s\over N_c g^2\log(1/g)T}\,,\quad c_s(|\vp|)=-{s\over 2}\chi_s(|\vp|)\,,\ee
 which gives our result for $\xi_5^c$,
 \bear
 \xi_5^c&=&-{g^2 N_F\over 3}\int {d^3\vp\over (2\pi)^3}\sum_{s=\pm}{dn_+(p^0)\over dp^0}\bigg|_{p^0=s|\vp|}{1\over |\vp|}\left(s c_s(|\vp|)-{1\over 2}\chi_s(|\vp|)\right)\nonumber\\
 &=& -{2\over 3\pi}{N_F\over N_c}{1\over \log(1/g)}{\mu_A\over T}\,.
 \eear 
The same resummation should also be achievable in the Bodecker's approach presented in Ref.\cite{Akamatsu:2014yza} by going to the next order in time derivatives.

It is utmost important to implement the correct value of chiral magnetic current in the presence of time-varying magnetic field in any realistic simulation of
chiral magnetic effect (or any other anomaly induced transport phenomena) in heavy-ion collisions. Our result should be an important step at weak coupling picture toward taking into account time-varying nature of the magnetic field in heavy-ion collisions, and will be instrumental in the quantitative studies of the chiral anomaly induced phenomena in the experiments at RHIC and LHC.

\vskip 1cm \centerline{\large \bf Acknowledgment} 
A.J. would like to thank Nuclear Theory Group at UIC for hospitality during his visit and Francisco Pena for useful comments. H.U.Y. thanks Dima Kharzeev, Daisuke Satow, and Misha Stephanov for discussions.
A.J. has been supported by FPU fellowship AP2010-5686, Plan Nacional de Altas Energias FPA2009-07890, Consolider Ingenio 2010 CPAN CSD200-00042 and Severo Ochoa award SEV-2012-0249.

\section*{Appendix 1: Sum rules for the P-odd part of HTL photon spectral density }

Let us start from the thermal relation
\be\label{kms}
G^{rr}_{\mu\nu}(q)=\left({1\over 2}+n_B(q^0)\right)\left(G^{ra}_{\mu\nu}(q)-G^{ar}_{\mu\nu}(q)\right)\,,
\ee
where $G^{ab}_{\mu\nu}(x)\equiv \langle A^{a}_\mu(x) A^{b}_\nu(0)\rangle_{SK}$ ($a,b=r,a$) are correlation functions in the Schwinger-Keldysh path integral, 
and
\be \label{eq2}
G^{ar}_{\mu\nu}(x)=\langle A^a_\mu(x) A^r_\nu(0)\rangle_{SK}=G^{ra}_{\nu\mu}(-x)\,,
\ee
where we use translational invariance of the system. 
Since what we encounter in writing down our integral equations in the main text is the combination $\left(G^{ra}_{\mu\nu}(q)-G^{ar}_{\mu\nu}(q)\right)$, let us naturally define the photon spectral density (including possible P-odd contributions in general)
\be
\rho_{\mu\nu}^{\rm ph}(q)\equiv G^{ra}_{\mu\nu}(q)-G^{ar}_{\mu\nu}(q)\,.
\ee
We will show that $\rho^{\rm ph}_{\mu\nu}(q)$ is in general a hermitian matrix in terms of $\mu\nu$ indices. For diagonal components that come from the usual P-even contributions, $\rho^{\rm ph,P-even}_{\mu\nu}(q)$ is therefore real. For P-odd contribution which turns out to be anti-symmetric in spatial $ij$ indices (there is no P-odd contribution to time-like component, at least up to linear order in $\mu$), we thus have $\rho^{\rm ph,P-odd}_{ij}(q)$ purely imaginary. 

To show that $\rho^{\rm ph}_{\mu\nu}(q)$ is a hermitian matrix, recall that the usual retarded propagator is defined as
\be 
G^R_{\mu\nu}(x)=-i \theta(x^0)\langle [A_\mu(x),A_\nu(0)]\rangle=-i G^{ra}_{\mu\nu}(x)\,.
\ee
It is not difficult to show, using the hermiticity of $A_\mu$, that $G^R_{\mu\nu}(x)$ is real valued, and this is what it should be since the retarded propagator gives the response of the system in real time which must be real valued. Therefore, in Fourier space, one has
\be
G^R_{\mu\nu}(-q)=\left(G^R_{\mu\nu}(q)\right)^*\,,
\ee
which in turn gives
\be 
G^{ra}_{\mu\nu}(-q)=-\left(G^{ra}_{\mu\nu}(q)\right)^*\,.\label{eq3}
\ee
On the other hand, from (\ref{eq2}) we have
\be 
G^{ar}_{\mu\nu}(q)=G^{ra}_{\nu\mu}(-q)=-\left(G^{ra}_{\nu\mu}(q)\right)^*\,,
\ee
where we use (\ref{eq3}) in the last equality. Therefore
\be
\rho^{\rm ph}_{\mu\nu}(q)=G^{ra}_{\mu\nu}(q)-G^{ar}_{\mu\nu}(q)=G^{ra}_{\mu\nu}(q)+\left(G^{ra}_{\nu\mu}(q)\right)^*\,,
\ee
which proves that $\rho^{\rm ph}_{\mu\nu}(q)$ is indeed a hermitian matrix.

The P-odd part of the retarded current-current correlation functions, that is the retarded photon self energy, in Hard Thermal Loop (HTL) limit has been recently computed in literature \cite{Son:2012zy,Manuel:2013zaa}. We will work in the Coulomb gauge where $G^{ra}_{0i}=0$ ($i=1,2,3$).
The P-odd contribution appears only in the spatial transverse part of the correlation functions, so we will discuss only the spatial transverse part of current correlation functions in the following. In matrix notation, the spatial part of the HTL resummed photon propagator is
\be
(G^{ra}(q))^{-1}=(G^{ra}_{(0)}(q))^{-1}-i\Sigma^R(q)\,,
\ee
where $G^{ra}_{(0)}(q)$ is the bare propagator which is given by
\be
G^{ra}_{(0)ij}(q)={-i P^T_{ij}(\vq)\over -(q^0+i\epsilon)^2+|\vq|^2}\,.
\ee
The HTL self-energy $\Sigma^R(q)$ including P-odd contribution is given by
\be
\Sigma^R_{ij}(q)=\Pi_T(q)P^T_{ij}(\vq)+i\,\Pi_{\rm odd}(q)\epsilon^{ijl}\vq^l\,,
\ee
where 
\bear\label{pis}
\Pi_T(q)&=&-{m_D^2\over 2}\left({(q^0)^2\over |\vq|^2}+\left({(q^0)^2\over |\vq|^2}-1\right){q^0\over 2 |\vq|}\log\left({q^0-|\vq|+i\epsilon\over q^0+|\vq|+i\epsilon}\right)\right)\,,\nonumber\\
\Pi_{\rm odd}(q)&=&-{e^2\mu\over 4\pi^2}\left(1-{(q^0)^2\over |\vq|^2}-\left({(q^0)^2\over |\vq|^2}-1\right){q^0\over 2 |\vq|}\log\left({q^0-|\vq|+i\epsilon\over q^0+|\vq|+i\epsilon}\right)\right)\label{pis}\,,
\eear
with $m_D^2=e^2(T^2/6+\mu^2/(2\pi^2))$. From this while keeping terms only up to linear in $\mu$, we have
\be
G^{ra}_{ij}(q)={-i P^T_{ij}(\vq)\over -(q^0)^2+|\vq|^2-\Pi_T(q)}+{\Pi_{\rm odd}(q)\over (-(q^0)^2+|\vq|^2-\Pi_T(q))^2}\epsilon^{ijl}\vq^l\,,
\ee
where the first term is the usual P-even HTL photon propagator, and the second term is the new P-odd contribution. 
The HTL photon spectral density $\rho^{\rm ph}_{ij}(q)$ is then given by
\be
\rho^{\rm ph}_{ij}(q)=\rho_T(q)P^T_{ij}(\vq)+i\rho_{\rm odd}(q)\epsilon^{ijl}\vq^l\,,
\ee
with
\be\label{withwith}
\rho_T(q)=2 \,{\rm Im}\left(1\over -(q^0)^2+|\vq|^2 -\Pi_T(q)\right)\,,\quad
\rho_{\rm odd}(q)= 2\, {\rm Im}\left(\Pi_{\rm odd}(q)\over (-(q^0)^2+|\vq|^2 -\Pi_T(q))^2\right)\,.
\ee
It is easy to see that $\rho_{\rm odd}(q)$ is an odd function in $q^0$, and what we need in the main text is the value of the integral
\be
J_n^{\rm odd}\equiv \int_{-|\vq|}^{|\vq|}{dq^0\over (2\pi)} (q^0)^{2n-1}\rho_{\rm odd}(q)\,,\quad n=0,1,2,\cdots\,.\label{jnodd}
\ee
One can compute them using the well-known sum-rule techniques exploring analytic property of the function $\Delta_{\rm odd}(q)$ defined by
\begin{equation}\label{deltaodd}
\Delta_{\text{odd}}(q)\equiv \frac{\Pi_{\rm odd}(q)}{(-(q^0)^2+|\vq|^2 -\Pi_T(q))^2}\,.
\end{equation}
We briefly sketch the procedure and present the results in two different regimes $|\vq|\ll m_D$ and $|\vq|\gg m_D$.

The starting point is the fact that $\Delta_{\rm odd}(q)$ in the complex $q^0$ plane is analytic in the upper half plane due to the causal nature of a retarded function. 
Thus, the integral
\be
\int_{-\infty}^\infty {dq^0\over 2\pi} {1\over q^0-\omega+i\epsilon}\Delta_{\rm odd}(q)=0\,,
\ee
vanishes for any real number $\omega$ by closing the contour with the upper hemi-circle at infinity (and $\Delta_{\rm odd}(q)\to 0$ sufficiently fast as $|q^0|\to\infty$). From $1/(q^0-\omega+i\epsilon)={\cal P} 1/(q^0-\omega)-i\pi\delta(q^0-\omega)$ where ${\cal P}$ is the principal integration, we have
\be
{\cal P}\int_{-\infty}^\infty {dq^0\over 2\pi} {1\over q^0-\omega}\Delta_{\rm odd}(q)-{i\over 2}\,\Delta_{\rm odd}(\omega,|\vq|)=0\,.
\ee
Considering the imaginary part of the above, we obtain one of the Kramers-Kronig dispersion relations for a retarded function (the real part gives the other dispersion relation),
\be
{\cal P}\int_{-\infty}^\infty {dq^0\over 2\pi} {1\over q^0-\omega}\rho_{\rm odd}(q)={\rm Re}\left[\Delta_{\rm odd}(\omega,|\vq|)\right]\,.\label{kkdis}
\ee
Setting $\omega=0$ and using $\Delta_{\rm odd}(0,|\vq|)=-e^2\mu/(4\pi^2|\vq|^4)$, one gets a sum rule
\be
\int_{-\infty}^\infty {dq^0\over 2\pi}\, {1\over q^0}\,\rho_{\rm odd}(q)=-{e^2\mu\over 4\pi^2|\vq|^4}\,.\label{sumr1}
\ee
Other sum rules are obtained from (\ref{kkdis}) by expanding both sides in $\omega\to\infty$. The left-hand side becomes
\be
-\sum_{n=0}^\infty {1\over \omega^{n+1}}\int_{-\infty}^\infty {dq^0\over 2\pi} { (q^0)^n}\rho_{\rm odd}(q)\,,
\ee while the right-hand side when expanded in large $\omega$ is $-e^2\mu/(12\pi^2 \omega^4)+{\cal O}(1/\omega^6)$, which gives other two sum rules,
\be
\int_{-\infty}^\infty {dq^0\over 2\pi}\, {q^0}\,\rho_{\rm odd}(q)=0\,,\quad \int_{-\infty}^\infty {dq^0\over 2\pi}\, {(q^0)^3}\,\rho_{\rm odd}(q)={e^2\mu\over 12\pi^2}\,.\label{sumr2}
\ee

The sum rules (\ref{sumr1}) and (\ref{sumr2}) are not precisely $J_n^{\rm odd}$ defined in (\ref{jnodd}), as the integration range for $J_n^{\rm odd}$ is $[-|\vq|,+|\vq|]$, not $[-\infty,+\infty]$. The imaginary part of $\Delta(q)$ (that is, $\rho_{\rm odd}(q)$)
consists of two distinct parts: one part coming from a branch cut just below the real line along the interval $q^0\in [-|\vq|,+|\vq|]$
from the logarithms in (\ref{pis}) (originated from the Landau damping), and the other part is from the two poles $\pm\omega_0$ satisfying $-(\omega_0)^2+|\vq|^2 -\Pi_T(\omega_0,|\vq|)=0$ corresponding to the time-like transverse photons in the medium.
The former has a continuous support in the interval $q^0\in [-|\vq|,+|\vq|]$, and hence contributes to $J_n^{\rm odd}$, while the latter pole contributions sit outside the interval, $\omega_0>|\vq|$, so do not contribute to $J_n^{\rm odd}$. Therefore, the only difference between the sum rule values in (\ref{sumr1}), (\ref{sumr2}) and the $J_n^{\rm odd}$ is simply the latter pole contributions which we can compute. 

Near the pole location $q^0\approx \omega_0-i\epsilon$, we have the expansion 
\be
(q^0)^{2n-1}2 \Delta_{\rm odd}(q)\approx {A\over (q^0-\omega+i\epsilon)^2}+{B\over q^0-\omega_0+i\epsilon}+{\rm \{regular\}}\,,
\ee
where
\be 
B={2(\omega_0)^{2n-1}\Pi_{\rm odd}(\omega_0)\over \left(2\omega_0+\Pi'_T(\omega_0)\right)^2}\left({\Pi'_{\rm odd}(\omega_0)\over \Pi_{\rm odd}(\omega_0)} +{(2n-1)\over\omega_0}-{2+\Pi''_T(\omega_0)\over 2\omega_0+\Pi'_T(\omega_0)}\right)\,,
\ee
with $\Pi'_T(q^0)\equiv d\Pi_T(q^0,|\vq|)/d q^0$, etc. The first double pole does not contribute to the imaginary part in $\epsilon\to 0$ limit while the second part contributes to $(q^0)^{2n-1}\rho_{\rm odd}(q)$ as $-\pi B\,\delta(q^0-\omega_0)\times 2=-2\pi B\,\delta(q^0-\omega_0)$ (the factor $2$ comes from having two poles $\pm\omega_0$), leading to the difference between the sum rules values in (\ref{sumr1}), (\ref{sumr2}) and the $J_n^{\rm odd}$ being given by $-B$, that is, $J_n^{\rm odd}$ is obtained by adding $B$ to the sum rule values in (\ref{sumr1}), (\ref{sumr2}).

In the case $|\vq|\ll m_D$, the pole location is $\omega_0\approx \sqrt{1/3}\,m_D(1+(9/5) |\vq|^2/m_D^2)+\cdots$
and an explicit computation of $B$ gives the values of $J_n^{\rm odd}$ in this regime as
\bear
J_0^{\rm odd}&\approx& -{e^2\mu\over 4\pi^2|\vq|^4}+{3 e^2\mu\over 4\pi^2 m_D^4}+{\cal O}(|\vq|^2/m_D^6)\,,\nonumber\\
J_1^{\rm odd}&\approx& -{3 e^2\mu|\vq|^2\over 5 \pi^2 m_D^4}+{\cal O}(|\vq|^4/m_D^6)\,,\nonumber\\
J_2^{\rm odd}&\approx& -{33 e^2\mu|\vq|^4\over 700\pi^2 m_D^4}+{\cal O}(|\vq|^6/m_D^6)\,.\label{jnoddfirst}
\eear
On the other hand, in the regime $|\vq|\gg m_D$, the poles are located in $\omega_0\approx |\vq|+(1/4)m_D^2/|\vq|+\cdots$, and we have the results in this regime $|\vq|\gg m_D$ as
\bear
J_0^{\rm odd}&\approx& {e^2\mu\over 8\pi^2|\vq|^4}\left(1+\log\left({m_D^2\over 8|\vq|^2}\right)\right)\,,\nonumber\\
J_1^{\rm odd}&\approx&{e^2\mu\over 8\pi^2|\vq|^2}\left(3+\log\left({m_D^2\over 8|\vq|^2}\right)\right)\,,\nonumber\\
J_2^{\rm odd}&\approx& {e^2\mu\over 8\pi^2}\left({11\over 3}+\log\left({m_D^2\over 8|\vq|^2}\right)\right)\,.\label{jnoddsecond}
\eear
The above results (\ref{jnoddfirst}) and (\ref{jnoddsecond}) will be used in the main text in section 3.

\section*{Appendix 2: Hard fermion damping rate }\label{app:HFDR}

In this appendix, we compute the damping rate of hard fermion including possible dependence on the chemical potential $\mu$.
Our primary objective is two-fold: we first would like to confirm that the integral that we have in (\ref{sp}) is indeed precisely equal to the damping rate induced by soft photon scatterings at full order in $e$ and $\mu$, which was instrumental in rewriting the integral equation to take the form (\ref{chieq2}) that contains only $e^4\log(1/e)$ terms.
Our second objective is to find a linear $s \mu$ dependence of the $\zeta_{\vp,s}^{\rm sf}$, that is, in the damping rate induced by soft fermion scatterings (or equivalently, fermion conversion-to-photon processes). This $s\mu$ dependence in $\zeta_{\vp,s}^{\rm sf}$ is important in finding the correct $\mu$ dependence in the solution of differential equation (\ref{diffchi}) for $\chi_s(|\vp|)$, which is crucial to get the correct result for $\xi_5$ as well as $\mu^2$ correction to usual electric conductivity.

\begin{figure}[h]
\centering
\includegraphics[width=460pt]{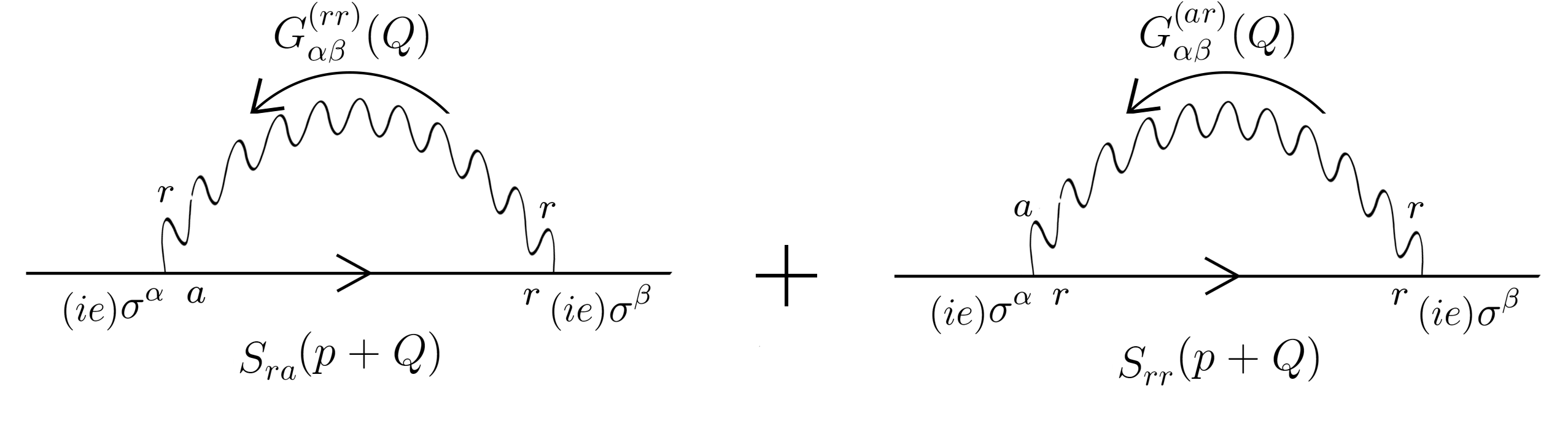}
\caption{Real-time Feynman diagrams for the retarded fermion self energy.\label{apfig1}}
\end{figure}
Let us start with the self energy resummed ra-propagator (which is equal to $i$ times of retarded propagator)
\be
S^{ra}(p)=S_{(0)}^{ra}(p)+S_{(0)}^{ra}(p)\Sigma^{ra}(p)S_{(0)}^{ra}(p)+\cdots=S_{(0)}^{ra}(p){1\over 1-\Sigma^{ra}(p)S_{(0)}^{ra}(p)}\,,
\ee
which gives
\be
(S^{ra}(p))^{-1}=(S^{ra}_{(0)}(p))^{-1}-\Sigma^{ra}(p)\,,\label{invsra}
\ee
where 
\be
S_{(0)}^{ra}(p)=\sum_s {i\over p^0-s|\vp|+i\epsilon}{\cal P}_s(\vp)\,,
\ee  is the bare ra-propagator, and the self-energy $\Sigma^{ra}(p)$ (which is a $2\times2$ matrix in the spinor space) at naive lowest order in coupling is given by two Feynman diagrams in Fig.\ref{apfig1} in real-time formalism with the expression
\be
\Sigma^{ra}(p)=(ie)^2\sigma^\beta \int{d^4Q\over (2\pi)^4}\left[G^{rr}_{\alpha\beta}(Q) S^{ra}(p+Q)+G^{ar}_{\alpha\beta}(Q) S^{rr}(p+Q)\right]\sigma^\alpha\,,
\ee
with the photon propagators $G^{ab}_{\alpha\beta}(Q)$ ($a,b=r,a$) (see the Appendix 1 for our notational conventions).
In the above expression, we haven't specified whether the propagators appearing in the loop are bare or HTL resummed ones, since depending on the situations we can
consider different approximations for them to get the right leading order quantities. For example, if the external momentum $p$ is soft and one is interested in the Hard Thermal Loop (HTL) approximation, it is enough to consider hard loop momentum $Q$ and the both propagators in the loop are the bare ones. On the other hand, in the case of damping rate with a hard momentum $p$, which is proportional to the imaginary part of the self-energy, the leading contribution comes from when one of the two loop propagators carries soft momentum (that is, either $Q$ or $p+Q$), and the soft propagator must then be the HTL resummed propagator while the other hard propagator is the bare one.

Rotational invariance dictates the self energy to take a form
\be
\Sigma^{ra}(p)=A(p^0,|\vp|)+B(p^0,|\vp|)\hat{\vp}\cdot\vec\sigma\equiv i\sum_{s=\pm} \Sigma^R_s(p){\cal P}_s(\vp)\,,\label{sigmar}
\ee
where 
\be
\Sigma^R_s(p) = -i\left(A(p^0,|\vp|)+s\, B(p^0,|\vp|)\right)\,.
\ee
From this and (\ref{invsra}) we have
\be
 S^{ra}(p)=\sum_s {i\over p^0-s|\vp|+\Sigma^R_s(p)}{\cal P}_s(\vp)\,.
\ee
In deriving above, we use the following properties of the projection operators to find the inverse of $S^{ra}(p)$, 
\be
{\cal P}_++{\cal P}_-={\bf 1}\,,\quad {\cal P}_\pm^2={\cal P}_\pm\,,\quad {\cal P}_+{\cal P}_-={\cal P}_-{\cal P}_+=0\,.
\ee
By comparing the above expression for $S^{ra}(p)$ with the one in (\ref{ra}), we see that the damping rate is given by the imaginary part of retarded self energy $\Sigma^R_s(p)$ at on-shell momentum $p^0=s|\vp|$,
\be
\zeta_{\vp,s}=2 \,{\Im}\left[\Sigma^R_s(p)\right]\bigg|_{p^0=s|\vp|}\,.
\ee
In the following, we will hence only concern about the imaginary part of $\Sigma^R_s(p)$.
From (\ref{sigmar}) and ${\rm tr}[{\cal P}_\pm]=1$, we have
\bear
\Sigma^R_s(p)&=&(-i)\,{\rm tr}\left[{\cal P}_s(\vp) \Sigma^{ra}(p)\right]\nonumber\\
&=& ie^2\,{\rm tr}\left[{\cal P}_s(\vp)\sigma^\beta
\int{d^4Q\over (2\pi)^4}\left[G^{rr}_{\alpha\beta}(Q) S^{ra}(p+Q)+G^{ar}_{\alpha\beta}(Q) S^{rr}(p+Q)\right]\sigma^\alpha\right]\,,\nonumber\\\label{stp}
\eear
which will be the starting point of our computation.

For a soft $p$, if one uses the HTL approximation to the retarded self energy $\Sigma^R_s(p)$, the result is the HTL fermion propagator. 
For a hard $p$, the HTL self energy is sub-leading in $e^2$ so can be negligible, and moreover its imaginary part at on-shell momentum $p^0=s|\vp|$, which would give a damping rate that could regularize pinch singularities, vanishes due to kinematic constraints. 
The leading contribution to the imaginary part of $\Sigma^R_s(p)$ at on-shell momentum arises when either $Q$ or $p+Q$ is soft, so that the corresponding propagator in the loop is the HTL resummed one. Calling the case of soft $Q$ the soft-photon contribution, $\zeta^{\rm sp}_{\vp,s}$, and the other case of soft $p+Q$ the soft-fermion contribution, $\zeta^{\rm sf}_{\vp,s}$, the total damping rate is the sum of the two, $\zeta_{\vp,s}=\zeta^{\rm sp}_{\vp,s}+\zeta^{\rm sf}_{\vp,s}$. 

Let us discuss $\zeta^{\rm sp}_{\vp,s}$ first. Since the fermion propagator is the bare one, we have after putting on-shell momentum $p^0=s|\vp|$
\bear
S^{ra}(p+Q)&=&\sum_{t=\pm}{i\over s|\vp|+q^0-t|\vp+\vq|+i\epsilon}{\cal P}_t(\vp+\vq)\nonumber\\
&=&\sum_t \left[P\,{i\over {s|\vp|+q^0-t|\vp+\vq|}}+\pi\delta\left(s|\vp+q^0-t|\vp+\vq|\right)\right]{\cal P}_t(\vp+\vq)\,,\nonumber\\\label{sraap}
\eear
where $P$ denotes principal value. Similarly,
\be
S^{rr}(p+Q)=\left({1\over 2}-n_+(s|\vp|+q^0)\right)\sum_t (2\pi)\delta(s|\vp|+q^0-t|\vp+\vq|){\cal P}_t(\vp+\vq)\,.\label{srrap}
\ee
Looking at the structure of (\ref{stp}), we have a spinor trace appearing
\be
{\rm tr}\left[{\cal P}_s(\vp)\sigma^\beta {\cal P}_t(\vp+\vq) \sigma^\alpha\right]\equiv H_{st}^{\beta\alpha}(\vp,\vq)\,,
\ee
which is a hermitian matrix in terms of $\alpha\beta$ indices (this can be shown easily using hermitian nature of $\sigma^\alpha$ and ${\cal P}_\pm$). Since 
\be
G^{rr}_{\alpha\beta}(Q)=\left({1\over 2}+n_B(q^0)\right)\left(G^{ra}_{\alpha\beta}(Q)-G^{ar}_{\alpha\beta}(Q)\right)\equiv
\left({1\over 2}+n_B(q^0)\right)\rho^{\rm ph}_{\alpha\beta}(Q)\,,
\ee
is also a hermitian matrix with $\left(G^{ar}_{\alpha\beta}(Q)\right)^*=-G^{ra}_{\beta\alpha}(Q)$ as shown in the Appendix 1, we see that $H^{\beta\alpha}_{st}(\vp,\vq)G^{rr}_{\alpha\beta}(Q)$ is a real number. Therefore, one sees that the imaginary part of $\Sigma^R_s(p)$ given in (\ref{stp}) arises only from the second $\delta$-function term in (\ref{sraap})
 when used in the first term of (\ref{stp}). Similarly, from
 \be
 \left(H^{\beta\alpha}_{st}(\vp,\vq)G^{ar}_{\alpha\beta}(Q)\right)^*=-H^{\beta\alpha}_{st}(\vp,\vq)G^{ra}_{\alpha\beta}(Q)\,,
 \ee
and 
\be
H^{\beta\alpha}_{st}(\vp,\vq)\rho^{\rm ph}_{\alpha\beta}(Q)=\left(H^{\beta\alpha}_{st}(\vp,\vq)G^{ra}_{\alpha\beta}(Q)-H^{\beta\alpha}_{st}(\vp,\vq)G^{ar}_{\alpha\beta}(Q)\right)\,,
\ee
we see that the real part of $H^{\beta\alpha}_{st}(\vp,\vq)G^{ar}_{\alpha\beta}(Q)$ is equal to $-(1/2)H^{\beta\alpha}_{st}(\vp,\vq)\rho^{\rm ph}_{\alpha\beta}(Q)$.
With (\ref{srrap}) the imaginary part of $\Sigma^R_s(p)$ from the second term in (\ref{stp}) only comes from the real part of $H^{\beta\alpha}_{st}(\vp,\vq)G^{ar}_{\alpha\beta}(Q)$, and therefore we can effectively replace $G^{ar}_{\alpha\beta}(Q)$ appearing in (\ref{stp}) with $-(1/2)\rho^{\rm ph}_{\alpha\beta}(Q)$ for the purpose of damping rate computation.
One then observes that the pieces in (\ref{sraap}) and (\ref{srrap}) that contribute to the damping rate are all proportional to the $\delta$-function $\delta(s|\vp|+q^0-t|\vp+\vq|)$ which has a non-zero support only for $t=s$ since $Q$ is assumed to be soft while $p$ is hard. 
After collecting all these pieces contributing to the imaginary part of $\Sigma^R_s(p)$, we finally have after some algebra,
\bear
\zeta^{\rm sp}_{\vp,s}&=&2\,{\rm Im}\left[\Sigma^R_s(p)\right]\nonumber\\
&=&e^2\int{d^4Q\over (2\pi)^4}\left(n_B(q^0)+n_+(s|\vp|+q^0)\right) H^{\beta\alpha}_{ss}(\vp,\vq)\rho^{\rm ph}_{\alpha\beta}(Q)(2\pi)\delta(s|\vp|+q^0-s|\vp+\vq|)\,,\nonumber\\
\eear
where
\be
H^{\beta\alpha}(\vp,\vq)_{ss}={\rm tr}\left[{\cal P}_s(\vp)\sigma^\beta {\cal P}_s(\vp+\vq) \sigma^\alpha\right]=\Sigma^\beta_s(\vp,\vq)\left(\Sigma_s^\alpha(\vp,\vq)\right)^* {1\over 2}\left(1+\hat{\vp}\cdot\widehat{\vp+\vq}\right)\,,
\ee
using the notations introduced in (\ref{functions}). From the fact that $F_s(\vp;\vq,0)$ introduced in (\ref{functions}) is equal to ${1\over 2}\left(1+\hat{\vp}\cdot\widehat{\vp+\vq}\right)$, and recalling our definition of kernel function in (\ref{kn})
\be
{\cal K}_s(\vp,0;Q)\equiv \Sigma^\beta_s(\vp,\vq) \left(\Sigma^\alpha_s(\vp,\vq)\right)^* \rho^{\rm ph}_{\alpha\beta}(Q) F_s(\vp;\vq,0)(n_+(s|\vp|+q^0)+n_B(q^0))\,,
\ee
we see that $\zeta^{\rm sp}_{\vp,s}$ is indeed equal to
\be
\zeta^{\rm sp}_{\vp,s}=e^2\int {d^4Q\over (2\pi)^4} {\cal K}_s(\vp,0;Q)(2\pi)\delta(s|\vp|+q^0-s|\vp+\vq|)\,,
\ee
which is precisely what appears in (\ref{sp}) and in the integral equation, which is crucial to have (\ref{chieq2}).

Although we don't need the value of $\zeta_{\vp,s}^{\rm sp}$ in this work, it is easy to compute it from the above expression. From (\ref{deltaid})
 \be
\int {d^4Q\over (2\pi)^4} (2\pi)\delta(s|\vp|+q^0-s|\vp+\vq|)=\int^\infty_0{d|\vq||\vq|\over (2\pi)}\int_{-|\vq|}^{|\vq|}{dq^0\over (2\pi)} \left(1+{s q^0\over |\vp|}\right)\bigg|_{\hat{\vp}\cdot\vq\to sq^0+{(q^0)^2-|\vq|^2\over 2|\vp|}}\,,\label{deltaid2}
\ee
and the small $Q$ expansion of ${\cal K}_s(\vp,0;Q)$,
\be 
  {\cal K}_s(\vp,0;Q)\approx {1\over\beta q^0}\left(\rho_L(Q)+\rho_T(Q)\left(1-{(q^0)^2\over|\vq|^2}\right)\right)\,,
  \ee
 where $\rho_{L/T}(Q)$ are P-even longitudinal and transverse photon spectral densities defined by 
 \be
 \rho^{\rm ph}_{00}(Q)=\rho_L(Q)\,,\quad \rho^{\rm ph}_{ij}(Q)=\rho_T(Q)\left(\delta_{ij}-{\vq_i\vq_j\over|\vq|^2}\right)+i\rho_{\rm odd}(Q)\epsilon^{ijl}\vq_l\,,
 \ee we have
 \be
 \zeta^{\rm sp}_{\vp,s}\approx e^2\int^\infty_0{d|\vq||\vq|\over (2\pi)}\int_{-|\vq|}^{|\vq|}{dq^0\over (2\pi)} {1\over\beta q^0}\left(\rho_L(Q)+\rho_T(Q)\left(1-{(q^0)^2\over|\vq|^2}\right)\right)\,.
 \ee
 The rest is to use the sum rules for $\rho_{L/T}(Q)$ that can be derived by the same way we derive the sum rules for the P-odd part in Appendix 1 \cite{Blaizot:2001nr}. The leading log arises from the momentum region $|\vq|\ll m_D$ and only from the transverse part for which we have
 \be
 \int^{|\vq|}_{-|\vq|}{dq^0\over (2\pi)} {1\over q^0}\rho_T(Q)={1\over |\vq|^2}+{\cal O}\left(1\over m_D^2\right)\,,
 \quad
 \int^{|\vq|}_{-|\vq|}{dq^0\over (2\pi)} {q^0}\rho_T(Q)={3|\vq|^2\over 5 m_D^2}+{\cal O}\left(|\vq|^4\over m_D^4\right)\,,
 \ee 
 and this gives
 \be
 \zeta^{\rm sp}_{\vp,s}\approx e^2{1\over\beta}\int_{e^2T}^{m_D}{d|\vq| |\vq|\over (2\pi)}\, {1\over |\vq|^2}
 \approx {e^2 \log(1/e)T\over 2\pi}\,,
 \ee
 where we put an IR cutoff of order $\Lambda_{\rm IR}\sim e^2T$. Strictly speaking, the $e^2T$ (or $g^2T$ for non-abelian theory) magnetic confinement scale exists only for non-abelian theory, while an abelian QED which becomes free at $Q\ll m_f$ does not possess any IR cutoff. In this case, the damping rate $\zeta_{\vp,s}^{\rm sp}$ is not a useful concept \cite{Blaizot:1996hd}, and the effective IR cutoff is provided by the time-scale one is looking at, so the hard fermions decay in time $t$ as \cite{Blaizot:1996hd}
 \be
 e^{-{\zeta^{\rm sp}_{\vp,s}} t/2}\bigg|_{\Lambda_{\rm IR}\sim {1/t}}\sim \left(m_D t\right)^{-{e^2 T\over 4\pi} t}
\,. \ee
 Since we are ultimately interested in QCD (see our discussion in section 6), we don't worry about this any more.
 Another aspect is that in realistic situations, the free nature of QED at $Q\ll eT$ means that this scale is never thermalized anyway. Since the damping rate arises from the scattering of fermion with thermally excited soft photons in this scale (recall $n_B(q^0)\sim 1/q^0$ term in the above), we wouldn't have these contributions in realistic situations in any case. This also justifies our use of $\Lambda_{\rm IR}\sim e^2T$ in the above.

Let us next compute the soft-fermion contribution to the damping rate, $\zeta^{\rm sf}_{\vp,s}$, with our main objective being to find a linear $s\mu$-dependence.
Since $p+Q$ is soft, it is convenient to shift the loop momentum $Q\to Q-p$ to have
 \bear
\Sigma^R_s(p)
= ie^2\,{\rm tr}\left[{\cal P}_s(\vp)\sigma^\beta
\int{d^4Q\over (2\pi)^4}\left[G^{rr}_{\alpha\beta}(Q-p) S^{ra}(Q)+G^{ar}_{\alpha\beta}(Q-p) S^{rr}(Q)\right]\sigma^\alpha\right]\,,\nonumber\\\label{stp2}
\eear
where now $Q$ is soft, and we need to use HTL resummed fermion propagators while the bare propagators are used for photon propagators.
The HTL resummed fermion ra-propagator is written as
\be
S^{ra}(Q)=\sum_t {i\over q^0-t|\vq|+\Sigma^{R,{\rm HTL}}_t(Q)}{\cal P}_t(\vq)\,,\label{htlf}
\ee 
where $\Sigma^{R,{\rm HTL}}_t(Q)$ is the HTL self-energy. An explicit computation gives (in fact, one uses the same expression (\ref{stp2}) with now both $Q$ and $p+Q$ being hard)
\be
\Sigma_t^{R,{\rm HTL}}(Q)=-{m^2_f\over 4|\vq|}\left(2t+\left(1-t{q^0\over |\vq|}\right)\log\left({q^0+|\vq|+i\epsilon\over q^0-|\vq|+i\epsilon}\right)\right)\,,
\ee
where 
\be
m^2_f={e^2\over 4}\left(T^2+{\mu^2\over \pi^2}\right)\,,
\ee
is the asymptotic thermal mass of fermions for a single Weyl fermion system. Since $\mu$ dependence is only quadratic for $\Sigma_t^{R,{\rm HTL}}(Q)$, we can ignore this dependence in the HTL self-energy to use $\mu=0$ result of $\Sigma_t^{R,{\rm HTL}}(Q)$.
This means that $S^{ra}(Q)$ and $S^{ar}=-(S^{ra}(Q))^\dagger$ can be replaced by their $\mu=0$ values up to linear in $\mu$ of our interest. On the other hand, the rr-propagator which is given by
\be
S^{rr}(Q)=\left({1\over 2}-n_+(q^0)\right)\left(S^{ra}(Q)-S^{ar}(Q)\right)\,,
\ee
does contain a linear $\mu$ dependence via its statistical factor in front, $n_+(q^0)$. We will indeed find shortly that this will be the (only) source of the final $s\mu$-dependence of $\zeta_{\vp,s}^{\rm sf}$.

Since the HTL resummed ra-propagator (\ref{htlf}) is analytic in the upper $q^0$ complex plane, one can introduce real spectral densities $\rho_\pm(Q)$ by
\be
{1\over q^0-t|\vq|+\Sigma^{R,{\rm HTL}}_t(Q)}=\int^\infty_{-\infty}{d\omega\over(2\pi)} {\rho_t(\omega,\vq)\over q^0-\omega+i\epsilon}=P\int^\infty_{-\infty}{d\omega\over(2\pi)} {\rho_t(\omega,\vq)\over q^0-\omega}-{i\over 2}\rho_t(q^0,\vq)\,,
\ee
or equivalently
\be
\rho_t(Q)=-2\,{\rm Im}\left[{1\over q^0-t|\vq|+\Sigma^{R,{\rm HTL}}_t(Q)}\right]\,,
\ee
in terms of which we have
\be
S^{rr}(Q)=\left({1\over 2}-n_+(q^0)\right)\sum_t \rho_t(Q){\cal P}_t(\vq)\,.
\ee
Introducing 
\be
L_{st}^{\beta\alpha}(\vp,\vq)\equiv {\rm tr}\left[{\cal P}_s(\vp)\sigma^\beta {\cal P}_t(\vq) \sigma^\alpha\right]\,,
\ee
which is a hermitian matrix, a similar discussion to that we have above leads us to replace
\be
S^{ra}(Q)\to {1\over 2} \rho_t(Q){\cal P}_t(\vq)\,,
\ee 
for computing the imaginary part of $\Sigma^R_s(p)$ in (\ref{stp2}).

On the other hand, the hard photon propagators in (\ref{stp2}) are bare ones. In the Coulomb gauge we have
\be
G^{ra}_{00}(p)={i\over |\vp|^2}\,,\quad G^{ra}_{ij}(p)={-i P^T_{ij}(\vp)\over -(p^0+i\epsilon)^2+|\vp|^2}\,,
\ee
where $P^T_{ij}(\vp)=\delta_{ij}-\hat{\vp}_i\hat{\vp}_j$ is the transverse projection operator, from which we have the bare photon spectral density as
\be
\rho^{\rm ph}_{00}(p)=0\,,\quad \rho^{\rm ph}_{ij}(p)=(2\pi) P^T_{ij}(\vp){\rm sign}(p^0)\delta\left((p^0)^2-|\vp|^2\right)\,,
\ee
with $G^{rr}_{\alpha\beta}(p)=\left({1/2}+n_B(p^0)\right) \rho^{\rm ph}_{\alpha\beta}(p)$. As before, for the imaginary part of $\Sigma^R_s(p)$ in (\ref{stp2}), we can replace
\be
G^{ar}_{\alpha\beta}(Q-p)\to -{1\over 2}\rho^{\rm ph}_{\alpha\beta}(Q-p)\,.
\ee
Collecting all these elements, the expression for $\zeta^{\rm sf}_{\vp,s}$ becomes
\bear
\zeta^{\rm sf}_{\vp,s}&=& e^2\sum_{t=\pm}\int {d^4 Q\over (2\pi)^4} L^{ji}_{st}(\vp,\vq) P^T_{ij}(\vq-\vp)\left(n_B(q^0-s|\vp|)+n_+(q^0)\right)\rho_t(Q)\nonumber\\&\times&{\rm sign}(q^0-s|\vp|)(2\pi)\delta((s|\vp|-q^0)^2-|\vp-\vq|^2)\,.
\eear
Since $Q$ is soft while $p$ is hard, we have ${\rm sign}(q^0-s|\vp|)=-s$ and
\be
\delta((s|\vp|-q^0)^2-|\vp-\vq|^2)={1\over 2(|\vp|-s q^0)}\delta(s|\vp|-q^0-s|\vp-\vq|)\,,
\ee
and using the identity
\be
n_B(q^0-s|\vp|)+n_+(q^0)=(-s)\left(n_B(|\vp|-sq^0)+n_{-s}(-s q^0)\right)\,,
\ee
we have
\bear
\zeta^{\rm sf}_{\vp,s}&=& e^2\sum_{t=\pm}\int {d^4 Q\over (2\pi)^4} L^{ji}_{st}(\vp,\vq) P^T_{ij}(\vq-\vp)\left(n_B(|\vp|-sq^0)+n_{-s}(-sq^0)\right)\rho_t(Q)\nonumber\\&\times&{1\over 2(|\vp|-sq^0)}(2\pi)\delta(s|\vp|-q^0-s|\vp-\vq|)\,.
\eear
It is straightforward to compute the leading log part of the above integral by expanding the integrand in powers of $Q/T$ or $Q/|\vp|$, the same procedure we use several times before. 
From 
$
L^{ji}_{st}(\vp,\vq) P^T_{ij}(\vq-\vp)\approx 1-st {\hat\vp\cdot\vq\over |\vq|}+{\cal O}(Q)=1-t{q^0\over |\vq|}+{\cal O}(Q)
$
and $n_B(|\vp|-sq^0)+n_{-s}(-sq^0)\approx n_B(|\vp|)+n_{-s}(0)+{\cal O}(Q)$,
the leading log comes from the expression
\be
\zeta_{\vp,s}^{\rm sf}={e^2\over 2|\vp|}\left(n_B(|\vp|)+n_{-s}(0)\right)\sum_t\int^\infty_{0}{d|\vq||\vq|\over 2\pi}\int^{|\vq|}_{-|\vq|}{dq^0\over 2\pi}
\left(1-t{q^0\over |\vq|}\right)\rho_t(Q)\,,
\ee
and using the sum rules\footnote{We point out that what is called $m_f^2$ in Ref.\cite{Aarts:2002tn} is in fact plasmino frequency $\omega_f^2$ which is equal to $m_f^2/2$ in terms of asymptotic thermal mass $m_f^2$.} \be
J_0^\pm={m_f^2\over 4|\vq|^2}\left(\log{4|\vq|^2\over m_f^2} -1\right)\,,\quad
J_1^\pm=\pm {m_f^2\over 4|\vq|}\left(\log{4|\vq|^2\over m_f^2} -3\right)\,,
\ee
where 
\be
J_n^\pm\equiv \int^{|\vq|}_{-|\vq|}{dq^0\over 2\pi} (q^0)^n \rho_\pm(Q)\,,
\ee
we finally have
\be
\zeta_{\vp,s}^{\rm sf}={e^2 \over 4\pi}{m_f^2\log(1/e)\over |\vp|}\left(n_B(|\vp|)+n_{-s}(0)\right)\,.\label{sfresult}
\ee

There exists $s\mu$ dependence in the result (\ref{sfresult}) via $n_{-s}(0)=1-n_s(0)\approx 1/2-(1/4)s\beta\mu$, which can be easily understood as follows.
The soft-fermion contribution to the damping rate comes from the process where a hard fermion (of type $s$) becomes a soft-fermion (of the same type $s$) by emitting a hard photon with almost the same momentum. The rate is proportional to $(n_B(|\vp|)+1)(1-n_s(0))$ where $(1-n_s(0))$ is the Pauli blocking factor of the final soft-fermion state, where we can put zero for soft-momentum at leading order in coupling. A similar process is where a hard fermion (of type $s$) meets with a soft-antifermion (of the type $-s$) to become a hard photon, and this rate is proportional to $(n_B(|\vp|)+1)n_{-s}(0)$ where $n_{-s}(0)$ is the number density of soft-antifermion. 
The time reversed processes also add up to the damping rate, which is a property of fermionic case. These are each proportional to $n_B(|\vp|)n_s(0)$ and $n_B(|\vp|)(1-n_{-s}(0))$. Using $n_{-s}(0)+n_s(0)=1$, the total sum can be found to be
\bear
&&(n_B(|\vp|)+1)(1-n_s(0))+(n_B(|\vp|)+1)n_{-s}(0)+n_B(|\vp|)n_s(0)+n_B(|\vp|)(1-n_{-s}(0))\nonumber\\
&&=2\,(n_B(|\vp|)+n_{-s}(0))\,,
\eear
which nicely explains our result (\ref{sfresult}).

\section*{Appendix 3: Expression of $F_s(\vp,\vq;\vk)$}

The function $F_s(\vp,\vq;\vk)$ is given by 
\be
F_s(\vp,\vq;\vk)={A\over 4\left(1+\hat\vp\cdot\widehat{\vp+\vk}\right)|\vp||\vp+\vq||\vp+\vk||\vp+\vq+\vk|}\,,
\ee
where 
\bear
A&=& \left(|\vp||\vp+\vq|+|\vp|^2+\vp\cdot\vq\right)\left(|\vp+\vk||\vp+\vk+\vq|+|\vp+\vk|^2+(\vp+\vk)\cdot\vq\right)\nonumber\\
&+&\left(|\vp||\vp+\vk|+|\vp|^2+\vp\cdot\vk\right)\left(|\vp+\vq||\vp+\vq+\vk|+|\vp+\vq|^2+(\vp+\vq)\cdot\vk\right)\nonumber\\
&-&\left(|\vp||\vp+\vq+\vk|-\vp\cdot(\vp+\vq+\vk)\right)\left(|\vp+\vq||\vp+\vk|-(\vp+\vq)\cdot(\vp+\vk)\right)\nonumber\\
&+&is \left(|\vp|+|\vp+\vq|+|\vp+\vk|+|\vp+\vq+\vk|\right)\epsilon^{ijl}\vp^i\vq^j\vk^l\,.
\eear
For the reasons mentioned in the main text, we are only interested in this quantity to linear order in the external momentum $\vk$ and to second order in the loop momentum $\vq$. To this order the function $F_s(\vp,\vq;\vk)$ is given by

\begin{align}
&F_s(\vp,\vq;\vk)\sim 1 + \frac{1}{4|\vp|^2}\left( (\vp\cdot \vq)^2 - |\vq|^2  \right)\nonumber\\
&+\frac{1}{2|\vp|^3} s i \epsilon^{ijl}\vp_i \vq_j \vk_l+\frac{1}{4|\vp|^4}\left( \vp \cdot \vk \left(|\vq|^2-2(\vp \cdot \vq)^2\right) +(\vp \cdot \vk)(\vp \cdot \vq) -3 s i \epsilon^{ijl}\vp_i \vq_j \vk_l  (\vp \cdot \vq)  \right)\,.
\end{align}

\section*{Appendix 4: $\mu^2$ correction to electric conductivity}

Our analysis in this work contains all the necessary ingredients to compute the full $\mu^2$ correction to the usual P-even electric conductivity at leading log order. The electric conductivity is given from $\chi_s(|\vp|)$ by
\be
\sigma=-{e^2\over 3}\int {d^{3}\vp\over (2\pi)^{3}}\,\,\sum_{s=\pm} \left(dn_+(p^0)\over dp^0\right)\bigg|_{p^0=s|\vp|}\chi_s(|\vp|)\,,
\ee
where $\chi_s(|\vp|)$ satisfies the second order differential equation written in (\ref{diffchi}),
\be
\zeta_{\vp,s}^{\rm sf}\chi_s(|\vp|)=s-{e^2 m_D^2\log(1/e)\over 4\pi}\left({1\over \beta|\vp|^2}\chi_s(|\vp|)-\left({1\over \beta|\vp|}+n_s(|\vp|)-{1\over 2}\right)\chi_s'(|\vp|)-{1\over 2\beta}\chi_s''(|\vp|)\right)\,,\label{diffchi2}
\ee 
where the soft-fermion contribution to the damping rate $\zeta^{\rm sf}_{\vp,s}$ is given by (\ref{sfresult}),
\be
\zeta_{\vp,s}^{\rm sf}={e^2 \over 4\pi}{m_f^2\log(1/e)\over |\vp|}\left(n_B(|\vp|)+n_{-s}(0)\right)\,.\label{sfresult2}
\ee
To correctly take into account $\mu^2$ corrections, we need to restore full expressions for $m_D^2$ and $m_f^2$ including $\mu^2$ corrections,
\be
m_D^2={e^2\over 6}\left(T^2+{3\mu^2\over\pi^2}\right)\,,\quad m_f^2={e^2\over 4}\left(T^2+{\mu^2\over\pi^2}\right)\,.
\ee
Also, we have to expand $n_s(|\vp|)$ in (\ref{diffchi2}) and $n_{-s}(0)$ in (\ref{sfresult2}) up to second order in $\mu$,
\be
n_s(|\vp|)=n_0(|\vp|-s\mu)\approx n_0(|\vp|)-s\mu \,n'_0(|\vp|)+{1\over 2}n_0''(|\vp|)\mu^2\,,\quad \ee
and $n_{-s}(0)\approx1/2-s\beta\mu/4+{\cal O}(\mu^3)$, where $n_0(x)\equiv1/(e^{\beta x}+1)$.
The resulting $\chi_s(|\vp|)$ should be found up to $\mu^2$ order as
\be
\chi_s(|\vp|)=s\chi_{(0)}(|\vp|)+\mu\,\chi_{(1)}(|\vp|)+s\mu^2\,\chi_{(2)}(|\vp|)+{\cal O}(\mu^3)\,,
\ee
where $(\chi_{(0)},\chi_{(1)},\chi_{(2)})$ can be obtained from (\ref{diffchi2}) by solving it order by order in $\mu$.

\vskip 0.5cm

\vfil


\begin{thebibliography}{99} \frenchspacing





  \bibitem{Kharzeev:2007tn}
  D.~Kharzeev and A.~Zhitnitsky,
  ``Charge separation induced by P-odd bubbles in QCD matter,''
  Nucl.\ Phys.\ A {\bf 797}, 67 (2007).


\bibitem{Kharzeev:2007jp}
  D.~E.~Kharzeev, L.~D.~McLerran and H.~J.~Warringa,
  ``The Effects of topological charge change in heavy ion collisions: 'Event by event P and CP violation',''
  Nucl.\ Phys.\ A {\bf 803}, 227 (2008).
  
\bibitem{Fukushima:2008xe}
  K.~Fukushima, D.~E.~Kharzeev and H.~J.~Warringa,
  ``The Chiral Magnetic Effect,''
  Phys.\ Rev.\ D {\bf 78}, 074033 (2008).

\bibitem{Son:2004tq}
  D.~T.~Son and A.~R.~Zhitnitsky,
  ``Quantum anomalies in dense matter,''
  Phys.\ Rev.\ D {\bf 70}, 074018 (2004).
  
  \bibitem{Metlitski:2005pr}
  M.~A.~Metlitski and A.~R.~Zhitnitsky,
  ``Anomalous axion interactions and topological currents in dense matter,''
  Phys.\ Rev.\ D {\bf 72}, 045011 (2005).


\bibitem{Son:2009tf} 
  D.~T.~Son and P.~Surowka,
  ``Hydrodynamics with Triangle Anomalies,''
  Phys.\ Rev.\ Lett.\  {\bf 103}, 191601 (2009).
  
\bibitem{Erdmenger:2008rm} 
  J.~Erdmenger, M.~Haack, M.~Kaminski and A.~Yarom,
  ``Fluid dynamics of R-charged black holes,''
  JHEP {\bf 0901}, 055 (2009).
  
\bibitem{Banerjee:2008th} 
  N.~Banerjee, J.~Bhattacharya, S.~Bhattacharyya, S.~Dutta, R.~Loganayagam and P.~Surowka,
  ``Hydrodynamics from charged black branes,''
  JHEP {\bf 1101}, 094 (2011).
  
  
  
\bibitem{Kharzeev:2009pj} 
  D.~E.~Kharzeev and H.~J.~Warringa,
  ``Chiral Magnetic conductivity,''
  Phys.\ Rev.\ D {\bf 80}, 034028 (2009).
  
\bibitem{Landsteiner:2011cp} 
  K.~Landsteiner, E.~Megias and F.~Pena-Benitez,
  ``Gravitational Anomaly and Transport,''
  Phys.\ Rev.\ Lett.\  {\bf 107}, 021601 (2011).
  
\bibitem{Golkar:2012kb} 
  S.~Golkar and D.~T.~Son,
  ``(Non)-renormalization of the chiral vortical effect coefficient,''
  JHEP {\bf 1502}, 169 (2015).
  

  
\bibitem{Satow:2014lva} 
  D.~Satow and H.~U.~Yee,
  ``Chiral Magnetic Effect at Weak Coupling with Relaxation Dynamics,''
  Phys.\ Rev.\ D {\bf 90}, no. 1, 014027 (2014).


\bibitem{Yee:2009vw} 
  H.~-U.~Yee,
  ``Holographic Chiral Magnetic Conductivity,''
  JHEP {\bf 0911}, 085 (2009).

\bibitem{Rebhan:2009vc} 
  A.~Rebhan, A.~Schmitt and S.~A.~Stricker,
  ``Anomalies and the chiral magnetic effect in the Sakai-Sugimoto model,''
  JHEP {\bf 1001}, 026 (2010).
  
\bibitem{Gynther:2010ed} 
  A.~Gynther, K.~Landsteiner, F.~Pena-Benitez and A.~Rebhan,
  ``Holographic Anomalous Conductivities and the Chiral Magnetic Effect,''
  JHEP {\bf 1102}, 110 (2011).
  
\bibitem{Hoyos:2011us} 
  C.~Hoyos, T.~Nishioka and A.~O'Bannon,
  ``A Chiral Magnetic Effect from AdS/CFT with Flavor,''
  JHEP {\bf 1110}, 084 (2011).
  
    
\bibitem{Amado:2011zx} 
  I.~Amado, K.~Landsteiner and F.~Pena-Benitez,
  ``Anomalous transport coefficients from Kubo formulas in Holography,''
  JHEP {\bf 1105}, 081 (2011).
  
  
\bibitem{Buividovich:2009wi} 
  P.~V.~Buividovich, M.~N.~Chernodub, E.~V.~Luschevskaya and M.~I.~Polikarpov,
  ``Numerical evidence of chiral magnetic effect in lattice gauge theory,''
  Phys.\ Rev.\ D {\bf 80}, 054503 (2009).

\bibitem{Abramczyk:2009gb} 
  M.~Abramczyk, T.~Blum, G.~Petropoulos and R.~Zhou,
  ``Chiral magnetic effect in 2+1 flavor QCD+QED,''
  PoS LAT {\bf 2009}, 181 (2009).
  
\bibitem{Yamamoto:2011gk} 
  A.~Yamamoto,
  ``Chiral magnetic effect in lattice QCD with a chiral chemical potential,''
  Phys.\ Rev.\ Lett.\  {\bf 107}, 031601 (2011).
  
\bibitem{Buividovich:2013hza} 
  P.~V.~Buividovich,
  ``Anomalous transport with overlap fermions,''
  Nucl.\ Phys.\ A {\bf 925}, 218 (2014).

\bibitem{Bali:2014vja} 
  G.~S.~Bali, F.~Bruckmann, G.~Endr�di, Z.~Fodor, S.~D.~Katz and A.~Sch�fer,
  ``Local CP-violation and electric charge separation by magnetic fields from lattice QCD,''
  JHEP {\bf 1404}, 129 (2014).

\bibitem{Abelev:2009ac} 
  B.~I.~Abelev {\it et al.}  [STAR Collaboration],
  ``Azimuthal Charged-Particle Correlations and Possible Local Strong Parity Violation,''
  Phys.\ Rev.\ Lett.\  {\bf 103}, 251601 (2009).


  
\bibitem{Wang:2012qs} 
  G.~Wang [STAR Collaboration],
  ``Search for Chiral Magnetic Effects in High-Energy Nuclear Collisions,''
  Nucl.\ Phys.\ A904-905 {\bf 2013}, 248c (2013).
  
    
\bibitem{Ke:2012qb} 
  H.~Ke [STAR Collaboration],
  ``Charge asymmetry dependency of $\pi^{+}/\pi^{-}$ elliptic flow in Au + Au collisions at $\sqrt{s_{NN}}$ = 200 GeV,''
  J.\ Phys.\ Conf.\ Ser.\  {\bf 389}, 012035 (2012).
  
\bibitem{Shou:2014cua} 
  Q.~-Y.~Shou {\it et al.}  [the STAR Collaboration],
  ``Charge asymmetry dependency of �/K anisotropic flow in U+U  = 193 GeV and Au+Au   =  200 GeV collisions at STAR,''
  J.\ Phys.\ Conf.\ Ser.\  {\bf 509}, 012033 (2014).



\bibitem{Adamczyk:2015kwa} 
  L.~Adamczyk {\it et al.}  [STAR Collaboration],
  ``Observation of charge asymmetry dependence of pion elliptic flow and the possible chiral magnetic wave in heavy-ion collisions,''
  arXiv:1504.02175 [nucl-ex].


  
\bibitem{Selyuzhenkov:2011xq} 
  I.~Selyuzhenkov [ALICE Collaboration],
  ``Anisotropic flow and other collective phenomena measured in Pb-Pb collisions with ALICE at the LHC,''
  Prog.\ Theor.\ Phys.\ Suppl.\  {\bf 193}, 153 (2012).
  
  
\bibitem{Kharzeev:2010gr} 
  D.~E.~Kharzeev and D.~T.~Son,
  ``Testing the chiral magnetic and chiral vortical effects in heavy ion collisions,''
  Phys.\ Rev.\ Lett.\  {\bf 106}, 062301 (2011).
  
\bibitem{Jiang:2015cva} 
  Y.~Jiang, X.~G.~Huang and J.~Liao,
  ``Chiral vortical wave and induced flavor charge transport in a rotating quark-gluon plasma,''
  arXiv:1504.03201 [hep-ph].
  
\bibitem{Kharzeev:2010gd} 
  D.~E.~Kharzeev and H.~-U.~Yee,
  ``Chiral Magnetic Wave,''
  Phys.\ Rev.\ D {\bf 83}, 085007 (2011).
  
  
\bibitem{Newman:2005hd} 
  G.~M.~Newman,
  ``Anomalous hydrodynamics,''
  JHEP {\bf 0601}, 158 (2006).
  
  
\bibitem{Burnier:2011bf} 
  Y.~Burnier, D.~E.~Kharzeev, J.~Liao and H.~-U.~Yee,
  ``Chiral magnetic wave at finite baryon density and the electric quadrupole moment of quark-gluon plasma in heavy ion collisions,''
  Phys.\ Rev.\ Lett.\  {\bf 107}, 052303 (2011).
  
\bibitem{Gorbar:2011ya} 
  E.~V.~Gorbar, V.~A.~Miransky and I.~A.~Shovkovy,
  ``Normal ground state of dense relativistic matter in a magnetic field,''
  Phys.\ Rev.\ D {\bf 83}, 085003 (2011).
  
\bibitem{Yee:2013cya} 
  H.~-U.~Yee and Y.~Yin,
  ``Realistic Implementation of Chiral Magnetic Wave in Heavy Ion Collisions,''
  Phys.\ Rev.\ C {\bf 89}, 044909 (2014).




\bibitem{Li:2014bha} 
  Q.~Li, D.~E.~Kharzeev, C.~Zhang, Y.~Huang, I.~Pletikosic, A.~V.~Fedorov, R.~D.~Zhong and J.~A.~Schneeloch {\it et al.},
  ``Observation of the chiral magnetic effect in ZrTe5,''
  arXiv:1412.6543 [cond-mat.str-el].
  
\bibitem{Kharzeev:2011ds} 
  D.~E.~Kharzeev and H.~-U.~Yee,
  ``Anomalies and time reversal invariance in relativistic hydrodynamics: the second order and higher dimensional formulations,''
  Phys.\ Rev.\ D {\bf 84}, 045025 (2011).
  
\bibitem{Loganayagam:2011mu} 
  R.~Loganayagam,
  ``Anomaly Induced Transport in Arbitrary Dimensions,''
  arXiv:1106.0277 [hep-th].
  
\bibitem{Loganayagam:2012pz} 
  R.~Loganayagam and P.~Surowka,
  ``Anomaly/Transport in an Ideal Weyl gas,''
  JHEP {\bf 1204}, 097 (2012).
  
\bibitem{Yee:2014dxa} 
  H.~U.~Yee,
  ``Chiral Magnetic and Vortical Effects in Higher Dimensions at Weak Coupling,''
  Phys.\ Rev.\ D {\bf 90}, no. 6, 065021 (2014).
  
\bibitem{Bhattacharyya:2013ida} 
  S.~Bhattacharyya, J.~R.~David and S.~Thakur,
  ``Second order transport from anomalies,''
  JHEP {\bf 1401}, 010 (2014).
  
\bibitem{Megias:2013joa} 
  E.~Megias and F.~Pena-Benitez,
  ``Holographic Gravitational Anomaly in First and Second Order Hydrodynamics,''
  JHEP {\bf 1305}, 115 (2013).
  
  
\bibitem{Son:2012wh} 
  D.~T.~Son and N.~Yamamoto,
  ``Berry Curvature, Triangle Anomalies, and the Chiral Magnetic Effect in Fermi Liquids,''
  Phys.\ Rev.\ Lett.\  {\bf 109}, 181602 (2012).
  
\bibitem{Stephanov:2012ki} 
  M.~A.~Stephanov and Y.~Yin,
  ``Chiral Kinetic Theory,''
  Phys.\ Rev.\ Lett.\  {\bf 109}, 162001 (2012).
  
\bibitem{Gao:2012ix} 
  J.~-H.~Gao, Z.~-T.~Liang, S.~Pu, Q.~Wang and X.~-N.~Wang,
  ``Chiral Anomaly and Local Polarization Effect from Quantum Kinetic Approach,''
  Phys.\ Rev.\ Lett.\  {\bf 109}, 232301 (2012).
  
  
\bibitem{Chen:2014cla} 
  J.~Y.~Chen, D.~T.~Son, M.~A.~Stephanov, H.~U.~Yee and Y.~Yin,
  ``Lorentz Invariance in Chiral Kinetic Theory,''
  Phys.\ Rev.\ Lett.\  {\bf 113}, no. 18, 182302 (2014).
  
\bibitem{Chen:2015gta} 
  J.~Y.~Chen, D.~T.~Son and M.~A.~Stephanov,
  ``Collisions in Chiral Kinetic Theory,''
  arXiv:1502.06966 [hep-th].
  
\bibitem{Baym:1990uj} 
  G.~Baym, H.~Monien, C.~J.~Pethick and D.~G.~Ravenhall,
  ``Transverse Interactions and Transport in Relativistic Quark - Gluon and Electromagnetic Plasmas,''
  Phys.\ Rev.\ Lett.\  {\bf 64}, 1867 (1990).
  
\bibitem{Arnold:2000dr} 
  P.~B.~Arnold, G.~D.~Moore and L.~G.~Yaffe,
  ``Transport coefficients in high temperature gauge theories. 1. Leading log results,''
  JHEP {\bf 0011}, 001 (2000).
  
  
\bibitem{Huang:2013iia} 
  X.~G.~Huang and J.~Liao,
  ``Axial Current Generation from Electric Field: Chiral Electric Separation Effect,''
  Phys.\ Rev.\ Lett.\  {\bf 110}, no. 23, 232302 (2013).
  
\bibitem{Jeon:1994if} 
  S.~Jeon,
  ``Hydrodynamic transport coefficients in relativistic scalar field theory,''
  Phys.\ Rev.\ D {\bf 52}, 3591 (1995).
  
  
\bibitem{ValleBasagoiti:2002ir} 
  M.~A.~Valle Basagoiti,
  ``Transport coefficients and ladder summation in hot gauge theories,''
  Phys.\ Rev.\ D {\bf 66}, 045005 (2002).
  
\bibitem{Aarts:2002tn} 
  G.~Aarts and J.~M.~Martinez Resco,
  ``Ward identity and electrical conductivity in hot QED,''
  JHEP {\bf 0211}, 022 (2002).
  
\bibitem{Gagnon:2006hi} 
  J.~-S.~Gagnon and S.~Jeon,
  ``Leading order calculation of electric conductivity in hot quantum electrodynamics from diagrammatic methods,''
  Phys.\ Rev.\ D {\bf 75}, 025014 (2007)
  [Erratum-ibid.\ D {\bf 76}, 089902 (2007)].
  
  
\bibitem{Pisarski:1988vd} 
  R.~D.~Pisarski,
  ``Scattering Amplitudes in Hot Gauge Theories,''
  Phys.\ Rev.\ Lett.\  {\bf 63}, 1129 (1989).
  
\bibitem{Braaten:1989kk} 
  E.~Braaten and R.~D.~Pisarski,
  ``Resummation and Gauge Invariance of the Gluon Damping Rate in Hot QCD,''
  Phys.\ Rev.\ Lett.\  {\bf 64}, 1338 (1990).
  
  
\bibitem{Selikhov:1993ns} 
  A.~Selikhov and M.~Gyulassy,
  ``Color diffusion and conductivity in a quark - gluon plasma,''
  Phys.\ Lett.\ B {\bf 316}, 373 (1993).
  
\bibitem{Bodeker:1998hm} 
  D.~Bodeker,
  ``On the effective dynamics of soft nonAbelian gauge fields at finite temperature,''
  Phys.\ Lett.\ B {\bf 426}, 351 (1998).
  
\bibitem{Arnold:1998cy} 
  P.~B.~Arnold, D.~T.~Son and L.~G.~Yaffe,
  ``Effective dynamics of hot, soft nonAbelian gauge fields. Color conductivity and log(1/alpha) effects,''
  Phys.\ Rev.\ D {\bf 59}, 105020 (1999).
  add color conductivity references here
  

  
\bibitem{Matsuo:2009xn} 
  Y.~Matsuo, S.~J.~Sin, S.~Takeuchi and T.~Tsukioka,
  ``Magnetic conductivity and Chern-Simons Term in Holographic Hydrodynamics of Charged AdS Black Hole,''
  JHEP {\bf 1004}, 071 (2010).
  
\bibitem{Sahoo:2009yq} 
  B.~Sahoo and H.~U.~Yee,
  ``Holographic chiral shear waves from anomaly,''
  Phys.\ Lett.\ B {\bf 689}, 206 (2010).
  
\bibitem{Blaizot:2001nr} 
  J.~P.~Blaizot and E.~Iancu,
  ``The Quark gluon plasma: Collective dynamics and hard thermal loops,''
  Phys.\ Rept.\  {\bf 359}, 355 (2002).
  
  
\bibitem{MartinezResco:2000pz} 
  J.~M.~Martinez Resco and M.~A.~Valle Basagoiti,
  ``Color conductivity and ladder summation in hot QCD,''
  Phys.\ Rev.\ D {\bf 63}, 056008 (2001).
  
\bibitem{Akamatsu:2014yza} 
  Y.~Akamatsu and N.~Yamamoto,
  ``Chiral Langevin theory for non-Abelian plasmas,''
  Phys.\ Rev.\ D {\bf 90}, no. 12, 125031 (2014).
  
  
  
\bibitem{Son:2012zy} 
  D.~T.~Son and N.~Yamamoto,
  ``Kinetic theory with Berry curvature from quantum field theories,''
  Phys.\ Rev.\ D {\bf 87}, no. 8, 085016 (2013).
  
\bibitem{Manuel:2013zaa} 
  C.~Manuel and J.~M.~Torres-Rincon,
  ``Kinetic theory of chiral relativistic plasmas and energy density of their gauge collective excitations,''
  Phys.\ Rev.\ D {\bf 89}, 096002 (2014).

\bibitem{Blaizot:1996hd} 
  J.~P.~Blaizot and E.~Iancu,
  ``Lifetime of quasiparticles in hot QED plasmas,''
  Phys.\ Rev.\ Lett.\  {\bf 76}, 3080 (1996).
  
    
\end{thebibliography}
\end{document}